\documentclass[journal,comsoc]{IEEEtran}
\usepackage[dvipsnames]{xcolor}
\usepackage[T1]{fontenc}
\usepackage{amsmath}
\usepackage[cmintegrals]{newtxmath}
\usepackage{cite}
\usepackage[breaklinks,colorlinks=true,citecolor=blue,pdftex]{hyperref}
\usepackage{url}
\usepackage{bm}
\usepackage{graphicx}
\usepackage{lipsum,adjustbox}
\usepackage{tikz}
\usetikzlibrary{trees}
\usetikzlibrary{positioning}
\usetikzlibrary{arrows.meta}
\usetikzlibrary{decorations.markings}
\usepackage{ragged2e}
\usepackage{makecell}
\usepackage{array}
\usepackage{subfigure}
\usepackage{multirow}

\tikzstyle{every node}=[draw=black,thick,anchor=west]
\tikzstyle{level 2}=[draw=black]
\tikzstyle{level 2.1}=[draw=black]
\tikzstyle{level 3}=[draw=black]
\tikzstyle{level 0}=[draw=black]

\newcommand{\techTochallenge}[2]{
    \raisebox{-0.3\totalheight}{\begin{tikzpicture}[font=\normalfont,box/.style={draw,rounded corners,text width=4cm,align=center},border/.style={draw,dashed,rounded corners,text width=4.5cm},numred/.style={draw=none,color=white,fill=Red,rounded corners},numblue/.style={draw=none,color=white,fill=Blue,rounded corners},numblack/.style={draw=none,color=white,fill=Green,rounded corners}]
    \node[numblack] at (0,0) (a) {#1};
    \node[numred] at (1,0) (b) {#2};
    \draw[->,line width = 1pt] (a.east) -- (b.west);
    \end{tikzpicture}}
}

\newcommand{\techToobjective}[2]{
    \raisebox{-0.3\totalheight}{\begin{tikzpicture}[font=\normalfont,box/.style={draw,rounded corners,text width=4cm,align=center},border/.style={draw,dashed,rounded corners,text width=4.5cm},numred/.style={draw=none,color=white,fill=Red,rounded corners},numblue/.style={draw=none,color=white,fill=Blue,rounded corners},numblack/.style={draw=none,color=white,fill=Green,rounded corners}]
    \node[numblack] at (0,0) (a) {#1};
    \node[numblue] at (1,0) (b) {#2};
    \draw[->,line width = 1pt] (a.east) -- (b.west);
    \end{tikzpicture}}
}

\begin{document}

\title{Datacenter Traffic Control:\\ Understanding Techniques and Trade-offs}

\author{
    Mohammad~Noormohammadpour~~~~~~~~~Cauligi~S.~Raghavendra\\
    \textit{noormoha@alumni.usc.edu~~~~~~~~~~~~~~~~~~raghu@usc.edu}\\
    Ming~Hsieh~Department~of~Electrical~Engineering\\
    University~of~Southern~California\\
    Los Angeles, CA 90089
}

\markboth{Datacenter Traffic Control Techniques and Trade-offs}{}

\maketitle

\begin{abstract}
Datacenters provide cost-effective and flexible access to scalable compute and storage resources necessary for today's cloud computing needs. A typical datacenter is made up of thousands of servers connected with a large network and usually managed by one operator. To provide quality access to the variety of applications and services hosted on datacenters and maximize performance, it deems necessary to use datacenter networks effectively and efficiently. Datacenter traffic is often a mix of several classes with different priorities and requirements. This includes user-generated interactive traffic, traffic with deadlines, and long-running traffic. To this end, custom transport protocols and traffic management techniques have been developed to improve datacenter network performance.

In this tutorial paper, we review the general architecture of datacenter networks, various topologies proposed for them, their traffic properties, general traffic control challenges in datacenters and general traffic control objectives. The purpose of this paper is to bring out the important characteristics of traffic control in datacenters and not to survey all existing solutions (as it is virtually impossible due to massive body of existing research). We hope to provide readers with a wide range of options and factors while considering a variety of traffic control mechanisms. We discuss various characteristics of datacenter traffic control including management schemes, transmission control, traffic shaping, prioritization, load balancing, multipathing, and traffic scheduling. Next, we point to several open challenges as well as new and interesting networking paradigms. At the end of this paper, we briefly review inter-datacenter networks that connect geographically dispersed datacenters which have been receiving increasing attention recently and pose interesting and novel research problems.

\end{abstract}


\IEEEpeerreviewmaketitle

\section{Introduction} \label{intro}
Datacenters provide an infrastructure for many online services such as on-demand video delivery, storage and file sharing, web search, social networks, cloud computing, financial services, recommendation systems, and interactive online tools. Such services may dynamically scale across a datacenter according to demands enabling cost-savings for service providers. Moreover, considering some degree of statistical multiplexing, better resource utilization can be achieved by allowing many services and applications to share datacenter infrastructure. To reduce costs of building and maintaining datacenters, numerous customers rely on infrastructure provided by large cloud companies \cite{google, azure, aws} with datacenters consisting of hundreds of thousands of servers.

\begin{figure}[t]
	\centering
	\includegraphics[width=\columnwidth]{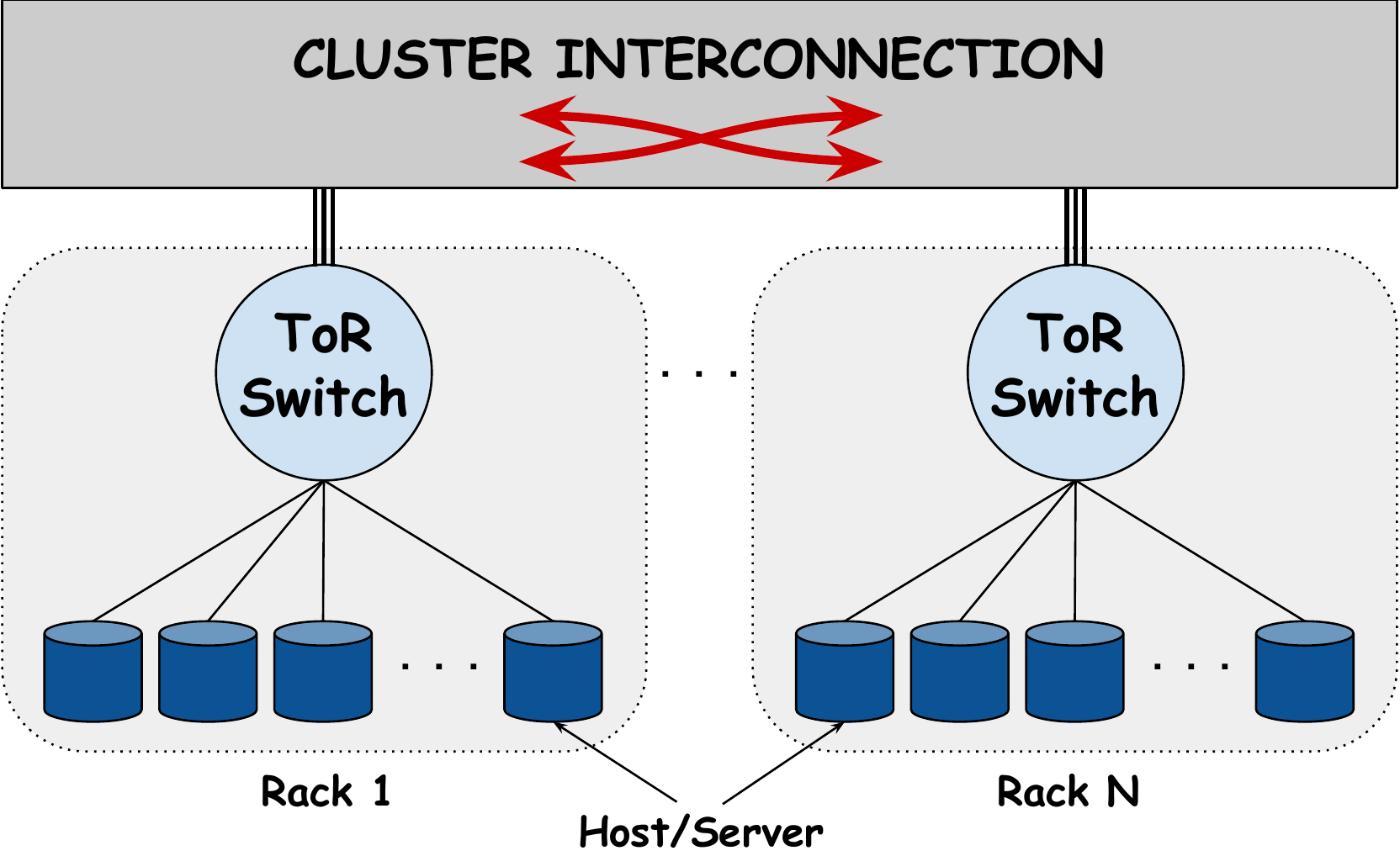}
	\caption{A typical datacenter cluster}
	\label{fig:datacenter-design}
\end{figure}

Figure \ref{fig:datacenter-design} shows the structure of a typical datacenter cluster network with many racks. A datacenter often hosts multiple such clusters with thousands of machines per cluster. A cluster is usually made up of up to hundreds of racks \cite{jupiter, vl2, server-per-rack-facebook}. A rack is essentially a group of machines which can communicate at line rate with minimum number of hops. All the machines in a rack are connected to a \textbf{Top of Rack (ToR)} switch which provides non-blocking connectivity among them. Rack size is typically limited by maximum number of ports that ToR switches provide and the ratio of downlink to uplink bandwidth. There is usually about tens of machines per rack \cite{jupiter, vl2, server-per-rack-facebook}. ToR switches are then connected via a large interconnection allowing machines to communicate across racks. An ideal network should act as a huge non-blocking switch to which all servers are directly connected allowing them to simultaneously communicate with maximum rate. 

Datacenter network topology plays a significant role in determining the level of failure resiliency, ease of incremental expansion, communication bandwidth and latency. The aim is to build a robust network that provides low latency, typically up to hundreds of microseconds \cite{pitfall, dcqcn, hull}, and high bandwidth across servers. Many network designs have been proposed for datacenters \cite{fattree, vl2, hyperx, dcell, leaf-spine, xpander, fbtopology, jellyfish}. These networks often come with a large degree of path redundancy which allows for increased fault tolerance. Also, to reduce deployment costs, some topologies scale into large networks by connecting many inexpensive switches to achieve the desired aggregate capacity and number of machines \cite{clos, jupiter}. Although the majority of these topologies are symmetrical, in practice, datacenter networks turn out to be often a-symmetrical due to frequent failures of network elements (switches, links, ports, etc.) \cite{wcmp, f10, evolve}. In contrast to fixed networks, reconfigurable topologies involve optical circuit switching, wireless or a combination of both to adapt to traffic demands \cite{circuit_dc, firefly, tm_circuit, projector}. These topologies rely on fast algorithms that take into account the reconfiguration latency.

Many applications may need to span over multiple racks to access required volume of resources (storage, compute, etc.). This increases the overall volume of traffic across racks. A datacenter network with full bisection bandwidth allows for flexible operation and placement of applications across clusters and improves overall resource utilization and on-demand scale out for applications \cite{jupiter, fattree, vl2, fbtopology}. This allows resources of any machine to be used by any application which is essential for hyper-scale cloud providers \cite{google, azure, aws}.

Designing networks for full bisection bandwidth is costly and unnecessary for smaller companies or enterprises. As a result, many datacenters may be over-subscribed, meaning the total inter-rack network capacity may be less than sum of intra-rack capacities across all racks. The underlying assumption is that applications are mostly rack local. Increasing the over-subscription ratio affects performance of different topologies differently. For instance, over-subscribed Fat-Trees provide less flexible communication across machines compared to over-subscribed Jellyfish networks \cite{fatfree}.

There is growing demand for datacenter network bandwidth. This is driven by faster storage devices, rising volume of user and application data, reduced cost of cloud services and ease of access to cloud services. Google reports 100\% increase in their datacenter networking demands every 12 to 15 months \cite{jupiter}. Cisco forecasts a 400\% increase in global datacenter IP traffic and 2.6 times growth in global datacenter workloads from 2015 to 2020 \cite{cisco-growth}.

Applications running on datacenters determine their traffic characteristics and the communication patterns across machines. Some popular applications include web services, cache followers, file stores, key-value stores, data mining, search indexing and web search. Many datacenters, especially cloud providers, run a variety of applications that results in a spectrum of workloads. Some applications generate lots of internal datacenter traffic, such as scatter-gather (also known as partition-aggregate) \cite{nature,d2tcp, dctcp, detail} and batch computing tasks \cite{mapred, dryad}. As a result, the total traffic volume within a datacenter is often much more than that of entering or leaving it. Cisco reports this ratio to be greater than 3 which is expected to increase further by 2020 \cite{cisco-growth}. 
Traffic control is necessary to highly utilize network bandwidth, keep latency low, offer quality of service, and fairly share resources among many users and applications by managing flow of traffic across the datacenter. There is a significant body of work on traffic control for datacenters. In this tutorial paper, we aim to review concepts in design and operation of traffic control schemes. 

The rest of this paper is organized as follows. In \S \ref{related-works}, we present some related works. In \S \ref{datacenter-networks}, we review a variety of datacenter topologies, provide an overview of datacenter traffic patterns, set forth the challenges of traffic control for datacenter networks, and the objectives of datacenter traffic control. In \S \ref{tc-management}, we review management schemes that can be applied to datacenters and point to some research work that use different management schemes. Next, in \S \ref{traffic-control-all}, we present a variety of traffic control techniques, discuss them in detail and explore the benefits and drawbacks associated with them. In \S \ref{open-challenges}, we discuss some general open traffic control problems. In \S \ref{related-paradigms}, we point to rising paradigms related to datacenters. In \S \ref{perspective}, we introduce a new research area that is a result of global distribution of datacenters. Finally, in \S \ref{conclusion}, we conclude the paper.

\section{Related Works} \label{related-works}
In this section, we briefly present some survey articles related to datacenter traffic control. In \cite{survey-lowlatency-tcp}, authors provide a short survey of low latency datacenter networking by reviewing approaches taken to achieve low latency, namely reducing queuing delay, accelerating retransmissions, prioritizing mice flows and utilizing multi-path. In \cite{survey-incast}, authors survey the methods used to address the transport control protocol (TCP) incast problem (please see \S \ref{incast}). In \cite{survey-bandwidth-allocation}, authors survey bandwidth sharing in multi-tenant datacenters using techniques of static reservation, minimum guarantees and no guarantees (resources obtained in a best effort fashion). In \cite{survey-transport-ieee}, authors point out datacenter transport problems namely TCP incast, latency, challenges in virtualized environments, bandwidth sharing in multi-tenant environments, under-utilization of bandwidth, and TCP in lossless Ethernet environments also known as Converged Enhanced Ethernet (CEE). In \cite{survey-tcp-india}, authors discuss TCP issues in datacenters pointing to TCP incast, queue buildup and buffer pressure. In \cite{survey-dcn}, authors provide a comprehensive overview of datacenter networking for cloud computing discussing cloud computing network architectures, communication technologies used, topologies, routing and virtualization approaches. In \cite{survey-congestion-india}, authors discuss various congestion notifications for datacenter networks. Finally, in \cite{survey-transport-springer}, authors survey transport protocols proposed for datacenter networks and briefly explain each one of the variety of research efforts made in addressing the incast problem, outcast problem and in reducing latency for datacenter networks.

In this tutorial paper, we merely focus on traffic control concepts that can be applied to a variety of transport protocols including TCP. We also point to research efforts that use different techniques as examples so that readers can elevate their knowledge in case they are interested in further details. We try to uncover the (not obvious) trade-offs in terms of complexity, performance and costs. This paper is different from prior work in that it covers a variety of aspects in traffic control in a conceptual way while not focusing on any specific transport, network or data link layer protocol. In addition, we provide an overview of datacenter traffic properties, topologies as well as traffic control challenges, objectives and techniques and their relationships which has not been done in prior work to our knowledge. Finally, at the end of this paper, we point to a recent research direction that involves inter-datacenter networks and offer three areas that demand further attention from the research community.


\section{Datacenter Networks} \label{datacenter-networks}
In this section, we dive deeper into specifics of datacenter networks. We review a variety of topologies proposed and general traffic properties of datacenter networks, point to traffic control challenges in datacenters and explain several traffic control objectives sought by various parties (i.e., operators, tenants, etc.) involved in using datacenter networks.

\subsection{Topologies}
\begin{figure*}
\centering
\hspace{-3em}\subfigure[Fat-Tree \cite{fattree}]{\includegraphics[height = 2.4in]{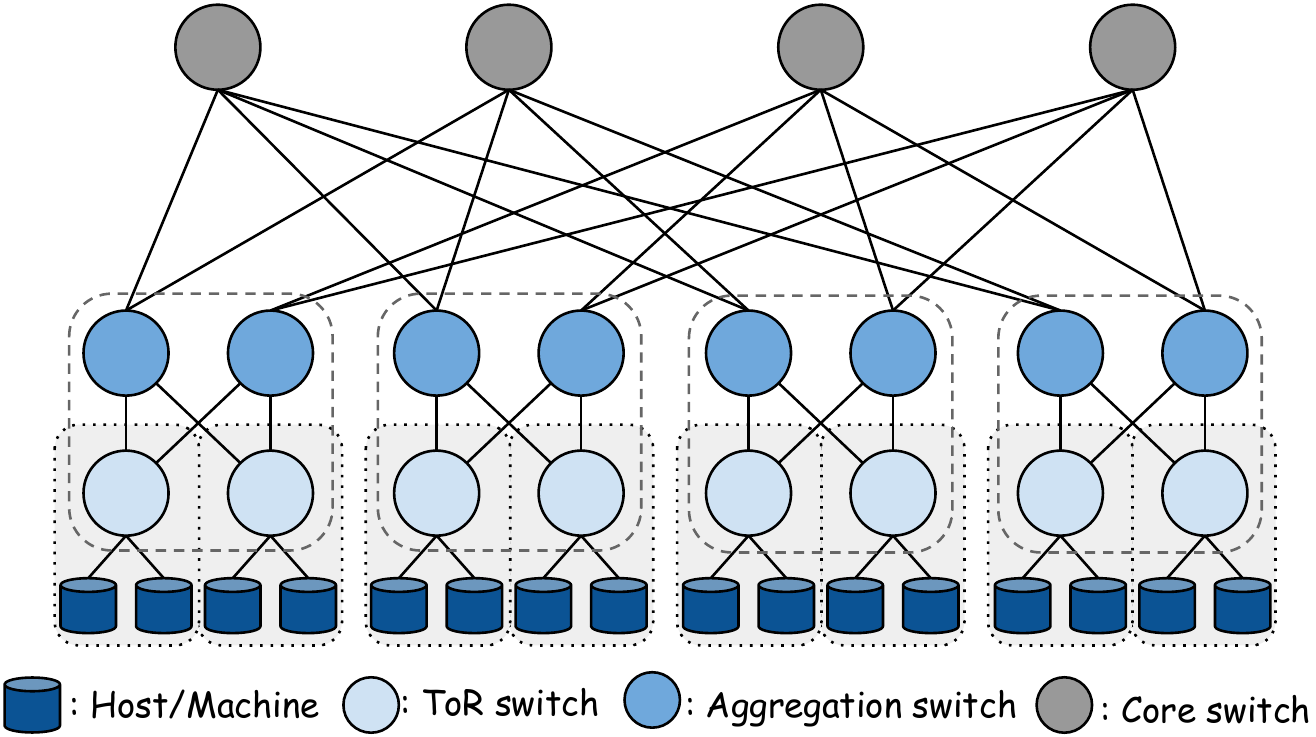} \label{fattree}}\hspace{6em}%
\subfigure[BCube \cite{bcube}]{\includegraphics[height = 2.4in]{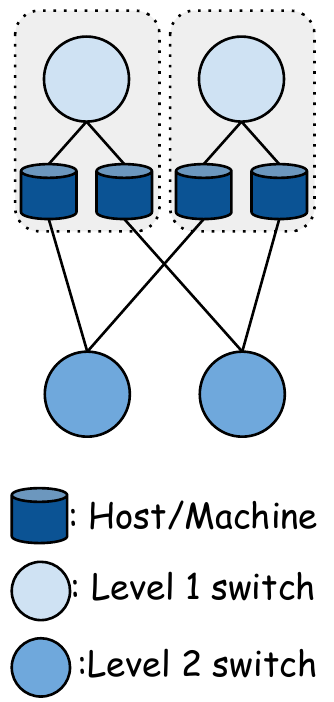} \label{bcube}} \vspace{2em}\\
\subfigure[VL2 \cite{vl2}]{\includegraphics[height = 2.4in]{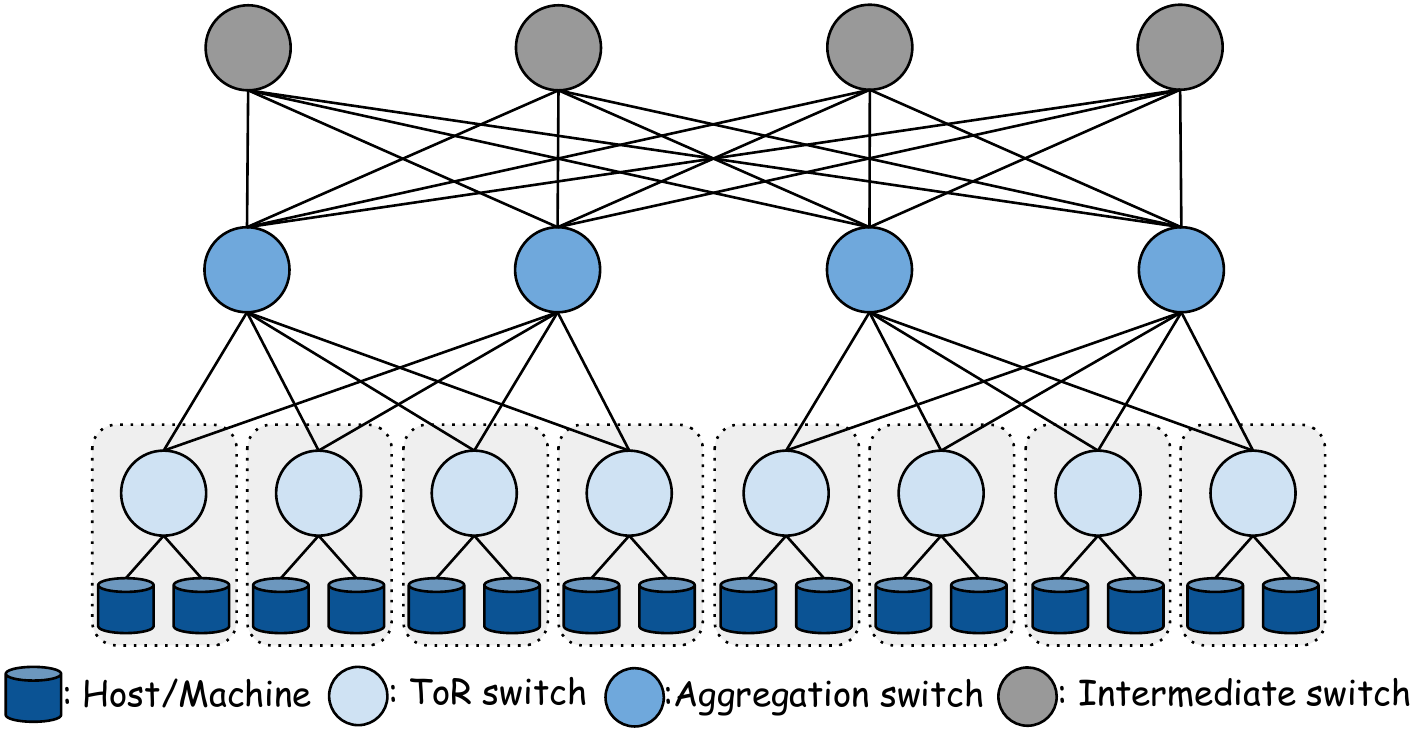} \label{vl2}}\hspace{1em}%
\subfigure[DCell \cite{dcell}]{\includegraphics[height = 2.4in]{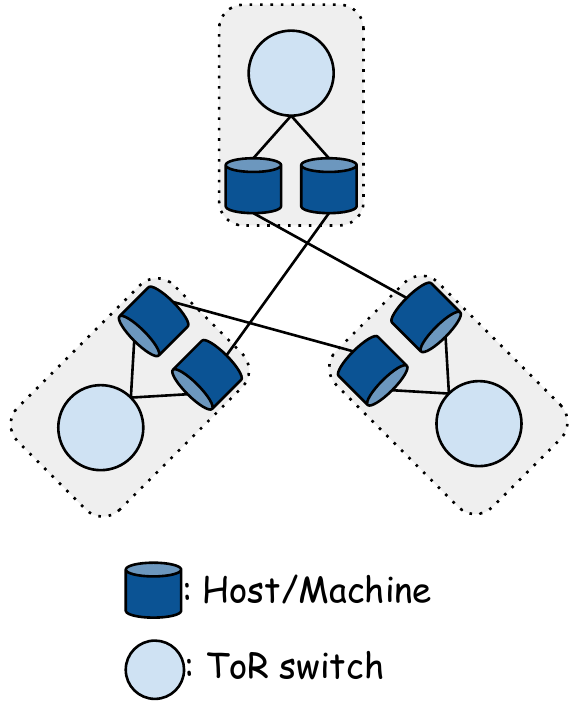} \label{dcell}} \vspace{2em}\\
\subfigure[JellyFish \cite{jellyfish}]{\includegraphics[height = 2.2in]{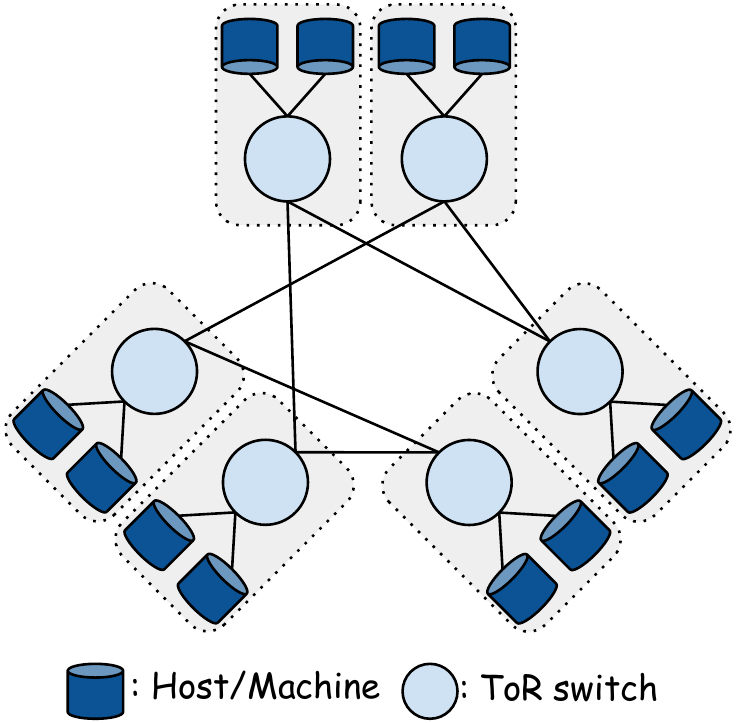} \label{jellyfish}}\hspace{1em}%
\subfigure[Xpander \cite{xpander}]{\includegraphics[height = 2.2in]{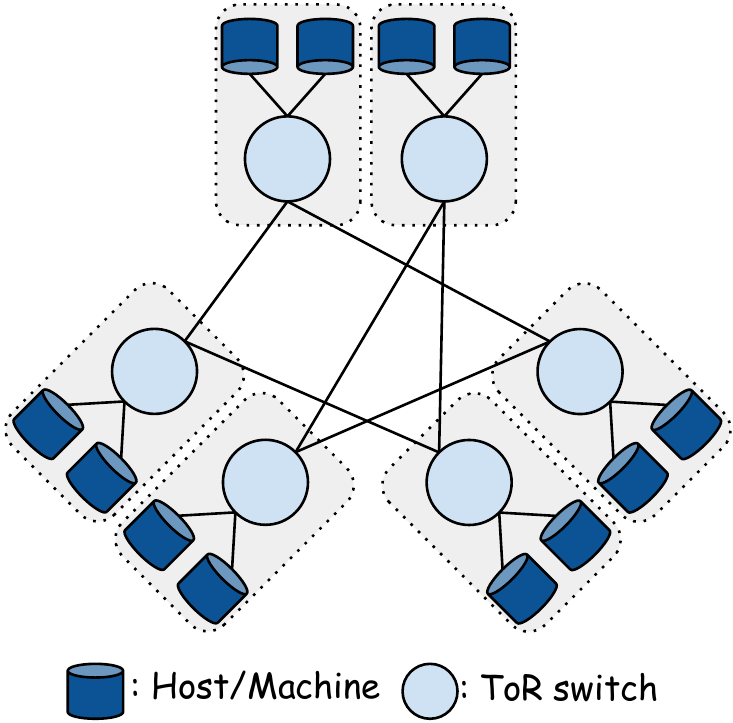} \label{xpander}}\hspace{1em}%
\subfigure[Leaf-Spine \cite{leaf-spine}]{\includegraphics[height = 2.2in]{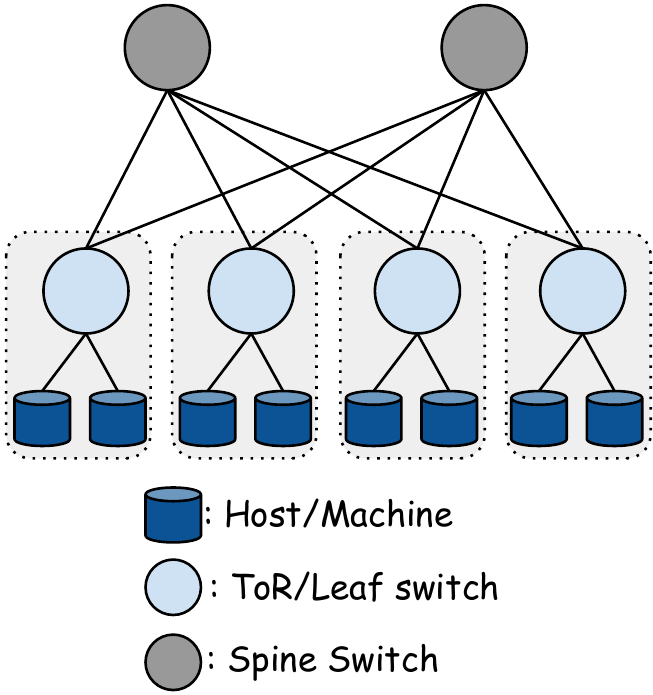} \label{leaf-spine}}%
\caption{Several examples of datacenter topologies with two machines per rack} \label{fig:intra-datacenter-topologies}
\end{figure*}

We shortly review popular physical datacenter topologies proposed and used in the literature either in testbed or simulation. Figure \ref{fig:intra-datacenter-topologies} shows examples of datacenter network topologies reviewed in the following (notice the network connecting ToR switches).

\subsubsection{Fat-Tree}
Fat-Tree \cite{fattree}, shown in Figure \ref{fattree}, is a multi-rooted tree topology where every root is called a core switch. It can be considered as a folded Clos \cite{clos} network \cite{folded-clos}. By increasing the number of roots, one can reduce the over subscription ratio (considering fixed capacity per link). This topology allows for high bisection bandwidth using a large number of less expensive switches allowing support for a large number of hosts at much less cost. There is an aggregate layer between the core and edge (ToR switches). The number of hops between any two servers attached to the same ToR switch is 2, to the same aggregate switch is 4 and otherwise is 6. A Fat-Tree topology built with $k$-port switches supports up to $\frac{k^3}{4}$ physical servers (assuming one physical port per server) and $\frac{k^2}{4}$ paths between any source and destination pair. As a result, it is possible to scale to huge clusters by interconnecting many inexpensive switches.

To effectively use a Fat-Tree, complex routing configurations may need to be done on switches to avoid creation of loops while using available paths for load balancing. For example, Portland \cite{portland} is a custom routing and forwarding protocol which works out of Layer 2 and improves on fault tolerance (link failures), scalability, and ease of management (e.g. moving VMs across physical servers). Portland uses a Fabric Manager that holds information on address mappings and a fault matrix that maintains per link health status.

\subsubsection{Leaf-Spine}
Leaf-Spine (or Spine-and-Leaf) \cite{leaf-spine-cisco, leaf-spine}, shown in Figure \ref{leaf-spine}, is a two tier network topology where leaf (i.e., ToR) switches are attached to servers and every spine switch is directly connected to all leaf switches similar to a bipartite graph. The links connected between the servers and leaf switches may have a different capacity from the ones connecting leaf switches to the spine. Leaf-Spine makes it easier to expand on capacity and ports (by adding more spine or leaf switches) and also allows straight-forward usage of Layer 3 routing with load balancing support without creation of loops. As a downside, in case high bisection bandwidth is intended, scaling to more than hundreds of servers in a cluster can lead to increased costs due to need for spine switches with many high capacity ports.

\subsubsection{VL2}
VL2 \cite{vl2}, shown in Figure \ref{vl2}, implements a complete routing and forwarding suite on top of 3-tier folded Clos networks (a multi-rooted tree) but differs from Fat-Tree in that switch-to-switch links are assigned much higher capacity than server-to-switch links. This requires less number of cables for connecting the aggregation and core layer. This topology has an intermediate tier that is connected to the aggregation trier in a bipartite graph topology. Each edge (ToR) switch is connected to two aggregation switches in a way that each aggregation switch gets connected to equal number of edge switches.

\subsubsection{JellyFish}
JellyFish \cite{jellyfish}, shown in Figure \ref{jellyfish}, is a topology where ToR switches are connected in a random setting and according to some rules: first, ports are randomly connected between switches until no more links can be added; next, for any switch $S$ with two or more free ports, an existing link $A-B$ is removed and two links are added between two of $S$'s free ports and two ends of the removed network link (i.e., $A-S$ and $B-S$), until no switch has more than one free port \cite{jellyfish-topo}. Since ToR switches are connected directly, the average path length (number of hops) is considerably smaller compared to 3-tier folded Clos. In addition, this topology is much easier to expand gradually. Authors show that with full bisection bandwidth, JellyFish supports more servers at the same cost compared to Fat-Tree. Also, with the same number of failed links, JellyFish offers a higher average throughput per server than Fat-Tree. One major problem with this topology is that the existing routing schemes cannot effectively use all paths since the number of hops across parallel paths changes by a large degree. Authors propose usage of $k$-shortest path routing for flows along with multipath TCP (MPTCP) \cite{mptcp}.

\subsubsection{DCell}
DCell \cite{dcell}, shown in Figure \ref{dcell}, is a hierarchical datacenter topology where a higher level DCell is built by putting together multiple lower level DCell structures. It can be incrementally expanded and does not have a single point of failure. DCell uses a custom routing algorithm that takes into account failures (DCell Fault-tolerant Routing) while aiming for near shortest path routing. Authors show that DCell offers higher network capacity compared to conventional tree topologies; however, it can perform much worse than multi-rooted trees such as Fat-Tree \cite{bcube}. Implementing DCell requires changes to the server networking protocol stack.

\subsubsection{BCube}
BCube \cite{bcube}, shown in Figure \ref{bcube}, is a leveled structure where a higher level is built by recursively attaching lower levels. Servers require multiple network ports to connect to switches and they also act as forwarding elements for other servers. BCube uses source routing and requires changes to the server networking protocol stack which can be done either in hardware (network card) or software. If forwarding is implemented in software, it can add CPU overhead especially at high rates. Authors show that BCube is more resilient to switch failures compared to Fat-Tree and almost as resilient to server failures.

\subsubsection{Xpander}
Xpander \cite{xpander}, shown in Figure \ref{xpander}, is a datacenter topology based on expander graphs \cite{expander-graphs} which offers all the performance improvements of JellyFish over Fat-Tree topology with higher throughput as network is incrementally expanded. This topology is made by connecting multiple meta nodes based on the following rules: first, each meta node is made up of equal number of ToR switches; second, no two ToRs are connected within the same meta node; third, same number of links is used to connect every pair of meta nodes. Compared to JellyFish, Xpander offers the benefit of being structured rather than random which improves implementation predictability. Authors test Xpander topology with similar routing and forwarding techniques used in JellyFish: $k$-shortest paths routing and MPTCP for multi-path load distribution.

\subsection{Traffic Properties}
Traffic characteristics of datacenters is highly dependant on applications and determines distribution of flow arrivals, flow sizes, and flow durations. For example, flows generated by web search queries are usually much smaller and shorter than that of batch computing jobs. This variety of applications could cause creation of long lived connections as well as short microbursts on the same network \cite{twitter}. There are limited publications on datacenter traffic characteristics. We review these works briefly focusing on applications and flow properties.

In \cite{nature}, authors collect a traffic dataset from a cluster running query processing applications on large volumes of data that run MapReduce like workloads under the hood to respond to queries. Authors find that more than 50\% of flows last less than 100~ms, 80\% of flows last less than 10 seconds, and almost all flows last less than 100 seconds. They also find that more than 50\% of traffic is carried by flows that last less than 25 seconds, less than 10\% by flows that last less than 10 seconds, and almost 1\% of traffic by flows that last less than 1 second. In terms of flow arrivals, authors find periodic short-term bursts of flows and periods of long silence.

In another work \cite{traffic_dc_char}, authors collect and process datasets from 19 datacenters with a wide range of workloads including web services, instant messaging and MapReduce. This work is focused on packet level traffic analysis and does not offer statistics on a per-flow basis. Authors observed an ON/OFF pattern for packet arrivals at ToR switches they monitored, meaning there was varying periods with no packet arrivals between bursts of packets.

In \cite{wild}, authors study 10 datacenters categorized as educational (storage, email, web, video), enterprise (custom applications in addition to storage, web and email) and cloud (instant messaging, web, search, indexing, video). They report the number of active flows less than 10000 per second per rack across all datacenters. More than 40\% of flows were less than 1~KB, more than 80\% of were less than 10~KB, and almost all flows were less than 10~MB. According to their results, durations are more dependant on the specific datacenter: the smallest median flow duration was about 500~$\mu$s in one datacenter while the largest median flow duration was 1 second in a different datacenter. The largest flow duration was between 50 seconds and 200 seconds across all datacenters. This work also confirms the ON/OFF behavior of packet arrivals.

In a recent paper \cite{social_inside}, Facebook shares statistics on their traffic characteristics. They report flow size distributions on a per application basis for three major applications they run. Median and tail flow sizes for Hadoop, Web Server and Cache applications are reported to be between about 100~KB and 100~MB, 3~KB and 10~MB, 1~KB and 10~KB within racks while 1~KB and 1~MB, 5~KB and 500~KB, 30~KB and 3~MB between racks, respectively. Regarding flow durations, Hadoop flows had a median of about 300~ms and tail of less than 1000 seconds, Web Server flows had a median of about 900~ms and a tail of about 200 seconds, and Cache flows had a median of almost 400 seconds and a tail of almost 800 seconds, respectively. Per server, the median inter arrival time of various flow types was between 1000~$\mu$s and 10000~$\mu$s and the tail was between 10000~$\mu$s and 100000~$\mu$s. Finally, authors did not observe an ON/OFF packet arrival pattern at the switches which is suggested to be due to a large number of concurrent destinations, since ON/OFF pattern was observed on a per destination basis.

In addition to per-flow properties, since racks constitute a main unit in datacenter networks, one may be interested in how much traffic stays within racks. This could be helpful in deciding the over subscription ratio of cluster interconnection. The ratio of rack local to inter-rack traffic is dependent on applications that run within the datacenter and how instances of such applications are deployed on the servers. As a result, some prior work report a highly rack local traffic pattern \cite{wild} while some find traffic neither rack local nor all to all \cite{social_inside},i.e., for some applications (e.g. Hadoop) traffic is mainly rack local while for others (e.g. Web Server and Cache) traffic is not at all rack local.

In summary, traffic characteristics, such as packet sizes, flow size distributions and flow inter-arrival times are highly correlated with applications. In addition, locality of traffic to racks is highly dependant on the way applications are deployed across the network and how such applications communicate with one another. For example, in \cite{social_inside}, authors report that servers of different types are deployed in separate racks and since most communication occurs between Web Servers and Cache Followers (different server types), there is less rack locality for these clusters. Given such strong dependency on applications, it is relatively hard to draw a general conclusion about datacenter traffic. Some common findings include the following. There is usually several orders of magnitude difference between median and maximum flow sizes (the actual difference varies according to applications). In addition, there can be a large number of flow arrivals per server (as many as thousands per second) with many concurrent flows. Finally, distributions of flow sizes and durations may be considerably different due to possibly uneven distribution of traffic across the network (flows may have to compete for bandwidth) and application of techniques like connection pooling which leads to long-lived connections that are not always transmitting.

\subsection{Traffic Control Challenges} \label{challenges}
We present some datacenter traffic control challenges frequently pointed to in the literature. Table \ref{dc-tc-challenges} provides a summary of these challenges.

\begin{table*}[t!]
\caption{Summary of datacenter traffic control challenges} \label{dc-tc-challenges}
\centering
\begin{tabular}{|p{3.5cm}|p{8cm}|p{5cm}|}
\hline
\textbf{Challenge} & \textbf{Description} & \textbf{Implications} \\
\hline
\hline
Unpredictable Traffic Matrix \S \ref{tm} & A traffic matrix represents the communication volume between pairs of end-points in a computer network. In datacenters, traffic matrix is varying and unpredictable. & Complicates traffic engineering and capacity planning. \\ 
\hline
Mix of Flow Types and Sizes \S \ref{mix} & Due to variety of applications that share the datacenter infrastructure, a mix of various flow types and sizes are generated. Flows may have deadline constraints or not and may be short latency-sensitive or large throughput-oriented. Also, for specific applications, flow sizes may be unknown. & Complicates flow scheduling to meet requirements of different flows over a shared medium. \\ 
\hline
Traffic Burstiness \S \ref{burst} & Burstiness has two different aspects. Traffic per flow could be highly bursty and flow arrival itself could be bursty as well. Burstiness intensity may change according to where traffic is measured, i.e., at end-point interfaces, at ToR switch ports, and so on. & Large buffer space at the switches to absorb bursts, careful and responsive traffic control to minimize average buffer space usage and react to bursts quickly. \\ 
\hline
Packet Reordering \S \ref{reordering} & Can be caused while applying some traffic control schemes to improve load balancing or increase throughput via using parallel paths at the same time. At high rates, reordering can exhaust end-point CPU and memory resources for buffering and putting segments back in order. & Efficient methods to put packets in order at receivers with minimal memory and CPU overhead and careful transmission methods at senders to minimize reordering when packets arrive at the receiver. \\ 
\hline
Performance Isolation \S \ref{isolation} & In cloud environments with many tenants where network resources are shared across tenants, mechanisms should be put in place to make sure tenants' use of resources cannot impact other tenants. & Allocation of bandwidth on a per tenant basis rather than per flow taking into account possibility of selfish or malicious behavior from tenants. \\ 
\hline
The Incast Problem \S \ref{incast} & A variety of applications, such as search, use the partition-aggregate communication pattern which can lead to a large number of incoming flows to end-points. If not properly handled, this can lead to congestion, dropped packets and increased latency. & Larger available switch buffer space, responsive and careful traffic control to keep switch buffer occupancy low and avoid dropped packets. \\ 
\hline
The Outcast Problem \S \ref{outcast} & Is observed in switches that use DropTail queues and when a disproportionate number of flows arrive at different incoming ports and exit the same switch output port. This problem is caused by synchronous dropped packets under high utilization. & Responsive and careful traffic control to keep switch buffer occupancy low and avoid dropped packets. \\ 
\hline
\end{tabular}
\end{table*}

\subsubsection{Unpredictable Traffic Matrix} \label{tm}
A variety of applications usually run on a datacenter creating many flows with different properties. Most datacenter flows are short \footnote{Short flows are usually considered to be less than 1~MB \cite{dctcp, d3, dctcp_better}} (a few packets at most) \cite{dctcp, pfabric, wild, traffic_dc_char, nature, social_inside, vl2, pitfall} and many short flows may be only one packet long \cite{pfabric}. However, most bytes are delivered by large flows \cite{dctcp, vl2, pitfall}. \cite{vl2} reports 80\% of flows to be less than 10~KB and 99\% of flows to be less than 100~MB.

Datacenter applications also determine the flow arrival rates and distributions. The median flow inter-arrival time for a single machine was reported between 2~ms to 10~ms (100 to 500 flows per second) for different servers in Facebook datacenters \cite{social_inside} (this median is calculated by measuring number of arriving flows per machine per second as samples averaged over number of machines). \cite{wild} reports between 100 to 10000 flow arrivals per switch in a one second bin in different educational and private datacenters. Finally, \cite{nature} finds the median arrival rate of flows in a cluster to be 100 flows per millisecond.

High flow arrival rate majority of which being short can create an unpredictable and fluctuating traffic matrix which makes it hard to perform traffic engineering or capacity planning in longer time scales to improve performance \cite{nature, vl2, tm_joint, tm_estimation}.

\subsubsection{Mix of various flow types/sizes} \label{mix}
Traffic in datacenters is a mix of various flow types and sizes \cite{vl2, d3, dctcp, karuna, wild, traffic_dc_char}. Knowledge of various flow requirements and characteristics can help us design and tune transport protocols to more efficiently use network resources. Size and priority of a flow are usually determined by the application that initiates it. For some applications, flow sizes maybe unknown upon initiation.

\textit{Interactive} flows which are created as a result of user interactions (for example generated by soft real time applications such as web search) can generate latency-sensitive flows that are usually short and should be delivered with high priority. Examples include queries (2 to 20~KB) and short messages (100~KB to 1~MB) \cite{dctcp}. Size of these flows is usually known apriori \cite{d3}. Responsiveness of online services depends on how interactive flows are handled which can impact the number of users for an online service in the long run. In a study by Google, 400~ms increase in delay reduced the number of searches by 0.6\% per user per day \cite{speed-google}. Also, 100~ms added latency could reduce Amazon sales by 1\% \cite{amazon_100_ms}.

\textit{Throughput-oriented} flows are not as sensitive to delay, but need consistent bandwidth \cite{dctcp}. They may range from moderate transfers (1~MB to 100~MB) such as ones created by data parallel computation jobs (e.g. MapReduce), to background long-running flows that deliver large volumes of data such as transfers that move data across datacenter sites for data warehousing and geo-replication \cite{calendaring}. For these flows, it is still preferred to minimize the transfer time.

\textit{Deadline} flows have to be completed prior to some deadlines. Their size is either known \cite{karuna} or a good estimate can typically be drawn \cite{d3}. Both latency sensitive and throughput oriented flows might have deadlines. Deadlines can be either soft or hard which implies how value of its completion drops as time passes \cite{amoeba}. A soft deadline implies that it is still profitable to complete the flow and that the value decreases according to a utility function as we move past and away from the deadline. A hard deadline means zero value once the deadline has passed. For example, in interactive scatter-gather applications, if a query is not replied by its deadline (usually less than 300~ms \cite{dctcp, dynamo}), the final answer has to be computed without it \cite{bing} (i.e., zero value for that flow), while if a backup process is not completed in time, it is still valuable to finish it, although it might increase the risk of user data loss due to failures.

\subsubsection{Traffic Burstiness} \label{burst}
Several studies find datacenter traffic bursty \cite{bullet, wild, social_inside, dctcp, hull, traffic_dc_char}. Theoretically, bursty traffic has been shown to increase packet loss and queuing delay while deceasing throughput \cite{bursty_is_bad}. In bursty environments, packet losses have been found more frequent at the network edge due to higher burstiness \cite{traffic_dc_char}. Burstiness can also lead to higher average queue occupancy in the network leading to increased flow completion times (FCT) for many flows \cite{hull} and increased packet drops due to temporary creation of full buffers in the network \cite{ictcp, dctcp}. In addition, highly bursty traffic can cause buffer space unfairness in shared memory switches if a switch port exhausts shared memory due to receiving long bursts \cite{dctcp}.

Several causes can lead to creation of bursty traffic. Hardware offloading features, such as Large Send Offload (LSO), that reduce CPU load, can lead to higher burstiness. Interrupt Moderation (Coalescing) \cite{interrupt_moderation}, which reduces CPU load and increases throughput by processing packets in batches, can lead to bursty traffic at the sender. Transport control features in software can create bursty traffic by scheduling a large window of packets to be sent together such as TCP slow start. Applications can increase burstiness by sending large pieces of data at once \cite{bullet}.

\subsubsection{Packet Reordering} \label{reordering}
Out of order arrival of packets can increase memory and CPU utilization and latency at the receiver especially at high rates \cite{juggler, mptcp-hard}. Some transport protocol features, such as fast retransmit \cite{rfc2001}, may mistake reordering with packet loss. Different hardware and software features are involved in putting packets back in order including Large Receive Offloading (LRO) \cite{LRO}, Receive Side Scaling (RSS) \cite{RSS} and Generic Receive Offloading (GRO) \cite{GRO}. LRO and GRO are usually implemented as part of driver software. Some NICs provide hardware support for these features. LRO focuses mostly on TCP/IPv4 stack while GRO is general purpose.

To understand the extend to which reordering can affect performance, we point to a prior work on improving the performance of handling out of order packet arrivals \cite{juggler}. Authors performed experiments with Vanilla Linux kernel and realized that at high rates (e.g. gigabits), significant reordering can increase CPU utilization to 100\% and limit server interface link utilization to 70\%. Even after applying optimizations at the driver and network stack level, CPU load increased by about 10\% with server interface link utilization at 100\%.

\subsubsection{Performance Isolation} \label{isolation}
Performance isolation is necessary in cloud environments where multiple tenants use shared network resources \cite{eyeq, seawall, elasticswitch}. Isolation prevents selfish or malicious behavior that aims to either unfairly obtain more resources, such as by creating many flows or using custom aggressive transport protocols \cite{acdc, vcc}, or to cause disruptions.

Enforcing performance isolation over a shared infrastructure is hard. To effectively isolate the effect of tenants and users, mechanisms need to be put in place in various layers and parts of the network. For example, a queue management scheme will need to divide buffer space according to users and tenants, bandwidth needs to be fairly divided, computational and memory overheads due to network communication needs to be controlled on a per user or per tenant basis, and all of this need to be done according to service level agreements between the operator, users and tenants.

\subsubsection{The Incast Problem} \label{incast}
When many end-hosts send traffic to one destination concurrently, the link attached to the destination turns into a bottleneck resulting in queue buildups, large queuing delays, and dropped packets \cite{ictcp, dctcp, iatcp, sfs, phost, japan_incast_problem}. This problem becomes more serious in high bandwidth low latency networks \cite{rto_min_random} and with shallow buffer switches \cite{dibs}.

For example, the incast problem could appear in clusters running applications such as search and batch computing jobs like MapReduce that follow the partition-aggregate processing model. In search applications, a server may query a large number of other servers the results to which may be returned to that server at the same time creating sudden increase in incoming traffic. In a MapReduce job, a Reduce instance may download outputs of many Map instances for reduction which can cause a similar situation. Both scenarios follow the many-to-one communication pattern.

\subsubsection{The Outcast Problem} \label{outcast}
This problem occurs due to synchronized drops of packets from an input port of a switch which is referred to as port blackout \cite{outcast}. This eventually leads to unfairness. For such synchronous drops to happen, two predicates have to be present. First, there needs to be contention between a large group of flows coming from one input port and a small group of flows coming from a different input port for access to the same output port. Second, the output port uses queues that follow TailDrop policy (if queue is full, the arriving packet is discarded). Port blackout occurs for the small group of flows and is observed temporarily over short periods of time. When the output queue is full, any arriving packet (during this window) is dropped which leads to consecutive drops for incoming flows. Such consecutive drops affect the smaller set of flows more than they affect the larger set (a smaller number of flows in the larger set are affected). The intensity of this problem increases as the ratio of flows in the larger over smaller group increases. This problem is called the ``outcast'' problem because some flows are cast aside (they cannot obtain bandwidth).

This could simply occur in tree-based topologies such as Fat-Tree and in partition-aggregate communication scenarios where many flows from different servers could be returning results to one server (many-to-one communication pattern). A disproportionate number of flows from incoming servers may end up on different input ports of the ToR switch attached to the receiver which could lead to flows on some port receiving less average throughput.

\subsection{Traffic Control Objectives} \label{goals}
Datacenter environments involve operators, tenants, and end-users with different objectives. Operators would like to use available resources as much as possible to provide higher volume of services, accommodate more tenants and eventually increase revenues. In addition, datacenter infrastructure may be shared by several tenants running various types of applications to offer different services. Tenants would like to receive a fair share of resources according to their service level agreement (SLA) which determines the cost of services. Finally, many end-users may rely upon services offered by datacenter tenants. They would like such services to be responsive and highly available. These possibly conflicting objectives are presented in the following. Table \ref{dc-tc-objectives} provides a summary of traffic control objectives in datacenters.

\begin{table*}[t!]
\caption{Summary of datacenter traffic control objectives} \label{dc-tc-objectives}
\centering
\begin{tabular}{|p{6cm}|p{11cm}|} 
\hline
\textbf{Objective} & \textbf{Description} \\
\hline
\hline
Minimizing Flow Completion Times (FCT) \S \ref{fct} & Faster completion times reduces the communication delay for distributed applications and improves their end-to-end responsiveness. \\
\hline
Minimizing Deadline Miss Rate or Lateness \S \ref{dmr} & For time constrained applications, it is important to meet specific deadline requirements. For some applications, only transactions that complete prior to their deadlines are valuable in which case deadline miss rate is the right performance metric. For some applications transactions completed past the deadlines are still valuable or even necessary in which case lateness (i.e., by how much we miss deadlines), is the right metric. \\
\hline
Maximizing Utilization \S \ref{utilization} & To maximize performance, it is desired to use available resources as much as possible. \\
\hline
Fairness \S \ref{fairness} & Resources should be fairly divided across tenants and users while paying attention to their class of service and service level agreements (SLAs). \\
\hline
\end{tabular}
\end{table*}

\subsubsection{Minimizing Flow Completion Times} \label{fct}
Flow Completion Time (FCT) is the time from the initiation of a flow to its completion. Depending on applications, FCT can directly or indirectly affect the quality of experience and service offered to end users \cite{rcp, dctcp, bing}. Major factors that determine FCT in datacenters include queuing and packet loss \cite{pitfall, dctcp}. Occasionally, retransmission of lost segments can significantly increase latency due to the time it takes to identify and retransmit the lost data. For some applications, it may be more helpful to focus on minimizing mean or median latency (e.g. static web applications) \cite{rcp, dctcp, pdq, racs, phost}, while for other applications tail latency may be more important (e.g. partition-aggregate) \cite{detail, bing}.

\subsubsection{Minimizing Deadline Miss Rate or Lateness} \label{dmr}
Many applications require timely delivery of data that can be viewed as flows with deadlines. In such applications, as long as pieces of data are delivered prior to the deadlines, customer SLAs are satisfied. The quality of services decrease as the fraction of missed deadlines increases. For some applications, delivery after the deadlines is still valuable and necessary. As a result, sometimes minimizing the amount by which we miss deadlines is more important which is referred to as ``lateness'' (e.g. synchronization of search index files).

\subsubsection{Maximizing Utilization} \label{utilization}
Increasing resource utilization can reduce provisioning costs and increase revenues. By better utilizing network bandwidth, operators can accommodate more tenants or improve the quality of service for existing ones. Effectively utilizing the network depends partly on network topology and design parameters, and partly on network traffic control scheme.

\subsubsection{Fairness} \label{fairness}
Many flows share datacenter resources such as link bandwidth and buffer space. In multi-tenant datacenters, it is necessary to make sure tenants receive fair share of network resources according to their service level agreement (SLA). Enforcing fairness also mitigates the starvation problem and prevents malicious behavior.

There are several definitions of fairness in networking context including max-min fairness, proportional fairness, and balanced fairness \cite{fairness_diff_types, fairness-axiomatic}. Fairness criteria determine how link bandwidth or buffer space is divided among flows. Max-Min Fairness (MMF) \cite{mmf}, which aims at maximizing the minimum share, is the most widely used. In general, fairness can be considered over multiple dimensions each representing a different kind of resource (e.g., bandwidth, CPU cycles, and memory) \cite{drf, mr-fe}. We however focus on network bandwidth fairness in this paper.

Fairness should be considered among proper entities as defined by the fairness policy. For example, fairness may be across groups of flows as opposed to individual flows to prevent tenants from obtaining more resources by creating many flows. Strict fairness across all flows can also lead to increased number of missed deadlines \cite{d3} and sub-optimal FCTs \cite{pdq}. One approach is to first apply fairness across tenants according to their classes of service and then across flows of each tenant considering flow priorities.

In addition to the traffic control objectives we mentioned, there are other objectives followed by many datacenter operators. An important objective is reducing energy costs by increasing energy efficiency. Since datacenter networks usually connect a huge number of servers, they are made of a large number of network equipment including fiber optics cables and switches. Due to varying amount of computational and storage load, average network load may be considerably less than its peak. As a result, operators may reduce energy costs by turning off a fraction of networking equipment at off-peak times \cite{elastictree, energy-aware-routing, energy-aware-routing-link-disjoint} or by dynamically reducing the link bandwidths across certain links according to link utilizations \cite{energy-proportional-trees}. There are however several challenges doing so. First, the resulting system should be fast enough to increase network capacity as computational or storage load increases to prevent additional communication latency. Second, it may be unclear where to reduce network capacity either by turning equipment off or by reducing link bandwidths (correlation between network load and placement of computation/storage can be considered for additional improvement \cite{greendcn}). Third, the effectiveness of such systems depends on load/utilization prediction accuracy. Further discussion on reducing datacenter power consumption is out of the scope of this paper.


\section{Datacenter Traffic Control Management} \label{tc-management}
To enforce traffic control, some level of coordination is needed across the network elements. In general, traffic control can range from fully distributed to completely centralized. Here we review the three main approaches used in the literature namely distributed, centralized or hybrid. Table \ref{dc-tc-management} provides an overview of traffic control management schemes.

\begin{table*}[t!]
\caption{Summary of datacenter traffic control management} \label{dc-tc-management}
\centering
\begin{tabular}{|p{2.5cm}|p{7cm}|p{7cm}|} 
\hline
\textbf{Scheme} & \textbf{Benefits} & \textbf{Drawbacks} \\
\hline
\hline
Distributed \S \ref{management-dist} & Higher scalability and reliability. Solutions can be completely end-host based or use explicit network feedback. More information in \S \ref{distributed_benefits}. & Access limited to local view of network status and flow properties. Limited coordination complicates enforcement of network-wide policies. More information in \S \ref{distributed_drawbacks}. \\
\hline
Centralized \S \ref{management-cent} & Access to global view of network status and flow properties. Central control and management can improve flexibility and ease of enforcing network-wide policies. More information in \S \ref{centralized_benefits}. & A central controller can become a single point of failure or a network hotspot. Latency overhead of communicating with a central entity and control plane inconsistencies are other concerns. More information in \S \ref{centralized_drawbacks}. \\
\hline
Hybrid \S \ref{management-hybr} & Potentially higher reliability, scalability, and performance. More information in \S \ref{hybrid_benefits}. & Higher complexity. Also, the final solution may not be as good as a fully centralized system. More information in \S \ref{hybrid_drawbacks}. \\
\hline
\end{tabular}
\end{table*}

\subsection{Distributed} \label{management-dist}
Most congestion management schemes coordinate in a distributed way as it is more reliable and scalable. A distributed scheme may be implemented as part of the end-hosts, switches, or both. Some recent distributed traffic control schemes include those presented in \cite{pdq, iatcp, presto, timely, dcqcn}.

Designs that can be fully realized using end-hosts are usually preferred over ones that need changes in the default network functions or demand additional features at the switches such as custom priority queues \cite{pfabric}, in-network rate negotiation and allocation \cite{pdq}, complex calculations in switches \cite{codel}, or per flow state information \cite{afd}. End-host implementations are usually more scalable since every server handles its own traffic. Therefore, popular transport protocols rely on this type of implementation such as \cite{reno, dctcp}.

As an example, the incast problem \S \ref{incast}, which is a common traffic control challenge, can be effectively addressed using end-host based approaches considering that incast congestion occurs at receiving ends. Some approaches are Server-based Flow Scheduling (SFS) \cite{sfs}, pHost \cite{phost}, NDP \cite{ndp} and ExpressPass \cite{credit-based-cc}. SFS uses the generation of ACKs to control the flow of data towards receivers and avoid congestion. The sender looks at the flows it is sending and schedules the higher priority flows first, while the receiver controls the reception by deciding on when to generate ACKs. pHost uses a pull-based approach in which the sender decides on reception schedule based on some policy (preemptive or non-preemptive, fair sharing, etc). A source dispatches a Request To Send (RTS) to a receiver. The receiver then knows all the hosts that want to transmit to it and can issue tokens to allow them to send. NDP limits the aggregate transmission rate of all incast senders by maintaining a PULL queue at the receiver that is loaded with additional PULL requests when new packets arrive from a sender (a PULL request contains a counter which determines number of packets its associated sender is allowed to send). PULL requests are then sent to senders in a paced manner to make sure the overall incoming transmission rate at the receiver is not larger than per interface line rate. ExpressPass manages congestion across the network by controlling the flow of credit packets at the switches and end-hosts according to network bottlenecks (a sender is allowed to send a new data packet when it receives a credit packet). 

Shifting more control to the network may allow for better resource management due to ability to control flows from more than a single server and availability of information about flows from multiple end-hosts. For example, flows from many hosts may pass through a ToR switch giving it further knowledge to perform scheduling and allocation optimizations.

Some examples of this approach include RCP \cite{rcp}, PDQ \cite{pdq}, CONGA \cite{conga}, Expeditus \cite{expeditus}, and RackCC \cite{rackcc}. RCP and PDQ perform in-network rate allocation and assignment by allowing switches and end-hosts to communicate using custom headers, CONGA gets help from switches to perform flowlet based load balancing in leaf-spine topologies, Expeditus performs flow based load balancing by implementing custom Layer 2 headers and localized monitoring of congestion at the switches, and RackCC uses ToR switches as means to share congestion information among many flows between the same source and destination racks to help them converge faster to proper transmission rates. To implement advanced in-network features, changes to the network elements might be necessary and switches may need to do additional computations or support new features.

\subsubsection{Benefits} \label{distributed_benefits} Higher scalability and reliability. Can be completely implemented using end-hosts. To further improve performance, network (i.e., switches, routers, etc.) can be involved as well. Completely end-host based approaches can operate simply by controlling the transmission rate according to implicit feedbacks from network (e.g. loss, latency). Network can offer explicit feedbacks (e.g. network queues' occupancy) to improve transmission rate management by allowing senders to make more accurate control decisions. For example, Explicit Congestion Notification (ECN) allows network to communicate high queue occupancy to end-hosts \cite{ecn, tcn} or trimming packet payloads in case of highly occupied network queues (instead of fully discarding them) can help receivers get a complete picture of transmitted packets \cite{cut-payload, ndp}.

\subsubsection{Drawbacks} \label{distributed_drawbacks} Usually just access to local view of network status and flow properties which allows for only locally optimal solutions. For example, while every end-point may strive to achieve maximum throughput for its flows by default (locally optimal), it may lead to a higher network wide utility if some end-points reduce their transmission rate in favor of other end-points with more critical traffic (globally optimal). Using distributed management, it may be harder to enforce new network wide policies (e.g. rate limits in a multi-tenant cloud environment) due to lack of coordination and limited view of network condition and properties. 

\subsection{Centralized} \label{management-cent}
In centralized schemes a central unit coordinates transmissions in the network to avoid congestion. The central unit has access to a global view of network topology and resources, state information of switches, and end-host demands. These include flow sizes, deadlines and priorities as well as queuing status of switches and link capacities. Scheduler can proactively allocate resources temporally and spatially (several slots of time and different links) and plan transmissions in a way that optimizes the performance and minimizes contentions. To further increase performance, this entity can translate the scheduling problem into an optimization problem with resource constraints the solution to which can be approximated using fast heuristics. For large networks, scheduler effectiveness depends on its computational capacity and communication latency to end-hosts.

TDMA \cite{tdma}, FastPass \cite{fastpass} and FlowTune \cite{flowtune} are examples of a centrally coordinated network. TDMA divides timeline into rounds during which it collects end-host demands. Each round is divided into fixed sized slots during which hosts can communicate in a contention-less manner. All demands are processed at the end of a round and schedules are generated and given to end-hosts. FastPass achieves high utilization and low queuing by carefully scheduling traffic packet by packet considering variation in packet sizes. FlowTune improves on scalability of FastPass using centralized rate-assignment and end-host rate-control.

There are several challenges using a fully centralized approach. The central controller could be a single point of failure since all network transmissions depend on it. In case of a failure, end-hosts may have to fall back to a basic distributed scheme temporarily \cite{fastpass}. There will be scheduling overhead upon flow initiation, that is the time it takes for the scheduler to receive, process the request, and allocate a transmission slot. Since majority of flows are short in datacenters, the scheduling overhead has to be tiny. In addition, this approach may only scale to moderate datacenters due to processing burden of requests and creation of a congestive hot-spot around the controller due to large number of flow arrivals. Bursts in flow arrivals \cite{nature} can congest the controller temporarily and create scheduling delays. It may be possible to apply general techniques of improving scalability for central management of larger networks such as using a hierarchical design.

\subsubsection{Benefits} \label{centralized_benefits} Can provide higher performance with a global view of network status and flow properties. Such information may include utilization of different network edges, their health status as well as flows' size, deadline and priority. With this information, one can potentially direct traffic carefully according to network capacity across a variety of paths while allowing flows to transmit according to their priorities to maximize utility. Central management can improve flexibility and ease of managing network policies. For example, new routing/scheduling techniques can be rolled out much faster by only upgrading the central fabric. Centralized schemes also increase ease of admission control in case strict resource management is necessary for guaranteed SLAs. 

\subsubsection{Drawbacks} \label{centralized_drawbacks} A central controller or management fabric can be a single point of failure or it may become a network hotspot in case of large networks. There is also latency and computational overhead of collecting network status and flow properties from many end-hosts (controller will have to process and understand incoming messages at high speed and act accordingly). Overhead of network resource allocation and scheduling (if central rate allocation is used). Finally, consistency of network updates can be an issue in large networks. For example, some updates my not be applied correctly at network elements (e.g. software bugs \cite{evolve}) or different updates may be applied with varying latency that can lead to transient congestion or packet losses which may hurt performance of sensitive services. 

\subsection{Hybrid} \label{management-hybr}
Using a hybrid system could provide the reliability and scalability of distributed control and performance gains obtained from global network management. A general hybrid approach is to have distributed control that is assisted by centrally calculated parameters.

Examples of this approach include OTCP \cite{otcp}, Fibbing \cite{fibbing}, Hedera \cite{hedera} and Mahout \cite{mahout}. OTCP uses a central controller to monitor and collect measurements on link latencies and their queuing extent using methods provided by Software Defined Networking (SDN) \cite{sdn} which we will discuss further in \S \ref{pfp}. For every new flow, the controller calculates parameters such as initial and minimum retransmission timeout and initial and maximum congestion window size considering which path is taken by flows which allows for fast convergence to steady state. Fibbing relies on a central controller to manipulate the costs associated with routes in the network or insert fake nodes into the routing tables of routers to force them to use or avoid some paths to control the distribution of load across network. Hedera and Mahout operate by initially allowing network to route all incoming flows using a distributed approach, then monitoring and detecting large flows that can be moved dynamically to different routes using a central controller with a global view of network status which allows for a more balanced distribution of load. While Hedera uses readily available switch capabilities (flow counters) to detect large flows, Mahout engineers and inserts a shim layer at the end-hosts to monitor and detect large flows to improve scalability. 

\subsubsection{Benefits} \label{hybrid_benefits} Reliability, scalability, and higher performance. A central controller can offer coarse grained solutions while fine-grained control is applied using a distributed scheme. Division of work between central and distributed components can be done in a way that minimizes the effects of centralized and distributed schemes' drawbacks. For example, in multi-tenant cloud datacenters, a hierarchical approach can be used where aggregate bandwidth given per tenant is calculated centrally while transmission control across flows per tenant is performed in a distributed fashion reducing central management overhead and increasing per tenant management scalability. 

\subsubsection{Drawbacks} \label{hybrid_drawbacks} Complexity may increase. Central controller now has limited control; therefore, the final solution may not be as good as a fully centralized system. Due to presence of centralized control, there still exists a single point of failure however with less impact in case of a failure compared to a fully centralized scheme. Also, the distributed component still operates on a locally optimal basis. For example, in multi-tenant cloud datacenters, if bandwidth per tenant is managed in a distributed fashion, due to limited local information per network element, it may be challenging to apply routing/scheduling policies that maximize utility according to flow properties.


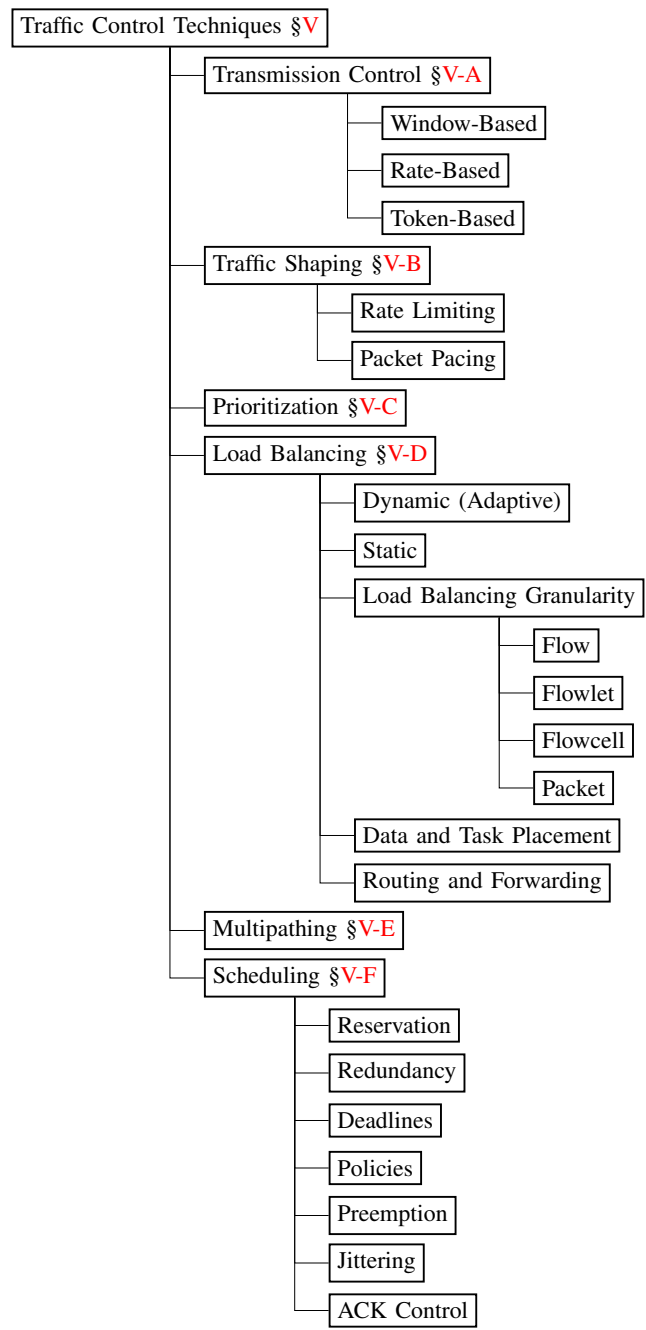
\begin{figure}[b!]
\centering
\begin{adjustbox}{width=0.95\columnwidth}
\begin{tikzpicture}[%
  grow via three points={one child at (0.5,-0.7) and
  two children at (0.5,-0.7) and (0.5,-1.4)},
  font=\normalfont,
  edge from parent path={(\tikzparentnode.south) |- (\tikzchildnode.west)}]
  \node [level 0] {Traffic Control Techniques \S \ref{traffic-control-all}}
    child { node {Transmission Control \S \ref{transmission}}
      child { node [level 2] {Window-Based}}
      child { node [level 2] {Rate-Based}}
      child { node [level 2] {Token-Based}}
    }
    child [missing] {}
    child [missing] {}
    child [missing] {}
    child { node {Traffic Shaping \S \ref{shaping}}
      child { node [level 2] {Rate Limiting}}
      child { node [level 2] {Packet Pacing}}
    }
    child [missing] {}
    child [missing] {}
    child { node {Prioritization \S \ref{priority}}}
    child { node {Load Balancing \S \ref{balancing}}
      child { node [level 2.1] {Dynamic (Adaptive)}}
      child { node [level 2.1] {Static}}
      child { node [level 2] {Load Balancing Granularity}
        child { node [level 3] {Flow}}
        child { node [level 3] {Flowlet}}
        child { node [level 3] {Flowcell}}
        child { node [level 3] {Packet}}
      }
      child [missing] {}
      child [missing] {}
      child [missing] {}
      child [missing] {}
      child { node [level 2] {Data and Task Placement}}
      child { node [level 2] {Routing and Forwarding}}
    }
    child [missing] {}
    child [missing] {}
    child [missing] {}
    child [missing] {}
    child [missing] {}
    child [missing] {}
    child [missing] {}
    child [missing] {}
    child [missing] {}
    child { node {Multipathing \S \ref{multipathing}}}				
    child { node {Scheduling \S \ref{scheduling}}
      child { node [level 2] {Reservation}}
      child { node [level 2] {Redundancy}}
      child { node [level 2] {Deadlines}}
      child { node [level 2] {Policies}}
      child { node [level 2] {Preemption}}
      child { node [level 2] {Jittering}}
      child { node [level 2] {ACK Control}}
    }
    child [missing] {}
    child [missing] {}
    child [missing] {}
    child [missing] {}
    child [missing] {}
    child [missing] {}
    child [missing] {};
\end{tikzpicture}
\end{adjustbox}
\caption{High level breakdown of traffic control techniques} \label{fig_topics}
\end{figure}

\begin{figure*}[ht!]
\centering
\begin{tikzpicture}[font=\normalfont,box/.style={draw,rounded corners,text width=4cm,align=center},border/.style={draw,dashed,rounded corners,text width=4.5cm},numred/.style={draw=none,color=white,fill=Red,rounded corners},numblue/.style={draw=none,color=white,fill=Blue,rounded corners},numblack/.style={draw=none,color=white,fill=Green,rounded corners},decoration={}]
\node[text=Green,text width=4.5cm,draw=none,align=center] at (6.25,1) {\Large \textbf{Techniques}};
\node[draw=Green!40,border,text height=7cm] at (6.25,-3) (x1) {};
\node[draw=Green,line width=1pt,box] at (6.5,0) (l1) {Transmission Control};
\node[draw=Green,line width=1pt,box] at (6.5,-1) (l2) {Rate Limiting};
\node[draw=Green,line width=1pt,box] at (6.5,-2) (l3) {Packet Pacing};
\node[draw=Green,line width=1pt,box] at (6.5,-3) (l4) {Prioritization};
\node[draw=Green,line width=1pt,box] at (6.5,-4) (l5) {Load Balancing};
\node[draw=Green,line width=1pt,box] at (6.5,-5) (l6) {Multipathing};
\node[draw=Green,line width=1pt,box] at (6.5,-6) (l7) {Scheduling};
%
\node[text=Blue,text width=4.5cm,draw=none,align=center] at (12.75,1) {\Large \textbf{Objectives}};
\node[draw=Blue!40,border,text height=7cm] at (12.75,-3) {};
\node[draw=Blue,line width=1pt,box,minimum height=1cm] at (13,0) (r1) {Reducing Latency};
\node[draw=Blue,line width=1pt,box,minimum height=1cm] at (13,-2) (r2) {Reducing Deadline Miss Rate / Lateness};
\node[draw=Blue,line width=1pt,box,minimum height=1cm] at (13,-4) (r3) {Maximizing Utilization};
\node[draw=Blue,line width=1pt,box,minimum height=1cm] at (13,-6) (r4) {Fairness};
\node[text=Red,text width=4.5cm,draw=none,align=center] at (-0.25,1) {\Large \textbf{Challenges}};
\node[draw=Red!40,border,text height=7cm] at (-0.25,-3) {};
\node[draw=Red,line width=1pt,box] at (0,0) (rd1) {Unpredictable Traffic Matrix};
\node[draw=Red,line width=1pt,box] at (0,-1) (rd2) {Traffic Burstiness};
\node[draw=Red,line width=1pt,box] at (0,-2) (rd3) {Mix of various flow types/sizes};
\node[draw=Red,line width=1pt,box] at (0,-3) (rd4) {Packet Reordering};
\node[draw=Red,line width=1pt,box] at (0,-4) (rd5) {Performance Isolation};
\node[draw=Red,line width=1pt,box] at (0,-5) (rd6) {The Incast Problem};
\node[draw=Red,line width=1pt,box] at (0,-6) (rd7) {The Outcast Problem};
%
\draw[postaction={decorate},line width=1pt] (l1.east) -- (r1.west);
\draw[postaction={decorate},line width=1pt] (l1.east) -- (r2.west);
\draw[postaction={decorate},line width=1pt] (l1.east) -- (r3.west);
\draw[postaction={decorate},line width=1pt] (l1.east) -- (r4.west);
\draw[postaction={decorate},line width=1pt] (l2.east) -- (r3.west);
\draw[postaction={decorate},line width=1pt] (l2.east) -- (r4.west);
\draw[postaction={decorate},line width=1pt] (l3.east) -- (r1.west);
\draw[postaction={decorate},line width=1pt] (l3.east) -- (r3.west);
\draw[postaction={decorate},line width=1pt] (l4.east) -- (r1.west);
\draw[postaction={decorate},line width=1pt] (l4.east) -- (r2.west);
\draw[postaction={decorate},line width=1pt] (l5.east) -- (r3.west);
\draw[postaction={decorate},line width=1pt] (l6.east) -- (r3.west);
\draw[postaction={decorate},line width=1pt] (l7.east) -- (r1.west);
\draw[postaction={decorate},line width=1pt] (l7.east) -- (r2.west);
\draw[postaction={decorate},line width=1pt] (l7.east) -- (r3.west);
\draw[postaction={decorate},line width=1pt] (l7.east) -- (r4.west);
\draw[postaction={decorate},line width=1pt] (l1.west) -- (rd2.east);
\draw[postaction={decorate},line width=1pt] (l1.west) -- (rd4.east);
\draw[postaction={decorate},line width=1pt] (l2.west) -- (rd2.east);
\draw[postaction={decorate},line width=1pt] (l2.west) -- (rd5.east);
\draw[postaction={decorate},line width=1pt] (l3.west) -- (rd2.east);
\draw[postaction={decorate},line width=1pt] (l4.west) -- (rd3.east);
\draw[postaction={decorate},line width=1pt] (l4.west) -- (rd5.east);
\draw[postaction={decorate},line width=1pt] (l5.west) -- (rd1.east);
\draw[postaction={decorate},line width=1pt] (l6.west) -- (rd4.east);
\draw[postaction={decorate},line width=1pt] (l7.west) -- (rd2.east);
\draw[postaction={decorate},line width=1pt] (l7.west) -- (rd3.east);
\draw[postaction={decorate},line width=1pt] (l7.west) -- (rd4.east);
\draw[postaction={decorate},line width=1pt] (l7.west) -- (rd6.east);
\draw[postaction={decorate},line width=1pt] (l7.west) -- (rd7.east);
\node[numred] at (0.1,0.5) {1};
\node[numred] at (0.1,-0.75) {2};
\node[numred] at (0.1,-1.5) {3};
\node[numred] at (0.1,-2.75) {4};
\node[numred] at (0.1,-3.75) {5};
\node[numred] at (0.1,-4.75) {6};
\node[numred] at (0.1,-5.75) {7};
\node[numblue] at (13.1,0.5) {1};
\node[numblue] at (13.1,-1.5) {2};
\node[numblue] at (13.1,-3.5) {3};
\node[numblue] at (13.1,-5.5) {4};
\node[numblack] at (6.6,0.25) {1};
\node[numblack] at (6.6,-0.75) {2};
\node[numblack] at (6.6,-1.75) {3};
\node[numblack] at (6.6,-2.75) {4};
\node[numblack] at (6.6,-3.75) {5};
\node[numblack] at (6.6,-4.75) {6};
\node[numblack] at (6.6,-5.75) {7};
\end{tikzpicture}
\caption{How traffic control techniques interact with challenges and objectives, every line shows a direct relationship (indirect relationships may exist between some blocks but they are not shown here)} \label{fig_relations}
\end{figure*}
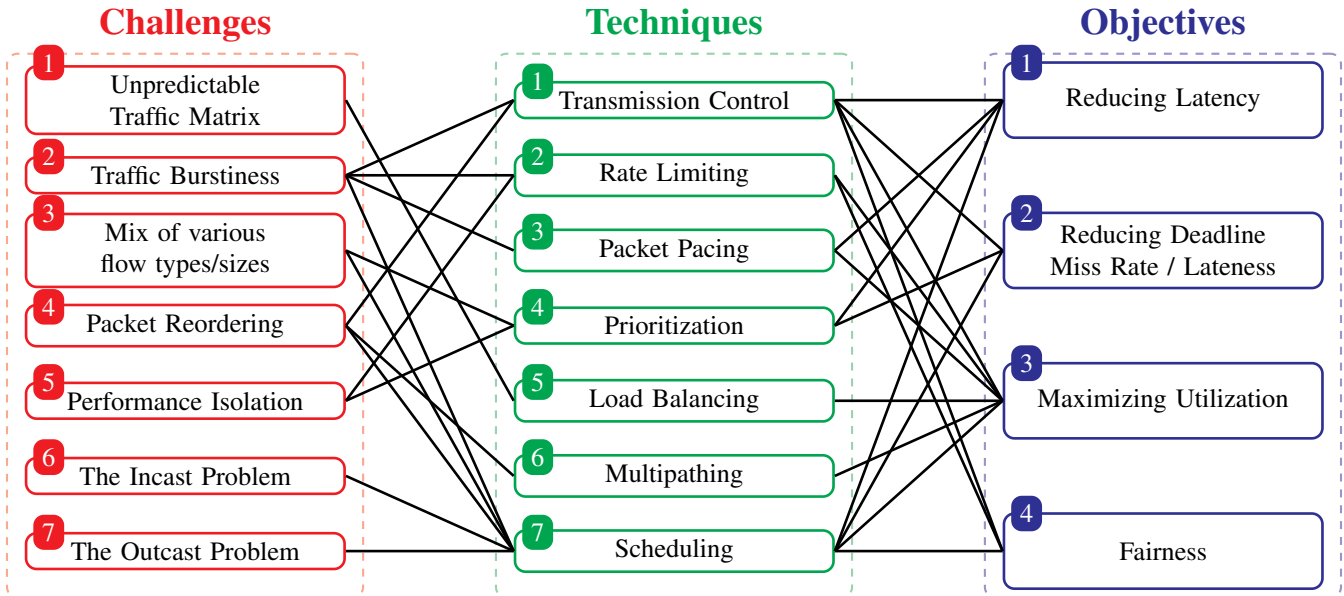

\section{Datacenter Traffic Control Techniques} \label{traffic-control-all}
Many design parameters and approaches can be considered in development of any traffic control scheme. Figure \ref{fig_topics} provides a high level breakdown of major datacenter traffic control techniques. In the following, we provide a list of these techniques and discuss some research efforts made regarding each. Figure \ref{fig_relations} provides an overview of the relationships between challenges, objectives and techniques. Details of each relationship is further explained in Tables \ref{table_relationship_tch_obj} and \ref{table_relationship_tch_cha}. More detailed information is provided in the following sections.

\subsection{Transmission Control} \label{transmission}
Transmission control is the mechanism that controls the flow of data sent to the network. There is typically a window of outstanding bytes for each flow at the sender determining the volume of data that can be transmitted before data reception is acknowledged. This allows for \textit{implicit} rate control. A larger window may increase the average transmission rate since it allows for more bytes in flight towards receiver. A significant body of work, including recent works, employ \textbf{window-based} rate control. Table \ref{dc-tc-transmission} provides a summary of this section.

Some recent window-based approaches include DCTCP \cite{dctcp}, D2TCP \cite{d2tcp}, L2DCT \cite{l2dct}, MCP \cite{mcp}, DAQ \cite{daq}, and PASE \cite{pase}. DCTCP uses explicit congestion signals from switches that is piggybacked on ACKs to change the window size according to the extent of congestion. Packets are marked to convey congestion signal according to instantaneous queue length upon their arrival at the queue. D2TCP modulates window size based on flow deadlines: flows with closer deadlines reduce window size less than others in response to network congestion signals. L2DCT modulates window size of a flow based on its estimated size which is dynamically calculated by counting the number of packets it has seen from a flow up to current time. MCP changes window size to approximate the solution to an stochastic optimization problem that minimizes long term mean packet delay using flow information and network state, such as queue length at the switches. DAQ allows ToR switches to calculate initial sender window for end-hosts to converge to proper transmission rates faster. PASE also gets help from network elements to decide on its initial coarse-grained window size, and then performs fine-grained adjustments to the window size using congestion signals received similar to DCTCP.

Although fairly simple, window-based approaches only allow for coarse-grained control of rate which can result in considerable variations and create highly bursty traffic. For example, a sender can release a full window of packets before it has to wait for the receiver to acknowledge. This may be followed by a batch of acknowledgements that move the window forward and allow for another full window of packets to be sent.

To decrease the transmission rate variations at the senders, \textbf{rate-based} mechanisms that employ \textit{explicit} rate control can be used. Packets in the outstanding window are carefully scheduled to achieve the proper bit rate which implies reduced burstiness of the traffic. Rate-control is necessary when using reservation based bandwidth control over shared links such as in \cite{flowtune}.

Examples of earlier rate-based protocols include TCP Friendly Rate Control (TFRC) \cite{tfrc_paper} and Rate Control Protocol \cite{rcp}. Several recent work also use rate-based methods in datacenters including PDQ \cite{pdq}, D3 \cite{d3}, TIMELY \cite{timely}, and RACKS \cite{racs}. TFRC calculates the allowed sending rate to compete fairly with TCP using an equation which captures various network parameters such as round trip time (RTT), loss rate, retransmission timeout, and segment size. RCP uses an equation based rate control to minimize average FCT with the help of network switches that divide bandwidth according to processor sharing policy. D3 uses flow deadlines as a factor in the rate calculation equation to reduce deadline miss rate (a closer deadline allows for a higher rate). RACS uses a similar approach to RCP but assigns weights to flows according to their priority, to emulate different scheduling algorithms at the switches according to weight assignment. PDQ improves on both RCP and D3 by rate allocation using switches and adding support for preemption. TIMELY calculates the sending rate as a function of network latency.

Explicit rate control can be applied both in hardware (e.g. NIC) and software (e.g. OS Kernel). While the former provides more precise rates, the latter is more flexible and allows for a larger number of flows to be rate controlled as well as more complex policies to be considered. Rate-control in software can result in creation of occasional bursts due to challenges of precise packet scheduling in OS kernel \cite{tfrc_paper}.

Complementary to these two approaches for rate control at the sender, one can also apply \textbf{token-based} transmission control which adds an extra control layer to make sure senders only transmit according to some quota assigned to them (this is sometimes referred to as \textit{pull-based} or \textit{credit-based} approach). For example, such quota could be assigned according to congestion status at receivers (to address the incast problem discussed earlier in \S \ref{incast}) or to implement a form of receiver based flow scheduling policy (more information in \S \ref{scheduling}).

\newcolumntype{P}[1]{>{\RaggedRight\hspace{-0.3pt}}p{#1}}

\begin{table*}[ht]
\caption{Description of relationships between \textbf{techniques} and \textbf{challenges} in Figure \ref{fig_relations}} \label{table_relationship_tch_cha}
\centering
\begin{tabular}{|p{1.5cm}|p{12.5cm}|P{2.6cm}|} 
\hline
\textbf{Relationship} & \textbf{Description} & \textbf{Some Related Works} \\
\hline
\hline
\techTochallenge{1}{2} & Transmission control at senders determines number of packets given to network driver for transmission. If a bunch of packets are sent to network interface for transmission, bursts may be created. &\cite{tcppacing_old}, \cite{tcppacing} \\
\hline
\techTochallenge{1}{4} & If transmission is performed over multiple interfaces for multipath delivery, careful transmission control is required to minimize packet reordering. This shall be done according to latencies of different paths to the receiver. &\cite{flowlet}, \cite{per_packet_load_balancing}, \cite{tinyflow}, \cite{presto} \\
\hline
\techTochallenge{2}{2} & Rate limiting can reduce burstiness by adding a protective layer after transmission control in case a large number of packets are given to the network driver for transmission. &\cite{nicpic}, \cite{carousel}, \cite{sf-rate-limiting} \\
\hline
\techTochallenge{2}{5} & Rate limiting can improve performance isolation across tenants and users by preventing any tenant/flow from taking too much bandwidth starving other tenants/flows (a tenant may initiate many flows). &\cite{eyeq}, \cite{gatekeeper} \\
\hline
\techTochallenge{3}{2} & Packet pacing reduces burstiness by adding time spacing between consecutive packets prior to transmission. &\cite{tcppacing_old}, \cite{tcppacing} \\
\hline
\techTochallenge{4}{3} & Prioritization helps allocate resources to flows accordingly and can improve the performance while scheduling a mix of flow types/sizes. For example, one could prioritize highly critical deadline flows over other flows. &\cite{dpp}, \cite{pias}, \cite{pfabric}, \cite{pase}, \cite{jump} \\
\hline
\techTochallenge{4}{5} & Prioritization can shield a tenant/application with high service guarantees from other tenants/applications with lower service guarantees. Priorities shall be determined according to tenant/application service level agreements (SLAs) or Quality of Service requirements (QoS). &\cite{secondnet}, \cite{trinity} \\
\hline
\techTochallenge{5}{1} & Datacenter topologies usually come with a large degree of path redundancy to provide low over-subscription ratio for better inter-rack communication. With load balancing, this capacity can be used to a higher extent giving the operators/tenants a better chance to handle any traffic matrix. &\cite{vl2}, \cite{fattree} \\
\hline
\techTochallenge{6}{4} & Multipathing can increase packet reordering. To reduce reordering while using multiple paths, one can perform careful packet scheduling according to path latencies at transmission control level so that packets arrive in the right order at the receiver (it is nearly impossible to eliminate reordering). Assignment of data segments to sub-flows is also an important factor in how well receivers can handle reordering. &\cite{mptcp-hard}, \cite{flowlet} \\
\hline
\techTochallenge{7}{2} & Transmission control, rate limiting, and packet pacing all depend on careful scheduling of packets at senders which can mitigate burstiness. & \cite{tfrc_paper}, \cite{nicpic}, \cite{carousel}, \cite{pacing_isi} \\
\hline
\techTochallenge{7}{3} & When a mix of flow types is present, scheduling of packets according to flow priorities, deadlines, sizes and arrival order can help us better meet traffic control objectives. & \cite{pdq}, \cite{pias}, \cite{pfabric} \\
\hline
\techTochallenge{7}{4} & If multiple paths are used for a flow, scheduling can reduce reordering by determining when packets should be sent at the sender over each path to arrive at the receiver with minimal reordering. & \cite{flowlet} \\
\hline
\techTochallenge{7}{6} & Scheduling can mitigate the incast problem by preventing all the incoming data (from many flows) from arriving at a receiver at the same time. Various techniques can be used such as adding random delays while initiating requests (jittering) and receiver based scheduling using either ACKs or receiver window size. & \cite{sfs}, \cite{japan_incast_problem} \\
\hline
\techTochallenge{7}{7} & Effective queuing disciplines which determine how packets in a switch queue are scheduled for transmission, such as Stochastic Fair Queuing (SFQ), can mitigate port blackout \S \ref{outcast}. & \cite{outcast} \\
\hline
\end{tabular}
\end{table*}

\begin{table*}[t]
\caption{Description of relationships between \textbf{techniques} and \textbf{objectives} in Figure \ref{fig_relations}} \label{table_relationship_tch_obj}
\centering
\begin{tabular}{|p{1.5cm}|p{12.5cm}|P{2.6cm}|} 
\hline
\textbf{Relationship} & \textbf{Description} & \textbf{Some Related Works} \\
\hline
\hline
\techToobjective{1}{1} & Transmission control can impact overall end-to-end latency by affecting queue occupancies at the network switches. & \cite{tcppacing_old}, \cite{tcppacing} \\
\hline
\techToobjective{1}{2} & Transmission control determines how many packets are sent by each flow which can be done according to flow deadlines to reduce deadline miss rate and/or lateness. & \cite{pdq}, \cite{d2tcp}, \cite{d3} \\
\hline
\techToobjective{1}{3} & Proper transmission control can maximize network bandwidth utilization while avoiding congestion (which usually leads to dropped packets and wasted bandwidth). & \cite{fastpass}, \cite{flowtune}, \cite{phost} \\
\hline
\techToobjective{1}{4} & Transmission control plays significant role in how fairly flows share the network. For example, if one of two equally important flows is given higher transmission quota over longer periods of time, this can lead to unfairness. & \cite{rcp} \\
\hline
\techToobjective{2}{3} & Rate limiting can prevent congestion and reduce dropped packets. As a result, it helps maximize utilization and minimize wasted bandwidth. & \cite{flowtune} \\
\hline
\techToobjective{2}{4} & Rate limiting can improve fairness by preventing selfish behavior of bandwidth hungry flows/tenants. & \cite{eyeq}, \cite{gatekeeper} \\
\hline
\techToobjective{3}{1} & Packet pacing can reduce average queue occupancy in the network (switches/routers) and therefore reduces end-to-end latency. & \cite{tcppacing_old}, \cite{tcppacing}, \cite{hull} \\
\hline
\techToobjective{3}{3} & Packet pacing can increase effective utilization by preventing bursty behavior of flows which can lead to higher buffer occupancy and dropped packets. It stabilizes transmission rates and reduces transmission spikes. & \cite{tcppacing}, \cite{hull} \\
\hline
\techToobjective{4}{1} & Prioritization can reduce average latency and flow completion times by giving higher priority to shorter flows. & \cite{pfabric}, \cite{pias} \\
\hline
\techToobjective{4}{2} & Prioritization according to flow deadlines can improve the overall deadline miss rate. For example, search queries with critical deadlines (e.g. 300~ms after arrival) can be given high priority and can be addressed before long running backup operations in a shared network environment. & \cite{daq} \\
\hline
\techToobjective{5}{3} & Load balancing allows us to make best use of large path diversity in datacenters and maximize utilization over all available paths. & \cite{hedera}, \cite{conga}, \cite{presto}, \cite{detail} \\
\hline
\techToobjective{6}{3} & Multipathing can increase utilization by allowing long running flows to use bandwidth of several available paths. & \cite{mptcp}, \cite{mptcp-hard}, \cite{mptcp-improving} \\
\hline
\techToobjective{7}{1} & Scheduling can reduce latency by applying scheduling disciplines that mimic Shortest Remaining Processing Time (SRPT). & \cite{racs} \\
\hline
\techToobjective{7}{2} & Deadline miss rate or lateness can be reduced by scheduling flows according to their deadlines such as by allotting more capacity for flows with closer deadlines. & \cite{d2tcp} \\
\hline
\techToobjective{7}{3} & Scheduling can improve utilization by reducing contention within the network for using available bandwidth. For example, by carefully deciding on when packets should be sent by end-hosts, we can avoid sudden arrival of many packets at the switches which can lead to dropped packets. & \cite{tdma}, \cite{fastpass} \\
\hline
\techToobjective{7}{4} & Scheduling can improve fairness by giving bandwidth to flows based on a fair sharing policy such as Max-Min Fairness (MMF) \cite{mmf}. &\cite{rcp}, \cite{racs} \\
\hline
\end{tabular}
\end{table*}

\begin{table*}[t!]
\caption{Overview of transmission control techniques} \label{dc-tc-transmission}
\centering
\begin{tabular}{|p{1.8cm}|p{2cm}|p{5.2cm}|p{7cm}|} 
\hline
\textbf{Scheme} & \textbf{Input} & \textbf{Benefits} & \textbf{Drawbacks} \\
\hline
\hline
Window-based & Maximum outstanding window size. & Simplicity, no careful packet scheduling overhead. & Coarse-grained control over transmission rate which can lead to occasional bursts of packets (e.g. when a bunch of ACKs arrive and suddenly the transmission window moves forward). \\
\hline
Rate-based & Desired transmission rate (in bits per second). & More accurate control over rate, reduced variations in rate, a more natural choice for better management of network bandwidth. & Overhead of packet scheduling either in software (less accurate, less costly) or in hardware (more accurate, requires hardware support). \\
\hline
Token-based & Transmission quota towards a specific receiver. & Receiver enforced quota prevents senders from congesting receivers. Can be applied complementary to window/rate-based control. & A sender needs to first coordinate with the receiver before it embarks on transmission which can add some latency overhead. \\
\hline
\end{tabular}
\end{table*}


\subsection{Traffic Shaping} \label{shaping}
We can improve network performance by making sure that it conforms to required profile and policy rules. This can reduce contention while using network resources. For example, traffic shaping can prevent some flows from hogging others. Shaping can be used to provide some level of resource isolation and guarantees in cloud environments with many users. Finally, traffic shaping can be useful in resource scheduling where senders follow rates specified in the schedule. A summary of traffic shaping techniques has been provided in Table \ref{dc-tc-shaping}.

\begin{table*}[t!]
\caption{Overview of traffic shaping techniques} \label{dc-tc-shaping}
\centering
\begin{tabular}{|p{2.5cm}|p{7cm}|p{7cm}|} 
\hline
\textbf{Scheme} & \textbf{Description} & \textbf{Limitations} \\
\hline
\hline
Rate Limiting \S \ref{rate-limiting} & Limits the transmission rate of outgoing packets from the rate limiter. A token bucket is usually used to limit maximum persistent packet transmission rate to token arrival rate. & Rate limiting has limited accuracy depending on how it is enforced, in software (usually less accurate but cheaper) or hardware (usually more accurate but needs hardware support). Also, various implementation techniques in software lead to different accuracy such as in the operating system kernel or using kernel bypass modules (see \S \ref{userspace-pp}). \\
\hline
Packet Pacing \S \ref{packet-pacing} & Inserts time spacing between consecutive packs to spread them uniformly across a window of round trip time (RTT). This reduces traffic burstiness and average network buffer occupancy, therefore improving end-to-end latency. & Packet pacing has limited accuracy and similar to rate-limiting, its accuracy depends on the implementation technique used. \\
\hline
\end{tabular}
\end{table*}

\subsubsection{Rate Limiting} \label{rate-limiting}
Rate limiting is usually applied by passing traffic through a Token Bucket filter. This ensures that average transmission rate does not exceed the token generation rate. Tokens are generated at a specific rate and an arriving packet can only be sent if there is a token available. Tokens are accumulated if there is no packet to be sent, but there is usually a cap on how many. This cap is to limit traffic burstiness in case the token bucket is idle for a while. Examples of works employing rate-limiting to provide resource isolation and guarantees include \cite{jump, flowtune, eyeq, gatekeeper, qcn, d3, scc}.

Rate-limiting can be done in OS Kernel, as part of the hypervisor in a virtualized environment, or via NIC. Scheduling of packets in software (Kernel or Hypervisor) is generally less precise and can be computationally intensive in high bandwidths and with a large number of flows. To reduce CPU utilization, OSes usually send packets to NIC in batches which can further reduce the scheduling precision. For example, Classful Queuing Disciplines (Qdisc) offered by Linux allows for coarse grained rate-control by determining the time and count of packets that NIC receives from RAM, however, the actual schedule of packets on the wire is determined by NIC.

To improve software rate-limiting performance one can use userspace packet processing tools some of which we point to in \S \ref{userspace-pp}. For example, Carousel \cite{carousel} is a rate-limiter that is implemented as part of a software NIC in userspace. Carousel uses a variety of techniques to reduce CPU and memory usage and improve scalability including deferred completion signalling (rate-limiter only signals completion to applications when data is actually transmitted to offer backpressure and minimize buffer space usage) and single queue shaping (using a timing wheel and by assigning timestamps to packets over the time horizon). 

Using NIC to schedule and send packets given rate limits reduces CPU load and traffic burstiness. Effective rate-limiting in hardware demands support for multiple queues and classes of traffic with hardware rate-limiters attached to them.

Hybrid rate-limiting approaches can be used to support a large number of priority classes while reducing hardware complexity and keeping scheduling precision high. NicPic \cite{nicpic} classifies and stores packets in queues located in RAM and labels them with proper rate limits by host CPU. Packets are then fetched by NIC via DMA and scheduled using hardware rate-limiters. NIC first decides on which flow's packets to send and then pulls them from RAM.

As the last option, rate-limits can also be applied at the application layer. Applications can do this by limiting the volume of data handed off to transport layer over periods of time. This approach is simple but requires changes to applications. Also, rate-limiting precision will be limited. Finally, this may lead to bursty traffic as incoming application data to transport layer may get buffered prior to transmission on the wire, i.e., applications have no control over how data is eventually sent.

\subsubsection{Packet Pacing} \label{packet-pacing}
Packet Pacing is the process of adding space between consecutive packets so that they do not arrive at the network back to back which reduces traffic burstiness. Burstiness can degrade network performance in several ways. Long bursts can overload switch buffer and create consecutive packet drops. Average latency of all packets then increases since they have to wait in longer queues. In addition, it creates transmission rate oscillations making it hard to do careful bandwidth allocation \cite{dctcp, hull}.

Earlier work \cite{tcppacing_old} experimenting with pacing in a general network setting has shown that it can considerably reduce queuing delay. Combined with other types of congestion signals, pacing can improve the performance by evenly distributing traffic across the timeline \cite{hull}. In addition, pacing should only be applied to long-running and throughput-oriented flows to reduce their impact on short latency-sensitive flows \cite{hull}. The benefit of pacing depends on the network bandwidth-delay product, buffer size, and the number of flows. Having so many flows that share a buffer can reduce the effectiveness of pacing due to inter-flow burstiness and creation of synchronized drops \cite{tcppacing}.

Pacing can be done in both hardware and software, but hardware pacing can be more effective due to higher scheduling precision, especially at high rates where spacing between packets is tiny \cite{hull}. Software pacing may be performed by end-hosts as part of a driver or a kernel module. In cloud environments, due to widespread use of virtualization, packets may be paced at the virtual interfaces in the hypervisor software. Pacing in software may be overridden by NIC offloading features such as LSO, if enabled, since NIC is the actual entity that sends packets out on the wire. In hardware pacing, packets are buffered at the NIC, each assigned a timer, and scheduled to be sent when their timer goes off.

Pacing can also be done at the network edges (e.g. ToR switches) as opposed to end-hosts. For example, Queue Length Based Pacing (QLBP) \cite{pacing_small_buffer_practical} uses a pacing controller attached to edge queues to determine when the next packet in the queue is supposed to be sent as a function of queue length.


\begin{table*}[t!]
\caption{Overview of prioritization techniques} \label{dc-tc-prioritization}
\centering
\begin{tabular}{|p{2cm}|p{2.5cm}|p{12cm}|} 
\hline
\textbf{Property} & \textbf{Scheme} & \textbf{Description (Benefits/Drawbacks)} \\
\hline
\hline
\multirow{2}{*}{Classification} & Static & A flow's priority is fixed once assigned. This approach can be applied when flow properties (e.g. size) are known apriori. \\
\cline{2-3}
 & Dynamic & A flow's priority may change over time according to its behavior, i.e., number of packets sent over time. This approach can be used if flow properties are unknown apriori. \\
\hline
\multirow{3}{*}{Criteria} & by flow size & Mimics the Shortest Remaining Processing Time (SRPT) scheduling discipline which aims at minimizing mean flow completion times. \\
\cline{2-3}
 & by flow deadline & To minimize deadline miss rate or lateness by first satisfying flows with closer deadlines. \\
\cline{2-3}
 & by class of service & If an application or a tenant has a higher service level requirement or agreement, flows associated with them can be prioritized accordingly to minimize the effect of other applications/tenants using the same physical network. \\
\hline
\multirow{2}{*}{Location} & at the switches & Switches can keep state information on flows passing through them and determine their priority according to flows' behavior. For example, in case of prioritization by flow size, switches can estimate flow sizes by counting the number of packets they have sent (mimicking least attained service discipline). Keeping state information at the switches may be costly when there are many flows. \\
\cline{2-3}
 & at the end-hosts & End-hosts can mark packets with priority tags allowing switches to simply enforce priorities according to tags. This reduces switch complexity but requires changes to the end-hosts' software or hardware protocol stack. \\
\hline
\multirow{3}{*}{Implementation} & at Layer 2 & Ethernet standard IEEE 802.1Q priority based  forwarding. \\
\cline{2-3}
 & at Layer 3 & Differentiated Services (DiffServ) can be used at the IP layer. \\
\cline{2-3}
 & Custom approaches & Can be used by adding support to switches and/or end-hosts, may require changes in software/hardware to switches and/or end-hosts. \\
\hline
\multirow{2}{*}{Enforcement} & Strict & A lower priority flow is only sent when there are no packets available from any of the higher priority flows. This minimizes the effect of lower priority flows on higher priority ones but can lead to starvation of lower priority flows. \\
\cline{2-3}
 & Non-strict & A lower priority flow can be sent even if there are packets available from higher priority flows. This occurs when a required volume of higher priority flows' demand is satisfied (e.g. one low priority packet is sent for every $K \ge 1$ high priority packets sent) and mitigates the starvation problem of lower priority flows. \\
\hline
\end{tabular}
\end{table*}

\subsection{Prioritization} \label{priority}
Table \ref{dc-tc-prioritization} provides an overview of this section. Classifying flows based on their priorities and treating them accordingly can improve performance. Such prioritization can be done in-network by using multiple queues at the switches and allowing higher priority traffic to go over lower priority traffic \cite{pias, pfabric, pase, jump}, and at the senders by performing rate-control according to priorities \cite{d3, d2tcp, pdq, l2dct}.

Priorities are usually assigned either based on flow size to minimize mean latency (by mimicking SRPT scheduling policy) \cite{pias, pfabric} or based on deadlines to minimize the number of deadline missing flows \cite{pase, phost}. Control traffic is naturally prioritized to improve the feedback timeliness and quality (e.g. ACKs in TIMELY \cite{timely} and Trimmed Packets in NDP \cite{ndp}) or decrease control loop delay (e.g. RTS in pHost \cite{phost}).

For many applications, flow sizes are either known or can be roughly estimated upon initiation \cite{pfabric, pase, pdq, d3} making it easy to assign priorities by size to reduce mean latency. In case flow sizes are unknown apriori, dynamic prioritization can be used where packets of a flow first receive the highest priority, but get demoted to lower priorities as more of them is seen.

For example, dynamic Packet Prioritization (DPP) \cite{dpp} uses two queues, an express queue which has higher priority and a normal queue. It reduces the priority of long running flows by counting their packets against a threshold. Having multiple queues allows for more precise classification of flows \cite{pias, pase}. However, recent work shows that most benefit in reducing mean FCT can be obtained using up to 8 queues \cite{pfabric}. Finding proper threshold values based on which flows are demoted to lower priorities may require solving an optimization problem that takes into account the flow arrival rates and flow size distribution. In datacenter environments with known traffic characteristics and workloads, such thresholds may be determined offline and applied in real-time \cite{pias}. It is also possible to virtually emulate the behavior of having infinite number of queues using only two actual queues per switch port, a high and a low priority queue \cite{epn}. This can be achieved by assigning flows with highest priority to high priority queue while the rest of flows to low priority queue and dynamically changing flows assigned to high priority queue when other flows with a higher priority complete.

Prioritization can be performed fully at switches by keeping state on flows passing through and using some priority assignment criteria such as the total number of packets sent. This simplifies end-hosts at the expense of higher computational and memory burden on the switches.

Another approach is for the end-hosts to tag flows with priority tags while switches just process tags and put packets in proper queues \cite{jump, pias}. End-host priority tagging can be done at the NIC, OS kernel, hypervisor, or even by applications before packets are sent to the network. In case of virtualization, if end-host VMs cannot be trusted to assign priorities properly, middleboxes can be used (e.g. at the hypervisor) that monitor and tag packets (e.g. using OpenVSwitch \cite{openvswitch}) which applies to both static \cite{pfabric} and dynamic \cite{pase, pias} prioritization.

Priorities can also be assigned in different ways. Ethernet standard IEEE 802.1Q priority based forwarding \cite{8021p} that provides 8 levels is supported by many switch vendors and can also be used in datacenters \cite{jump}. At the IP layer, Differentiated services (DiffServ) \cite{diffserv} can be used \cite{rdma_scale}. Custom queuing techniques and headers can also be used which may require changes to both switches and end-hosts \cite{pfabric}.

Strictly prioritizing flows can lead to starvation where lower priority flows cannot make progress due to large volume of higher priority traffic. A simple solution is to use weighted queuing instead of strict prioritization. For instance, DAQ \cite{daq} uses a weighted round-robin between long and short flows to make sure that throughput-oriented flows keep making progress. Aging can also be used to address starvation while minimally affecting critical flows. An aging-rate can be used to increase the priority of low priority flows as a function of their waiting time \cite{pdq}.


\subsection{Load Balancing} \label{balancing}
Datacenter topologies typically provide a large degree of path redundancy. Properly distributing load across these paths reduces contention among flows while increasing overall resource utilization. Without effective load balancing many links may not be utilized while some experiencing congestion \cite{traffic_dc_char, flyways}. Table \ref{dc-tc-balancing} provides an overview of general load balancing concepts and their trade-offs.

\begin{table*}[t!]
\caption{Overview of load balancing techniques} \label{dc-tc-balancing}
\centering
\begin{tabular}{|p{1.5cm}|p{2.5cm}|p{12.5cm}|} 
\hline
\textbf{Property} & \textbf{Scheme} & \textbf{Description (Benefits/Drawbacks)} \\
\hline
\hline
\multirow{3}{*}{Classification} & Static & A new flow is assigned to any of the available paths using some fixed criteria such as by hashing parts of its packets' header. This approach is simple but inflexible. For example, in case two throughput oriented flows are assigned to the same path, they cannot be moved to other less utilized paths later. \\
\cline{2-3}
& Dynamic (reactive) & Flows can be moved across any of the available paths according to available bandwidth. Offers a better performance in general but adds the complexity of measuring link utilizations, accounting for flows, and calculating best flow assignments accordingly. \\
\cline{2-3}
& Dynamic (proactive) & After a flow is assigned to one of the available paths according to some criteria, its assignment will remain fixed. The initial assignment is performed according to network conditions such as available bandwidth. This approach is somewhat between the previous two assignments above in terms of implementation overhead, flexibility and performance. \\
\hline
\multirow{4}{*}{Granularity} & per packet & Finest load balancing but can lead to high reordering. \\
\cline{2-3}
& per flow & Coarse load balancing but achieves minimal reordering. \\
\cline{2-3}
& per flowlet & A flowlet's size dynamically changes according to differences of latencies of candidate paths. At high rates and/or high latency difference between available paths, a flowlet can become significantly large. As a result, this can result in both fine and coarse grained load balancing (it is always somewhere between per packet and per flow). Flowlets have been found effective for load balancing over asymmetric (i.e., with different available bandwidth) paths \cite{letitflow}. As a drawback, flowlets may cause reordering of small flows and hurt their completion times. \\
\cline{2-3}
& per flowcell & A flowcell has a fixed size that is usually about tens of packets. Using flowcells simplifies load balancing compared to flowlets (no need to carefully measure path latencies and schedule accordingly) and reduces possible reordering of small flows. It can however significantly increase reordering for larger flows that will be broken into many flowcells. \\
\hline
\end{tabular}
\end{table*}

Load balancing can be static or dynamic (adaptive). Static approaches use a fixed criteria to assign traffic to available paths such as by hashing specific fields from packet headers. For example, ECMP \cite{ecmp} is a popular static load balancing technique that only distributes load across equal cost paths. Adaptive load balancing dynamically selects paths for traffic according to distribution of load to minimize hot-spots. Various criteria can be used for path assignment such as per-hop or per-path queue occupancies \cite{load_balancing_survey}. After choosing the initial path according to current load, some adaptive approaches keep monitoring the network status and distribution of traffic. They change direction of traffic to eliminate or reduce hot-spots. These approaches are referred to as reactive. If not applied with care, reactive load balancing might lead to oscillations.

Examples of reactive dynamic load balancing techniques include Planck \cite{planck}, Hedera \cite{hedera}, MPTCP \cite{mptcp}, DIBS \cite{dibs} and CONGA \cite{conga}. Planck uses a controller that monitors traffic and generates congestion events that include the transmission rate of flows passing through the congested link. It then routes traffic away from congested spots. Hedera initially places flows via hashing, and then uses a central controller to monitor the network, detect long running flows and reschedule such flows on a lightly loaded path to balance the load. MPTCP establishes multiple sub-flows from the beginning and then shifts load between them according to the level of congestion observed across each sub-flow. DIBS forwards packets that arrive at a full queue to one of the nearby switches instead of dropping them which will be forwarded towards the destination through another path. CONGA proposes a technique for leaf-spine topologies \cite{leaf-spine} based on lazy evaluation. A leaf switch has a table which holds the load seen along its outgoing paths. Such load information is collected by receiving switches and then piggybacked on traffic.

Some proactive adaptive approaches include DeTail \cite{detail}, Presto \cite{presto} and Expeditus \cite{expeditus}. DeTail uses a per-hop adaptive method and operates in lossless environments with layer 2 flow control \cite{why_lossless_ethernet, pfc, pfc_cisco}. At every hop, packets are forwarded to the egress port with minimal queuing. Presto breaks flows into small units called cells and sends them across all available paths. This implies that small flows are kept intact as long as they fit into one cell. Expeditus dynamically assigns flows to paths in 3-tier Clos topologies. It uses dedicated packet tags to communicate load information across switches for path selection upon arrival of a new flow. For path election, the upstream switch sends a message to its downstream peer expressing congestion at its egress ports. The receiving switch compares the upstream congestion metrics with the ones for its ingress ports choosing the ports that minimize the maximum congestion along the path.

Load balancing can be done per-packet, per-group of packets that belong to the same flow, and per flow. While considering these options, two important performance metrics are packet reordering and distribution of load across the network. Reordering is known to waste server resources and increase latencies \cite{juggler}.

\textit{per flow} load balancing minimizes packet re-ordering. Congested spots might be created in case multiple large flows are assigned to the same links. Even when carefully placing flows, per flow load balancing provides limited benefit if there is a large number of throughput-oriented flows \cite{load_balancing_survey}. To improve performance, one might reroute flows from their initially assigned paths according to network conditions. However, moving flows from their initial paths might still result in re-ordering. For example, Hermes \cite{load-balancing-wild} performs per flow load balancing added that it can perform fast rerouting of flows according to network conditions. It can potentially reroute flows per packet in case conditions change rapidly and continuously. One however needs to consider the stability of such schemes, especially as load increases, since many flows may be interacting in the network (in case of instability, such schemes may keep changing routes, which increases re-ordering, while not improving load balancing). Flier \cite{flier} is another flow-level load balancing scheme that reroutes flows according to level of congestion observed (via checking ECN markings) and failures (by paying attention to timeouts).

\textit{Per-packet} load balancing provides the finest balancing degree but leads to packet re-ordering. For every new packet, one of the available paths can be chosen either randomly, according to usage history, or adaptively based on the distribution of load. Valiant Load Balancing (VLB) \cite{vlb} can uniformly distribute packets across paths. Packet Spraying \cite{spray} (a.k.a. packet scatter) uses a Round Robin approach to multiplex packets of a flow across all possible next hops. Packet scatter can greatly balance load at the network core \cite{mptcp-improving, phost} and reduce latency \cite{pfabric, fastlane}. Another per-packet approach, DRB \cite{per_packet_load_balancing}, reduces network queuing and buffering required at the receiver by choosing the forwarding paths that avoid clustering of packets between same source-destination pairs. DRILL \cite{drill} determines the forwarding path of every packet of a flow independently by considering per port local queuing at the switches sending an arriving packet to the least occupied queue. It is argued that DRILL's packet reordering is minimal due to similar latencies across all the paths between every source and destination pair due to well-balanced switch queue occupancies.

Another option is to group several packets of a flow and perform load balancing \textit{per-group of packets}. Packets can be grouped according to their inter-packet delay or their accumulative size. In the former case, all packets of a flow whose inter-packet delay is less than some timeout threshold form a flowlet \cite{flowlet}. Flowlet scheduling essentially makes use of natural traffic burstiness for load balancing \S \ref{burst}. In the latter case, packets are grouped with a limit on total group volume to form flowcells \cite{tinyflow, presto}. Each flowlet or flowcell can be sent over a different path to distribute load.

In flowlet scheduling, smaller timeout values allow for finer load balancing while larger values reduce reordering. To minimize reordering, one can choose the timeout value to be greater than the difference between latencies of paths with minimum and maximum latencies. Flowlets have been found to effectively balance load while incurring minimal reordering in datacenters \cite{conga, flowtune}. However, dependence of flowlet switching on inter-packet intervals could lead to creation of arbitrarily large flowlets at high rates, which could lead to congestion in case they collide. Another drawback is the possibility that small flows are split into several flowlets which could lead to reordering and increased latency.

In flowcell scheduling, smaller flowcells balance load better while larger ones reduce reordering. Flows shorter than the cell size are guaranteed to be sent on a single path minimizing their latency and reordering. Previous work has used grouping thresholds of tens of kilobytes (10~KB at ToR switches \cite{tinyflow} and 64~KB at the hypervisor layer \cite{presto}) to effectively spread the load across the network and reduce creation of random hot-spots. As a drawback, this approach may lead to higher reordering for long flows compared to flowlets.

\subsubsection{Data and Task Placement} Many datacenter applications need to access data stored on multiple servers to perform computations or respond to queries. Therefore, placement of data determines what options are available to access them. Such data could be a value for a key in a distributed key-value store or an object in a replicated or erasure coded store. For example, any of the replicas in a replicated store can be accessed or any of the $k$ pieces out of $n$ pieces of data would allow data recovery in an $(n,~k)$ erasure coded storage system.

Pieces of data can be distributed across racks and servers (depending on topology) to allow wider load balancing options. For example, Ceph \cite{ceph} uses Selective Replication that distributes copies of the original data across the cluster according to their popularity. Ceph also distributes contents of large directories and ones with lots of writes across many servers to reduce hot-spots. HDFS \cite{hdfs} allows for similar features but also considers the network topology while distributing replicas. For example, there is better connectivity and usually higher available bandwidth within a rack.

Placement of tasks (execution) could be as important as placement of data. Task schedulers and resource managers can place computation in accordance with placement of data to reduce network usage, contention for network access and queuing \cite{corral, tetris, silo, ali-INFOCOM, ali-near-data-locality-scheduling, gb-pandas}. A task scheduler may consider the flow scheduling policy of the network (FCFS, SRPT, Fair Sharing, etc.) in addition to placement of data to improve overall task completion times \cite{neat}.

\subsubsection{Routing and Forwarding} \label{forwarding}
Switches forward packets according to Forwarding Information Base (FIB) which contains a set of rules that determine outgoing port(s) for incoming packets. FIB rules can be installed proactively or reactively. Proactive installation may result in a larger number of rules as not all of them may be used at all times while reactive installation of rules may incur setup time overhead. Such rules can be installed either directly or by a routing protocol that calculates the rules and installs them such as BGP, IS-IS or OSPF. In case of routers, FIB usually reflects a subset of Routing Information Base (RIB) which is a table of routes learned or calculated by a router. For load balancing, various forwarding techniques can be used to direct traffic across several paths.

Standard distributed protocols can be used for load balancing. As a Layer 2 solution, VLAN based load balancing puts same machines on several virtual networks allowing traffic to be spread across paths via using different VLAN tags. It however provides limited scalability due to creation of large broadcast domains. Layer 3 routing for large networks with support for load balancing can be used in case multiple next hops are available for a destination. For example, Equal Cost Multipathing (ECMP) statically selects the next hop by hashing packet header fields. For fast convergence, IGP routing protocols can be used such as IS-IS or OSPF. Load balancing using these protocols is challenging since path costs need to be exactly equal and the number of supported equal paths is limited. BGP provides higher flexibility for this purpose \cite{bgp_better_igp} and can be customized to converge fast \cite{bgp_pic}.

Load balancing can be performed using centralized techniques. In Layer 2, scalable forwarding can be built by replacing the default MAC address broadcast and discovery approach with a centralized one \cite{shadow_macs, past}. A controller can then setup Layer 2 forwarding that accounts for path diversity \cite{past}. Centrally implemented Layer 3 approaches allow FIBs to be calculated centrally using free routing software stacks such as Quagga \cite{quagga_flow}, and installed on switches to implement various protocol stacks such as OSPF and IS-IS with ECMP support which provides much higher flexibility \cite{monsoon, vl2, b4, portland}. For example, to help BGP converge faster, a central controller can calculate and update BGP tables in routers to achieve desired forwarding behavior \cite{jupiter}. Simpler centralized approaches in which new FIB rules are installed directly by a controller upon arrival of a new flow can also be used. Forwarding can be done in a way that distributes load, either by hashing or adaptively selecting least loaded paths. Using this approach, careful consideration of scalability is necessary.

Previous approaches relied mostly on network for routing. An end-host based approach is Source Routing which simplifies the network by moving the forwarding information to packets, and eliminating the need to disseminate updates of forwarding state \cite{source_routing}. To properly encode paths in packets, end-hosts need to be aware of network topology and paths. In addition, a mechanism is needed to detect and disseminate network status and failure information to the end-hosts. BCube \cite{bcube} uses probe packets to measure available bandwidth across paths and assign new flows to least loaded ones. PSSR \cite{secondnet} encodes a list of outgoing port numbers instead of addresses for next hops to decouple addressing from forwarding.

\subsubsection{Effect of Failures} \label{failures}
The scale of datacenter networks has made failures an important concept. Even using high quality and expensive equipment, the probability that some network element fails (e.g. a switch, a port or a link, etc.) can be considerably large at any moment \cite{f10}. When some network equipment fails, capacity in the network decreases accordingly; however, since datacenter networks are usually connected with large degrees of redundancy, network failures rarely lead to complete inability to reach parts of the network. However, the capacity loss due to failures in some parts of network can lead to complications in load balancing by affecting the effective capacity of different paths, i.e., creating capacity asymmetries. This may not have a significant impact on topologies that are inherently asymmetrical (e.g. JellyFish, Xpander, etc.); however, more basic load balancing techniques, such as ECMP, which are used in symmetric topologies (e.g. Fat-Tree, Leaf-Spine, VL2, etc.), will have a hard time effectively distributing load in case of failures \cite{wcmp, conga, letitflow}. This is noteworthy considering that many industry datacenters are based on symmetric topologies. 

A variety of solutions have been proposed to perform better load balancing in case of capacity asymmetries across paths examples of which include WCMP \cite{wcmp}, CONGA \cite{conga}, HULA \cite{hula}, Presto \cite{presto}, LetFlow \cite{letitflow}, DRILL \cite{drill} and Hermes \cite{load-balancing-wild}. WCMP mitigates asymmetries by extending ECMP and assigning weights to different paths proportional to their capacity referred to as weighted traffic hashing. Using WCMP, weights determine the number of hash entries per outgoing port at the switch which are selected proportional to capacity. CONGA and HULA operate by performing path-wise congestion monitoring and shifting traffic accordingly. As a result, if capacity of a path is reduced due to failures, its congestion metric will increase faster and it will automatically be assigned less traffic. Presto applies a weighted forwarding mechanism in a way similar to WCMP where weights are pushed to the end-host virtual switches. LetFlow is a simple approach where flowlets are used as means to dynamically adapt to varying path capacities. LetFlow relies on natural property of flowlets which allows them to shrink or expand (in size) according to available capacity over paths. DRILL performs load balancing over asymmetric topologies by first breaking them into groups of symmetric components and then performing load balancing on them per switch. Symmetric paths should have equal number of hops and their queues should be shared by same flows at every hop from source to destination. Hermes uses comprehensive sensing at the end-hosts to monitor path conditions and reroute flows affected by failures or congestion caused by asymmetries. Hermes considers monitoring of latency and ECN markings for detection of congestion while looking at frequent timeouts and retransmissions as signs of failures. In addition, to improve visibility into path conditions, Hermes uses active probing by sending small probe packets between end-host pairs periodically.


\subsection{Multipathing} \label{multipathing}
To improve the overall throughput for a single flow and provide better reliability in case of failures, multi-path techniques can be used \cite{mptcp, mptcp-improving, mptcp_cc, mptcp-hard}. A flow is split into multiple sub-flows each sent over a different path. To receive traffic on multiple paths, the receiver needs to buffer data received from each sub-flow and put them in order. Buffering is proportional to sum of throughput across paths times the latency of the longest path \cite{mptcp-hard, delayed_ack_mptcp}. Although latencies are quite small in datacenters, link bandwidths can be significantly high. Additional sub-flows can generally increase both memory and CPU utilization \cite{mptcp-hard}.

Depending on flow sizes, the applications may decide whether to use multiple sub-flows. Overhead of setup and tear-down for multiple sub-flows may be considerable for short flows. For long-running background flows, using every bit of bandwidth available through multipathing may improve their total average throughput.

Several examples of multipath transports include MPTCP \cite{mptcp-improving}, XMP \cite{xmp} and MMPTCP \cite{mmptcp}. MPTCP leverages ECMP to route sub-flows over various paths and increase total throughput while balancing load across paths by moving load across sub-flows. XMP approximates the solution to an optimization problem that maximizes utilization. It uses RED \cite{red, red_one_parameter} with ECN marking to keep the queue occupancies low. XMP achieves most benefit by using two sub-flows. MMPTCP aims to improve FCT for short flows by employing a two phase approach. First phase uses packet scatter by randomizing source port numbers and routing via ECMP to minimize FCT for short flows and second phase uses MPTCP to maximize throughput for long flows.

Although multipathing is generally helpful in increasing utilization, the benefit it offers is limited under various conditions. Under heavy workloads, multipathing may not improve throughput if most paths are already highly utilized \cite{pdq}. In addition, if paths have different characteristics, such as latency or capacity, multipathing may offer marginal increase (or even decrease) in throughput \cite{mptcp_cc}. This may occur in datacenters in case different communication technologies are used across machines, such as a combination of wireless and wired networks or if available paths have varying number of hops or different capacities. Another complication is caused by overlapping paths where a multihomed sender's paths to the receiver actually have common edges \cite{multipathing-issues}. This can occur in datacenters as well in case of link failures which reduce available paths or if sub-flows are mistakenly hashed to the the same links. Finally, multipathing may lead to unfairness if multipath flows share a bottleneck with regular flows \cite{mptcp_honda}. Unfairness may also arise when multipath flows with different number of sub-flows compete for bandwidth over a bottleneck.


\subsection{Scheduling} \label{scheduling}
Given a list of flows with their priorities and demands, the scheduling problem aims to optimize an utility function of several performance metrics such as utilization, fairness or latency. The objective is usually to maximize throughput for bandwidth-hungry flows and minimize FCT for latency-sensitive flows considering fairness among flows in each class of service. Different scheduling techniques can be used to reduce FCT, provide better bandwidth guarantees to long-running flows, and help deadline flows meet their deadlines. In general, scheduling a mix of flow types requires formulation of a complex optimization problem that is generally computationally expensive to solve. Table \ref{dc-tc-scheduling} offers an overview of scheduling techniques presented here.

\begin{table*}[t!]
\caption{Summary of scheduling techniques} \label{dc-tc-scheduling}
\centering
\begin{tabular}{|p{2.5cm}|p{7cm}|p{7cm}|} 
\hline
\textbf{Scheme} & \textbf{Description} & \textbf{Limitations} \\
\hline
\hline
Reservation \S \ref{reservation} & Reserving bandwidth prior to a sender's transmission minimizes congestion and queuing latency. & Reservation adds the latency overhead of calculating a transmission schedule before a flow is allowed to transmit. In addition, it is challenging to enforce a transmission schedule network wide. Inaccuracies in transmission rates may necessitate continues updates to the schedule which is costly. \\
\hline
Redundancy \S \ref{redundancy} & A flow can be replicated multiple times to reduce the effect of high tail latency. The fastest reply is obtained and then replicas are terminated. & Effective if there is a considerable gap between tail and median latency. In addition, it is less effective when network is heavily loaded. \\
\hline
Deadline-Awareness \S \ref{deadline-awareness} & Scheduling can be done according to deadlines to minimize deadline miss rate and lateness. & It may not be possible to meet all the deadlines in which case it should be determined whether deadline miss rate is more important (hard deadlines) or lateness (soft deadlines). In presence of deadlines, it is unclear how to effectively schedule traffic if flow sizes are not known apriori or cannot be estimated. \\
\hline
Disciplines \S \ref{disciplines} & A variety of scheduling disciplines can be applied according to desired traffic control objectives. For example, SRPT minimizes mean latency while Fair Queuing maximizes fairness. & Disciplines can usually optimize for only one performance metric. A mix of scheduling policies can hierarchically optimize for more than one objective. If a utility of objectives is desired, well-known policies may provide solutions far from optimal. \\
\hline
Preemption \S \ref{preemption} & Allows the network to update current schedule (along with already scheduled flows) according to new flow arrivals. & Preemption may offer limited benefit if all flows have similar properties (i.e., size, deadline, etc.). \\
\hline
Jittering \S \ref{jittering} & Prevents a sender from initiating many flows together. For example, this helps mitigate the incast problem \S \ref{incast}. & This approach only offers very coarse grained control over incoming traffic. \\
\hline
ACK Control \S \ref{ackcontrol} & By carefully controlling when ACKs are sent to senders, a receiver can control the incoming flow of traffic. & This approach only offers coarse grained control over incoming traffic. \\
\hline
\end{tabular}
\end{table*}

\subsubsection{Reservation} \label{reservation}
To provide better bandwidth guarantees and prevent creation of congestion spots resources may be first checked for availability and then allocated before an end-host can start transmitting a flow. Requesting resources can be done in different units and such requests might be processed centrally or in a distributed fashion.

In a fully centralized approach, end-hosts can report their demands to a central scheduler and ask for transmission slots or rates. Some examples are TDMA \cite{tdma}, FastPass \cite{fastpass}, FlowTune \cite{flowtune} and TAPS \cite{taps}. TDMA uses a coarse-grained centralized approach in which end-hosts send their demands to a fabric manager which then allocates them contention-less transmission slots. The scheduling is performed in a round by round basis where each round is several slots during which different hosts can communicate over the fabrics. FastPass allocates slices of time on a per-packet basis to improve utilization and considers variations in packet size. FlowTune performs centralized rate allocation on a per-flowlet basis. TAPS uses a central SDN controller (please refer to \S \ref{pfp}) to receive and process flow requests with known demands, verify whether deadlines can be met on a per-task basis, allocate contention-less slices of time for senders to transmit and install forwarding rules.

Distributed reservation can be done upon connection setup. A sender can request the rate at which it would like to transmit. This rate along with the available bandwidth is considered by network fabrics to determine the rate and path of the flow. Allocated rate can be piggybacked on ACKs. RCP \cite{rcp} and PDQ \cite{pdq} perform the rate allocation at switches with this approach. Receiver-based reservation can also be used. Receivers view multiple incoming flows and determine their transmission schedule. Senders can also communicate their demands to the receiver which calculates the transmission schedule \cite{incast_cross_schedule}. In addition, token-based techniques can be used where a receiver sends back tokens to allow senders to transmit a unit of data (e.g. packet) per token. The token-based approach has been found very effective in addressing the incast problem which is achieved by controlling the rate at which tokens are sent back to senders from the receiver to ensure that data arriving at the receiver conforms with available capacity \cite{phost, ndp, credit-based-cc}.

\subsubsection{Redundancy} \label{redundancy}
Tail latency is an important quality metric for many latency-sensitive applications, such as search, as it determines quality of user experience. Late flows can take much longer than median latency to finish \cite{dctcp, bing}. An effective approach is to replicate flows, use the fastest responding replica and terminate the rest. For simplicity, creation of replicates may be performed completely at the application layer. The probability of more than one replica being late is usually significantly small. Replicated flows can be scheduled on different paths and can even target different servers to reduce correlation.

One approach is to replicate every flow and then take the one whose handshaking is finished earlier and terminate the rest \cite{repflow_node_js}. Another approach is to only replicate slow requests by reissuing them \cite{bing}. It is necessary to judiciously decide on the number of redundant requests to balance resource usage and response time. Since only a tiny portion of all flows are usually laggards, additional resources needed to allow large improvements may be small \cite{bing}.

\subsubsection{Deadline-Awareness} \label{deadline-awareness}
For many applications, criticality of flows can be captured as deadlines. Scheduling techniques are expected to minimize deadline miss rate. In case of hard deadlines, in which case delivery after deadline is pointless, flows can be terminated early if their deadlines cannot be met. D2TCP \cite{d2tcp}, D3 \cite{d3}, PDQ \cite{pdq}, and MCP \cite{mcp} are examples of deadline-aware scheduling schemes that reduce deadline miss rate by performing rate control according to flow deadlines, either by explicit rate allocation (D3, PDQ) or implicitly by adapting sender's outstanding window size (D2TCP, MCP). Tempus \cite{calendaring} formulates a complex optimization problem to maximize the fraction of flows completed prior to their deadlines while considering fairness among flows. Amoeba \cite{amoeba} aims to guarantee deadlines by performing initial admission control via formulating an optimization problem. RCD \cite{rcd} and DCRoute \cite{dcroute} aim to quickly determine whether deadlines can be met by performing close to deadline scheduling and guarantee deadlines via bandwidth reservation. In addition, considering task dependencies in meeting deadlines can help reduce deadline miss rate of tasks \cite{varys, taps}.

While it is important to meet deadlines, finishing deadline flows earlier than necessary can hurt the FCT of latency-sensitive traffic \cite{karuna}. In general, we can form an optimization problem for scheduling flows \cite{mcp}. A study of how well-known scheduling policies perform under mix flow scenarios with varying fraction of deadline traffic can be found in \cite{mix-flow-mohammad}.

\subsubsection{Disciplines} \label{disciplines}
The following are some well-known scheduling disciplines. First Come First Serve (FCFS) is a simple policy where a task has to be completed before the next task can begin. FCFS provides bounded lateness when scheduling flows with deadlines \cite{FCFS-tardiness} and is close to optimal for minimizing tail completion times given light-tailed flow size distributions \cite{fcfs}. Processor Sharing (PS) divides available resources equally among flows by giving them access to resources in tiny time scales. Fair Queuing (FQ) approximates PS within transmission time of a packet and can be used to enforce max-min fairness. Earliest Deadline First (EDF) minimizes deadline miss rate for deadline flows. Shortest Job First (SJF) minimizes mean flow completion time in offline systems where all flows and their demands are known apriori. For online systems where requests can be submitted at any time, Shortest Remaining Processing Time (SRPT) which is preemptive, minimizes mean FCT \cite{srpt}. SRPT also offers close to optimal tail completion times while scheduling flows with heavy-tailed size distributions \cite{fcfs}. Least Attained Service (LAS) \cite{las}, which prioritizes less demanding flows, can be used to approximate SJF without apriori knowledge of flow demands \cite{non-clair}.

Many works use policies that approximate or implement well-known scheduling disciplines. RCP \cite{rcp} performs explicit rate control to enforce processor sharing across flows sharing the same links. PDQ \cite{pdq} combines EDF and SJF giving higher priority to EDF to first minimize deadline miss rate and then minimize FCT as much as possible. FastPass \cite{fastpass}, implements \textit{least recently allocated first} giving each user at least as much as their fair share to reach global user-level max-min fairness. Also, across flows of a user, \textit{fewest remaining MTUs first} is used to minimize FCT by emulating SJF. pFabric \cite{pfabric} and SFS \cite{sfs} follow the shortest remaining flow first which is essentially SRPT. PIAS \cite{pias} uses a dynamic priority assignment approach which approximates LAS by counting the number of packets sent by flows so far. RACS \cite{racs} uses weighted processor sharing with configurable weights to approximate a spectrum of scheduling disciplines such as SRPT and LAS.

Aside from research, one may be interested in what disciplines can be enforced using industry solutions. Switches by default provide support for FIFO queues per outgoing port which enforce FCFS scheduling policy. Some switches provide support for multiple levels of priority at the outgoing ports (multiple FIFO queues with different priorities). Using these queues, it is possible to mimic the SRPT scheduling policy by putting smaller flows into higher priority queues. For example, Cisco offers switches that support this feature using dynamic packet prioritization (DPP) \cite{dpp} which tracks flows as their packets arrive (estimating a flow's size according to LAS policy) and assigns them to priority queues according to their sizes. Weighted Fair Queuing (WFQ) is also supported by many switch vendors per ``class of service'' or per flow where weights determine the proportional importance of flows, i.e., a flow is assigned bandwidth proportional to its weight.

\subsubsection{Preemption} \label{preemption}
Many practical systems have to address arrival of requests in an online manner where requests can arrive at any time and have to be addressed upon arrival. Order of arrivals can impact the performance of scheduling algorithms due to race conditions which can lead to priority inversion. For example, in an online scenario with the objective of minimizing mean FCT, SJF might perform poorly if many short flows arrive shortly after a large flow. Preemptive scheduling policies (SRPT in this case) can be used to address this problem \cite{pdq}.

\subsubsection{Jittering} \label{jittering}
A server issuing many fetch requests can use jittering to desynchronize arrival of traffic from various flows and reduce peaks in traffic volume. Such peaks can lead to temporary congestion, dropped packets and increased latency. Jittering can be applied by adding random delays at the application layer when initiating multiple requests at the same time \cite{dctcp}.

\subsubsection{ACK Control} \label{ackcontrol}
ACKs may be used as part of the network traffic scheduling process since they determine how a sender advances its outstanding window of bytes. They can be thought of as permits for the sender to transmit more data. A receiver can stall a sender by intentionally holding back on ACKs or limit sender's rate by delaying them. For example, a receiver can pause ACKs for low priority flows upon arrival of higher priority traffic and can generate ACKs to control which flows are assigned more bandwidth \cite{sfs}. In addition, by reporting a small receiver window in ACKs, a receiver may limit the transmission rate of a sender \cite{iatcp, ictcp}. This approach has considerable similarities with the token-based transmission control applied in schemes such as pHost \cite{phost}, NDP \cite{ndp} and ExpressPass \cite{credit-based-cc}.

\section{Open Challenges} \label{open-challenges}
In this section, we point to a few open challenges with regards to traffic control. To find an optimal solution, these problems may be modeled as complex optimization scenarios that are computationally expensive to solve (large number of variables and constraints, presence of integer variables and/or non-linear constraints, complex objective functions) and practically hard to enforce (lack of hardware support, slow response time of software implementations, presence of failures and errors). Current approaches apply a variety of heuristics and simplifying assumptions to come up with solutions that are practical and attractive to industry. In addition, such optimization scenarios may be infeasible due to presence of contradictory constraints meaning it may not be possible to optimize for all objectives given available resources. Therefore, it becomes necessary to relax some requirements. In the following, we do not provide any optimization models, rather we point to cases where such complex models may appear. Table \ref{dc-tc-open} offers an overview of open challenges in this section.

\begin{table*}[t!]
\caption{Summary of open challenges} \label{dc-tc-open}
\centering
\begin{tabular}{|p{3cm}|p{14cm}|} 
\hline
\textbf{Open Challenge} & \textbf{Description} \\
\hline
\hline
Handling Mix Workloads & Datacenter environments house a variety of applications that generate flows with different properties and requirements. Effectively handling the mix of traffic workload requires clearly defining a utility function of performance metric variables (delay, utilization, deadline miss rate, lateness, fairness) and formulating an optimization problem. \\
\hline
Load Balancing vs. Packet Reordering & To minimize packet reordering, a sender needs to carefully schedule packets over all available paths while paying attention to other traffic workload. Enforcing a no reordering constraint can result in lower network utilization. As a result, to increase utilization while imposing acceptable level of reordering, one can consider a utility function of these factors and formulate an optimization problem solving which provides a desirable transmission schedule. \\
\hline
Achieving High Throughput and Low Latency & In distributed traffic control approaches, a feedback from network is provided to senders to adapt their transmission rate. While transmitting at high rates, the network may not provide feedbacks fast enough leading to network overload and congestion. Therefore, limited network responsiveness may increase average queuing delay and latency at high throughput. \\
\hline
Objective Mismatch & Due to resource constraints, it may not be possible to come up with a transmission schedule where all performance objectives are optimal. It then becomes necessary to define a utility of performance metric variables and aim to maximize utility by formulating an optimization scenario. \\
\hline
\end{tabular}
\end{table*}

\textbf{Handling Mix Workloads:} Datacenter environments are usually shared by a variety of applications that generate different network workloads with different performance metrics. For example, two applications of search and backup may be sharing the same network; while search network traffic requires low communication latency, backup network traffic demands high throughput. Some network workloads may have deadlines, either soft or hard, which should be considered along with non-deadline network traffic. Effectively scheduling such mix of network workload is an open problem. As a solution, one can formulate a complex optimization problem for such scheduling that considers maximizing some utility function. This utility could be a function of performance metric variables such as deadline miss rate, average lateness, link utilizations, and latency which represents the value achieved by tenants and operators. For example, one can examine how the objective presented in \cite{karuna}, i.e., to reduce per packet latency, translates into utility.

\textbf{Load Balancing vs. Packet Reordering:} In datacenters, there is usually a large number of paths between any two end-hosts which according to topology, could have equal or unequal lengths (number of hops and latency). To use all available paths, one could apply load balancing techniques. Per packet load balancing can provide the finest level of balancing but in general leads to significant packet reordering. In contrast, although per flow load balancing does not lead to packet reordering, it may not be effective in using available bandwidth over many parallel paths. In general, effective load balancing necessitates scheduling of packets over available paths according to available bandwidth. One could consider an additional constraint that allows sending packets on a path only if such scheduling does not lead to out of order arrival of packets at the receiver. But this might reduce utilization by decreasing packet transmission opportunities. As a result, effectively utilizing available bandwidth over all paths could be at odds with minimizing packet reordering. One could relax the no-reordering constraints by allowing reordering to some extent. This relationship can be formulated as a utility of reordering and bandwidth utilization maximizing which in general is an open problem. For example, the works discussed in \S \ref{balancing} and \S \ref{multipathing} offer several possible approaches.

\textbf{Achieving High Throughput and Low Latency:} There is a strong connection between traffic control and the control theory. A traffic control scheme depends on concepts from control theory in managing flow of traffic across the network: transmission rate of senders is a function of feedback received from the network, and the time it takes from when a sender changes its rate until it receives a feedback of that change constitutes the loop delay. Due to existence of this loop delay, a network has limited responsiveness due to processing and propagation latency as well as queuing delay. The former two factors are usually much smaller in datacenters and queuing delay determines responsiveness. When transmitting at high rates, it is easy for senders to overload the network (which increases queue occupancy and queuing delay) before a feedback is provided to senders to reduce their rate (which takes at least as much as response time of the network) \cite{proactive-cc}. As a result, in distributed traffic control, achieving maximum throughput with minimal latency is an open problem. Using centralized schemes that perform bandwidth reservation prior to allowing senders to transmit is a candidate solution \cite{fastpass}. There is however several trade-offs as discussed in \S \ref{dc-tc-management}.

\begin{figure}[t!]
	\centering
	\includegraphics[width=0.7\columnwidth]{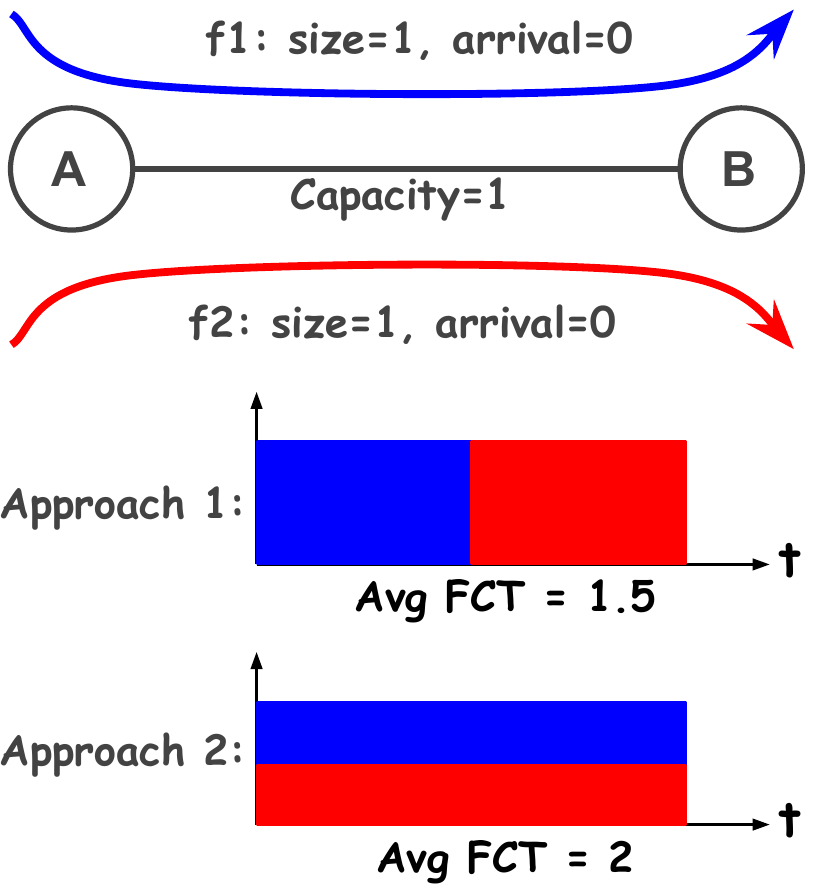}
	\caption{It may not be possible to optimize all performance metrics together (mismatch between fairness and mean FCT)}
	\label{fig_mismatch_1}
\end{figure}

\begin{figure}[t!]
	\centering
	\includegraphics[width=0.7\columnwidth]{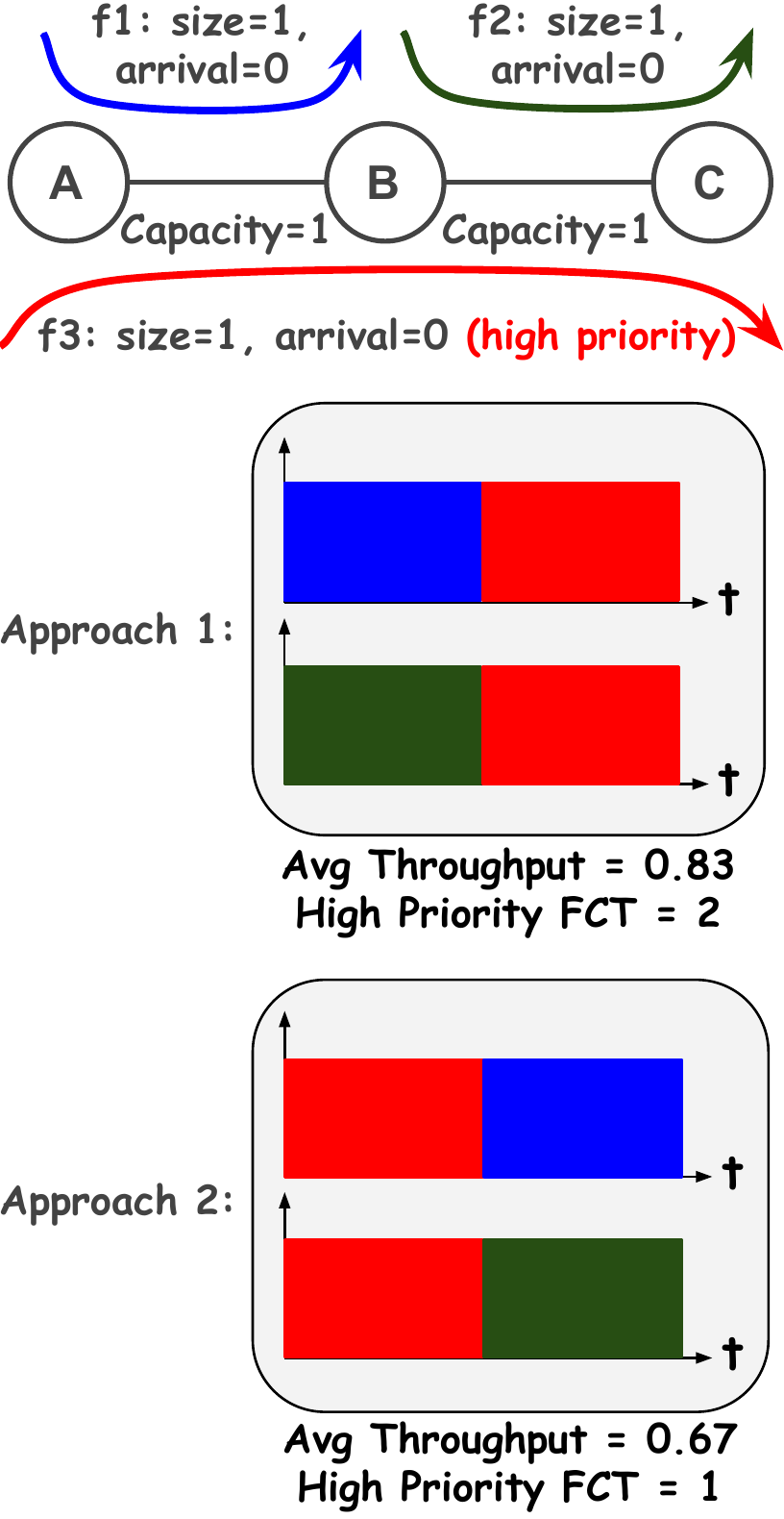}
	\caption{It may not be possible to optimize all performance metrics together (mismatch between average throughput of transfers and FCT of high priority traffic)}
	\label{fig_mismatch_2}
\end{figure}

\textbf{Objective Mismatch:} In \S \ref{dc-tc-objectives}, we presented a few objectives for traffic control. In practice, even for a single flow type, it may not be possible to optimize for all performance objectives since improving one may negatively impact the others. For example, maximizing fairness may be at odds with minimizing average latency as shown in Figure \ref{fig_mismatch_1}: the first approach is unfair but offers minimal FCT compared to second approach which is fair. Next, as shown in Figure \ref{fig_mismatch_2}, flows may have different priorities which determine their relative importance. The first approach only focuses on maximizing average throughput increasing completion time of high priority flow while the second approach minimizes FCT of high priority flow although it reduces average throughput. As another example, as stated in \cite{pdq}, fairness could be at odds with minimizing deadline miss rate. As a result, it is desirable to define a utility of these objectives and aim to maximize utility. In general, maximizing a utility of objective variables such as latency, utilization, deadline miss rate, lateness, and fairness is an open problem. Current research efforts mostly focus on maximizing performance with regards to one or two of these metrics. For example, the approach presented in \cite{calendaring} formulates an optimization scenario that considers both fairness and meeting deadlines.


\section{Related Paradigms} \label{related-paradigms}
We review a few networking paradigms that have affected the design and operation of datacenter networks. Essentially, networks have become more flexible and controllable giving operators more room for performance optimizations.

\subsection{Programmable Forwarding Planes} \label{pfp}
Programmable forwarding planes offer significant room for efficient control and management of datacenter networks. They make it possible to develop and implement custom policies and algorithms for network control and management. Forwarding plane can be programmed centrally using a controller that takes into account policies and resource constraints through some interface provided by forwarding elements. This is considered as part of Software Defined Networking (SDN) \cite{sdndef}. A comprehensive survey of SDN architecture and applications can be found in \cite{survey-sdn}.

The dominant framework in this realm is OpenFlow \cite{sdn, openflowspec} where forwarding elements, such as switches, can be managed via an open interface. An important benefit of an open interface is that switches built by different vendors can be operated in the same way allowing for cost effective expansion. Rolling out new updates to the network also becomes much easier as only controllers need to be patched. In general, there could be multiple controllers each managing a part of the network while coordinating together which is most applicable to large networks. There is two-way communication between switches and controllers: a controller can register for specific events at the switches and perform changes to the switches' forwarding table by adding new rules, modifying existing rules or removing them.

The forwarding process begins when a packet enters a switch from any port. The switch tries to match the packet to a forwarding rule according to its header fields and will forward it to the correct outgoing port. It is also possible to modify the packet contents (packet rewrite) before forwarding it. If the packet cannot be matched to any rule, it can be sent to the controller (if forwarding table is configured with the right table-miss entry) for further inspection and if necessary the controller will update the forwarding plane accordingly for forwarding of this packet and the rest of packets from the same flow. If there are multiple matches, the highest priority rule will be executed. More complex operations can be executed using Group Tables which allow for forwarding to multiple outgoing ports, selection of the outgoing port according to some hash of the packet (e.g. for load balancing), and switching to a connected output port for failover. Group Table features have been added since version 1.1 of OpenFlow \cite{openflowspec11} and currently version 1.5 has been released \cite{openflowspec15}.

By centrally managing forwarding rules, one can implement different routing protocol stacks in a centralized manner. For example, BGP protocol stack can be deployed on top of SDN \cite{sdnbgp}. In addition, one can implement a more sophisticated control plane protocol that understands and communicates with a variety of other protocols, such as legacy BGP routers, while running a custom protocol stack itself \cite{jupiter}.

\subsection{Programmable Data Planes}
Generally, data plane operations are implemented using Application-Specific Integrated Circuits (ASICs) at the hardware layer allowing for forwarding at maximum rate but only offering a set of fixed switch functions that can only be changed by replacing the ASICs. This introduces a few issues namely being prone to bugs as well as long and unpredictable time to implement new functions \cite{pdp-slides, pdp-next}. Programmable data planes (PDPs) allow the packet processing functions to be changed at the forwarding devices, i.e., switches can apply new forwarding functions at the line rate. This is orthogonal to programmable forwarding planes (e.g., SDN) where different forwarding rules can be selectively applied to packets. PDPs make it easier to deploy traffic control schemes that depend on custom in-network processing or feedback. For example, \cite{hull, pfabric, pdq} rely on custom packet headers and feedback from switches.

PDPs can be realized using Protocol-Independent Switch Architecture (PISA) using which new features can be introduced to switches or bug fixes can be applied in a short time \cite{tofino}. There are emerging proposals for hardware designs that allow for PISA \cite{rmt, domino}. Hardware prototypes (switch chips) have also been built that have made this possible \cite{tofino}. P4 \cite{p4, p4-web} is a high level language to program PISA switches which is vendor independent, and protocol independent (i.e., operates directly on header bits and can be configured to work with any higher layer protocol). P4 compiler can also compile P4 code to run on a general purpose processor as software switches.

\subsection{Advanced NICs}
NICs have been providing basic offloading features to OSes, such as segmentation offloading, for many years. Several vendors have been developing NICs with advanced offloading features that perform complex transport tasks and deliver the results without minimal involvement from OS and CPU. These features allow complex operations at high line rates of datacenters (40~Gbps and more) doing which at the OS may incur significant CPU overhead and additional communication latency.

Examples of offloading features include cryptography, quality of service, encapsulation, congestion control, storage acceleration, erasure coding, and network policy enforcement. Examples of such NICs include Mellanox ConnectX \cite{connectx} and Microsoft SmartNIC \cite{smart_nic, azure_smart_nic} developed as part of Open Compute Project (OCP) \cite{ocp}. SmartNIC relies on FPGA accelerators and can be used to apply SDN/Networking policies. It also makes low latency transport possible using Lightweight Transport Layer (LTL) that creates end-to-end transport connection between NICs (FPGA does all processing of segmentation, ordering, and ACKs).

\subsection{Userspace Packet Processing} \label{userspace-pp}
By default, packets pass through the Operating System networking stack before they are received by applications. When packets arrive at the NIC, interrupts are generated to invoke the Operating System routines that read and process them. Processing of packets is usually done in batches to reduce CPU utilization at high rates, for example by enabling Interrupt Moderation \cite{interrupt_moderation}.

To improve the packet processing performance (decrease packet processing latency and increase throughput), a different approach would be to bypass Operating System's networking stack and use polling instead of interrupts. This can be realized using kernel bypass modules, such as Netmap \cite{netmap, netmap-github}, Vector Packet Processing (VPP) \cite{vpp, vpp-github} and Data Plane Development Kit (DPDK) \cite{dpdk}, which have been shown to reduce the number of required cycles to process a packet by up to 20$\times$ on average \cite{netmap}. These modules allow userspace programs to directly access NIC buffers to read incoming packets or write packets for transmission.

Userspace networking stacks have been developed on top of kernel bypass modules. Sandstorm \cite{sandstorm} and mTCP \cite{mtcp}, implement TCP in userspace and rely on Netmap and VPP, respectively. SoftNIC \cite{softnic} is built on top of DPDK and allows developers to program custom NIC features in software. RAMCloud \cite{ramcloud} distributed key-value store and FastPass \cite{fastpass} make use of kernel bypass and polling to speed up Remote Procedure Calls (RPC). NDP \cite{ndp} is a datacenter transport protocol that is designed for ultra-low latency and operates on top of DPDK.

\subsection{Lossless Ethernet and RDMA}
TCP has been the dominant transport protocol across datacenters for the majority of applications. Since implemented as part of OS protocol stack, using TCP at high transmission rates can exhaust considerable CPU resources and impose notable amount of communication latency. Remote Direct Memory Access (RDMA) is a transport protocol that allows delivery of data from one machine to another machine without involving the OS networking protocol stack. RDMA operates on the NICs of machines communicating. Compared to TCP, RDMA offers higher bandwidth and lower latency at lower CPU utilization \cite{timely, farm, dcqcn}. RDMA can also be used for seamless offloading of large datasets to nearby machines (as opposed to using pagefiles) \cite{infiniswap}.

To use RDMA, the underlying network has to be lossless since RDMA does not support recovery from lost data by default. Ethernet which is the favorite choice of transport in datacenters, however, does not support reliability by default. Lossless Ethernet, also known as Converged Enhanced Ethernet (CEE), supports per-hop flow control at Layer 2. Backpressure is used as the mechanism to stop senders in case of a full buffer in the network. PAUSE messages are sent to previous hops, pausing the output ports that are connected to inputs with full buffers, until the ultimate senders receive a PAUSE message. The flow control mechanism used by lossless Ethernet is referred to as Priority Flow Control (PFC) and offers 8 priority levels for various classes of traffic \cite{pfc, pfc_cisco}.

For Layer 2 networks, RDMA can be deployed using RDMA over Converged Ethernet (RoCE) \cite{roce} which is based on lossless Ethernet and works across a single Layer 2 domain. For larger networks that span across Layer 3, RDMA can be deployed on top of IP and UDP using version 2 of RoCE (RoCEv2) \cite{roce2}. It can also be deployed on top of IP and TCP using iWARP \cite{iwarp} which implements a full TCP stack on end-host NIC to provide a lossless end to end transport. iWARP does not require a lossless infrastructure and can work on top of usual Ethernet, but is less performant and has limited capabilities compared to RoCE \cite{iwarp_vs_roce}.

Using lossless Ethernet can lead to a few performance issues in general. Some hindering issues include Layer 2 Head of Line (HOL) Blocking, unfairness (because Layer 2 has no understanding of upper layer notions such as flows), and deadlocks (due to per-port/class PAUSE feature and possible circular dependency of routes). HOL blocking might occur since pausing happens on a per-port/class basis. Therefore, a flow can overflow a port causing it to be blocked stopping other flows going through that port as well. As a result, it is necessary to prevent formation of full buffers. Furthermore, loss-based congestion control approaches are rendered useless since there is no packet loss in case of full buffers.

Quantized Congestion Notification (QCN) \cite{qcn_original}, which is fully implemented in Layer 2, can be used to reduce PAUSE messages by signaling senders before buffers are full. It sends notifications back to the sender's NIC from switches. Packets need to be tagged with a flow ID at the senders which will be used at the switches when notifications are generated to determine which flows should be slowed down. QCN is limited to boundaries of a single Layer 2 domain and therefore is insufficient for datacenters with large networks.

TCP Bolt \cite{bolt} and DCQCN \cite{dcqcn} operate across Layer 3 using RoCEv2. Both of these schemes use DCTCP \cite{dctcp} like ECN marking to reduce buffer occupancy and minimize PAUSE signals. To prevent deadlocks, TCP Bolt creates edge disjoint spanning trees (EDSTs) across the network with different PFC classes to prevent cyclic dependencies as flows are routed in the network. TIMELY \cite{timely} can also be used on lossless networks which uses a delay-based approach to detect increased buffer occupancy and manages its rate accordingly to reduce it.


\section{Broader Perspective} \label{perspective}
Cloud companies and content providers, such as Google \cite{google}, Microsoft \cite{azure}, Facebook \cite{facebook-dc-count-2016} and Amazon \cite{aws}, have built multiple datacenters in different continents and countries. Multiple datacenters offer a variety of benefits for distributed applications with geographically wide range of users such as email, multimedia (e.g., YouTube), social networks (e.g., Facebook, Instagram, and Google Plus) and online storage. These benefits include increased availability and fault-tolerance, global or regional load balancing, reduced latency to customers and reduced global bandwidth usage via caching. For example, to minimize user access latency, data can be placed on local datacenters close to users via replication.

To further improve reliability, load balancing and data availability, large datacenter operators (such as Amazon and Microsoft) operate in two hierarchies of zones and discrete datacenters. Each availability zone is usually made up of a few discrete datacenters that are close enough to communicate with negligible latency (e.g. less than 2 ms for Microsoft Azure, i.e., few tens of miles), while far enough to allow them to operate as distinct failure domains. Datacenters usually have rich connectivity within zones which themselves are connected using long haul fiber optics (usually hundreds of miles) \cite{microsoft-azure-backbone}. These links are either owned by datacenter operators or leased from a provider with existing backbone infrastructure. These links that connect multiple datacenters within regions and across them are referred to as inter-datacenter networks. Maintaining and operating inter-datacenter networks requires significant capital investment from datacenter companies which makes it imperative to efficiently use them \cite{b4, calendaring, swan}.

In general, we can categorize traffic that goes within and across datacenters into traffic that is a result of direct interaction with users and the business internal traffic that is a result of backend data processing or migration. Recent work points to significant increase in the overall business internal traffic (which includes both intra and inter-datacenter traffic) that is growing at a much faster pace than user generated traffic \cite{b4, jupiter, facebook-express-backbone}. Such increase not only demands higher interconnection bandwidth across servers within datacenters, but also higher network capacity across datacenters. Over the past decade, significant attention has been given to intra-datacenter networks to improve their performance and efficiency which was the topic of discussion in previous sections. However, similar attention has not been paid to increasing efficiency and performance of connectivity across datacenters.

To communicate across datacenters, many companies purchase bandwidth from ISP networks with present WAN infrastructure and are billed according to some usage criteria. A widely used pricing scheme calculates traffic costs by looking at 95 percentile of network bandwidth usage over some period of time \cite{95percentile}. A variety of research efforts focus on minimizing inter-datacenter traffic transit costs considering a similar pricing scheme by aiming to not increase the peak bandwidth usage or minimally increase it if necessary \cite{jetway, mbdt_initial, mbdt, grease, dtb, postcard, ecoflow, netstitcher, dynamic_pricing}.

Large datacenter operators take advantage of dedicated inter-datacenter connections over long haul optical networks. Google operates their own backbone network referred to as B4 \cite{b4, evolve}. Microsoft Global WAN \cite{swan-backbone} connects Microsoft datacenters over their private (dark) fiber network. Facebook has also developed their cross datacenter backbone network referred to as Express Backbone \cite{facebook-express-backbone}. These private dedicated networks offer a unique opportunity for further optimization of network resource management. Considering that all end-points of such networks are managed by one organization, we can improve efficiency by coordinating transmission across such end-points. In addition, despite their huge geographical scale, these networks are usually made up of tens to hundreds of links (that connect distant locations) making such coordination practically feasible. For example, B4 is currently managed centrally by a system called Bandwidth Enforcer \cite{bwe}, Microsoft Global WAN is managed by SWAN \cite{swan} and Express Backbone is also managed centrally by a traffic engineering controller \cite{facebook-express-backbone}.

\subsection{Private Dedicated Inter-datacenter Networks}
Bandwidth allocation is an effective approach for global traffic engineering over private dedicated networks \cite{bwe, swan}. To take into account latency overhead of bandwidth allocation, a hybrid approach is usually taken where an aggregate bandwidth is set aside for short latency-sensitive flows (mostly user generated) per link or such flows are assigned strictly higher traffic priority. Large long-running flows which we refer to as transfers will then become the focus of per flow bandwidth allocation. In addition, since the network environment is continuously changing as new transfers arrive or due to varying volume of higher priority traffic (from latency-sensitive flows), a slotted timeline is considered where transmission rates can be updated on a per timeslot basis.

\begin{figure*}
    \centering
    \includegraphics[width=0.8\textwidth]{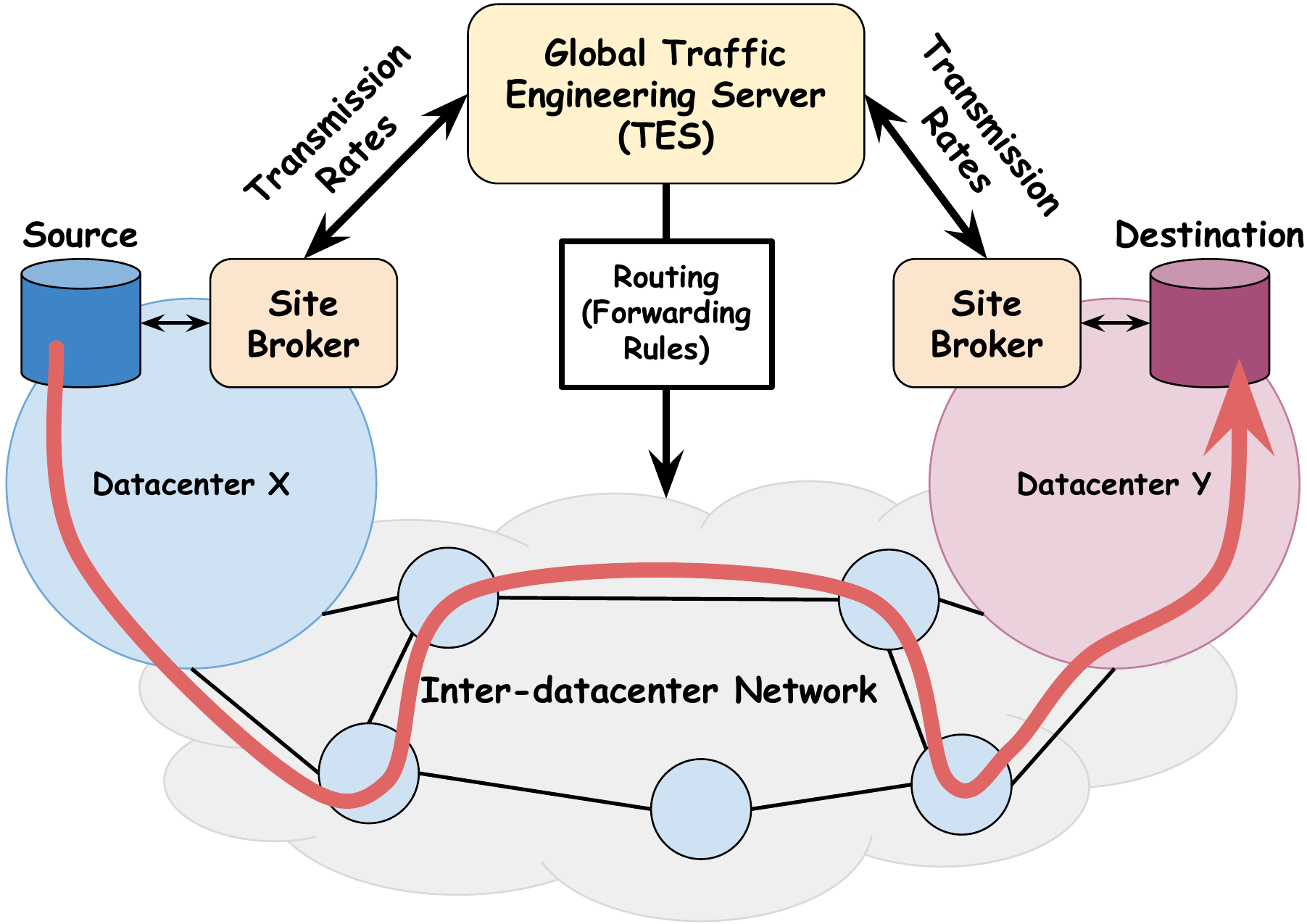}
    \caption{Central management of private dedicated inter-datacenter networks}
    \label{fig:central-management-inter}
\end{figure*}

Figure \ref{fig:central-management-inter} shows an example architecture for this purpose which is based on similar concepts as \cite{bwe, swan}. One way ro realize this architecture is using SDN. An inter-datacenter network that can be operated using SDN is sometimes referred to as Software Defined WAN (SDWAN). This allows data transmission and routing to be managed centrally according to transfer properties and requirements as well as network status and topology \cite{google-optical-network}.

A central Traffic Engineering Server (TES) calculates transmission rates and routes for submitted transfers as they arrive at the network. Rates are then dispatched to local agents that keep track of local transfers (initiated within the same datacenter) called site brokers. Senders first communicate with local site broker which in turn forwards transfer requests to TES. The local site broker then receives TES's response with the transmission rates and forwards it to the sender. Local site brokers add a level of indirection between senders and TES which can help in several ways. It reduces the request response overhead for TES by maintaining a persistent connection with brokers and possibly aggregating several transfer requests before relaying them to the server, which can be useful for hierarchical bandwidth allocation that is done by locally grouping many transfers and presenting them to TES as one (trading some traffic engineering accuracy for scalability). Site broker may also inform TES of network conditions within datacenters for more careful rate allocation and routing. Finally, site broker may modify TES's response according to varying local network conditions or allow senders to switch to a standby backup TES in case it goes offline. Other tasks in this system setup include the following.

\textbf{Rate-limiting:} We earlier discussed a variety of techniques for rate-limiting in \S \ref{rate-limiting}. TES can calculate transmission rates on a per transfer basis in which case end-hosts (e.g. virtual machines initiating the transfers) should comply with such rates by limiting rates before traffic is transmitted on the wire. In addition, applications that initiate the transfers can themselves apply rate-limiting by carefully controlling the amount of data handed off to the transport layer. Although simple, this approach demands changes to the applications. TES can also compute rates per groups of transfers which can improve scalability \cite{bwe}. Rate-limiting can then be applied at some intermediate network element via traffic shaping/policing \cite{shapingvspolicing}. Transfers in a group will then have to use a bandwidth sharing policy among themselves such as Fair Sharing.

\textbf{Routing:} Upon arrival of a transfer, forwarding routes are calculated for it by TES. Such routes are then installed in the network by adding proper forwarding rules to network's switching elements. To reduce setup overhead or save network forwarding state, TES can also reuse existing forwarding rules or aggregate several of them as one if possible. Per transfer, senders may have to attach proper forwarding labels to their packets so that they are correctly forwarded (like a Virtual LAN ID). Such labeling may also be applied transparent to the sender at an intermediate network entity (e.g. hypervisor virtual switches, backbone edge, etc).

\subsection{Research Directions}
In this section, we provide an overview of several research directions considering the scenario mentioned above.

\subsubsection{Inter-datacenter Global Rate-allocation and Routing} The majority of inter-datacenter related research is on global optimization of inter-datacenter networks. Metrics similar to ones we discussed in \S \ref{dc-tc-objectives} can be considered as objectives. For example, B4 \cite{b4} and SWAN \cite{swan} focus on maximizing utilization, Tempus \cite{calendaring} aims to meet transfer deadlines and maximizes minimal fraction of transfers that complete prior to deadlines in case there is not enough capacity, Amoeba \cite{amoeba} performs admission control for incoming traffic and only admits new transfers if their deadlines can be guaranteed and DCRoute \cite{dcroute} performs fast admission control to guarantee deadlines while minimizing packet reordering. The majority of prior work related to this problem either focus on delay tolerant transfers or aim at meeting deadlines while not considering completion times as a metric. For many long-running data transfer operations, completion times are important in increasing overall utility. For example, faster completion of backup operations may reduce the chance of data loss due to failures or speedy replication of data objects can improve average user's quality of experience. In addition, most prior work formulate complex optimization problems that are computationally expensive and slow especially if need be solved as transfers arrive and can increase scheduling latency. Further research is then necessary in developing global rate computation and routing algorithms that lead to solutions quickly and also consider completion times of transfers.

\subsubsection{Inter-datacenter One-to-Many Transfers} Many services run across several datacenters close to regional users to offer a better quality of experience by minimizing customer access latency (e.g. CDNs cache objects for local viewers \cite{utube, netflix, jetway, ecoflow, social_inside}). This approach also reduces overall inter-datacenter bandwidth consumption by keeping a copy of popular objects close to users. There is also need for propagating application state to multiple locations for synchronization (e.g. search index databases \cite{b4}) or making multiple distant data copies for higher reliability and availability \cite{google-optical-network}. All of these lead to data delivery from one datacenter to multiple datacenters referred to as Point to Multipoint (P2MP) transfers \cite{dccast}. P2MP data delivery over private dedicated inter-datacenter networks can be considered as a special case of multicasting for which a significant body of work is available including in-network multicasting \cite{ip_multicast, multicast-challenges} and using overlay networks \cite{nice, narada}. However, there is need for coordinated schemes that improve performance by carefully selecting multicast forwarding trees and assigning transmission slots to transfers. One should also note that centralized multicasting solutions proposed for intra-datacenter networks, such as \cite{avalanche, datacast}, may not work for inter-datacenter networks due to significant differences in topology (former is usually structured and regular while latter is not). New objectives can be considered in addition to ones proposed in \S \ref{dc-tc-objectives} for P2MP transfers, such as maximizing number of receivers per transfer that complete reception in a given period of time.

\subsubsection{Inter-datacenter Failure-aware Routing} Given their scale, inter-datacenter networks may be exposed to a variety of physical conditions and natural environments. As a result, different inter-datacenter links may have significantly different link failure probabilities (possibly by as much as three orders of magnitude \cite{failure-aware-routing}). In addition, inter-datacenter traffic in general is made up of a variety of classes with different quality of service requirements and priorities \cite{evolve, social_inside}. In general, transfers can be rerouted and rescheduled in reaction to failures. This however leads to possible service disruptions which can more seriously affect services that are sensitive to loss or latency. Given unique properties of private dedicated inter-datacenter networks, steps can be taken proactively as new transfers arrive to mitigate the effect of failures on overall utility. Several objectives can be considered per failure event including minimizing number of affected transfers upon failures considering their class of service, minimizing the amount by which latency increases across a variety of services, or maximizing utilization while leaving off spare capacity to perform quick in-network failover rerouting \cite{ffc}. The latter could be applied only to a fraction of inter-datacenter traffic (i.e., more important traffic). In addition, one could study these objectives for both one-to-one and one-to-many transfer scenarios (or a mix of them).




\section{Conclusions} \label{conclusion}
Many online services today depend on datacenter infrastructures to provide high availability and reliability along with scalable compute at minimal costs. Datacenter environments may be shared by many tenants and applications offering different services to their consumers (end-users). Traffic control is a necessary task in datacenter networks to efficiently use the resources and fairly share them.

In this tutorial paper, we reviewed several challenges operators, tenants and applications are faced with in datacenter traffic control. We discussed different elements of traffic control explaining the problems, challenges, proposed solutions and their trade-offs. Despite significant research efforts, the majority of traffic control proposals are in the state of research and far from adoption in the industry considering metrics of complexity, performance and cost altogether. Further research efforts are required to reach solutions that offer higher performance at same or lower costs while maintaining low complexity.

More broadly, at the end of this paper, we also pointed to inter-datacenter communication as an evolving research area that requires further attention. We proposed a centralized architecture that captures the essence of existing solutions and discussed how efficient traffic control can be realized using it. We also pointed to three major active research areas regarding inter-datacenter communication that need further attention from the research community.


\section*{Acknowledgement}
We would like to thank the anonymous reviewers of IEEE Communications Surveys and Tutorials for their valuable comments and suggestions that helped us significantly improve the quality of this paper. The first author would also like to especially thank David Oran, Joshua Gahm, Narayan Venkat and Atif Faheem who provided valuable intellectual support during an internship at Cisco Research Center located in Cambridge, Massachusetts, USA.


\bibliographystyle{IEEEtran}
\bibliography{citations.bib}

\begin{thebibliography}{100}
\providecommand{\url}[1]{#1}
\csname url@samestyle\endcsname
\providecommand{\newblock}{\relax}
\providecommand{\bibinfo}[2]{#2}
\providecommand{\BIBentrySTDinterwordspacing}{\spaceskip=0pt\relax}
\providecommand{\BIBentryALTinterwordstretchfactor}{4}
\providecommand{\BIBentryALTinterwordspacing}{\spaceskip=\fontdimen2\font plus
\BIBentryALTinterwordstretchfactor\fontdimen3\font minus
  \fontdimen4\font\relax}
\providecommand{\BIBforeignlanguage}[2]{{%
\expandafter\ifx\csname l@#1\endcsname\relax
\typeout{** WARNING: IEEEtran.bst: No hyphenation pattern has been}%
\typeout{** loaded for the language `#1'. Using the pattern for}%
\typeout{** the default language instead.}%
\else
\language=\csname l@#1\endcsname
\fi
#2}}
\providecommand{\BIBdecl}{\relax}
\BIBdecl

\bibitem{google}
\BIBentryALTinterwordspacing
\emph{{Google data center locations}}. [Online]. Available:
  \url{https://google.com/about/datacenters/inside/locations}
\BIBentrySTDinterwordspacing

\bibitem{azure}
\BIBentryALTinterwordspacing
\emph{{Azure Cloud Services by Location or Region}}. [Online]. Available:
  \url{https://azure.microsoft.com/en-us/regions/}
\BIBentrySTDinterwordspacing

\bibitem{aws}
\BIBentryALTinterwordspacing
\emph{{AWS Global Infrastructure}}. [Online]. Available:
  \url{https://aws.amazon.com/about-aws/global-infrastructure/}
\BIBentrySTDinterwordspacing

\bibitem{jupiter}
A.~Singh, J.~Ong, A.~Agarwal, G.~Anderson, A.~Armistead, R.~Bannon, S.~Boving,
  G.~Desai, B.~Felderman, P.~Germano \emph{et~al.}, ``{Jupiter Rising: A decade
  of Clos topologies and centralized control in Google's datacenter network},''
  \emph{ACM SIGCOMM Computer Communication Review}, vol.~45, no.~4, pp.
  183--197, 2015.

\bibitem{vl2}
\BIBentryALTinterwordspacing
A.~Greenberg, J.~R. Hamilton, N.~Jain, S.~Kandula, C.~Kim, P.~Lahiri, D.~A.
  Maltz, P.~Patel, and S.~Sengupta, ``{VL2: A Scalable and Flexible Data Center
  Network},'' \emph{Commun. ACM}, vol.~54, no.~3, pp. 95--104, March 2011.
  [Online]. Available: \url{http://doi.acm.org/10.1145/1897852.1897877}
\BIBentrySTDinterwordspacing

\bibitem{server-per-rack-facebook}
\BIBentryALTinterwordspacing
S.~Subramanian, \emph{{Network in Hyper-scale data centers - Facebook}}, 2015.
  [Online]. Available:
  \url{http://www.innervoice.in/blogs/2015/03/30/network-in-hyper-scale-data-centers-facebook/}
\BIBentrySTDinterwordspacing

\bibitem{pitfall}
G.~Judd, ``{Attaining the Promise and Avoiding the Pitfalls of TCP in the
  Datacenter},'' \emph{12th USENIX Symposium on Networked Systems Design and
  Implementation (NSDI 15)}, pp. 145--157, 2015.

\bibitem{dcqcn}
\BIBentryALTinterwordspacing
Y.~Zhu, H.~Eran, D.~Firestone, C.~Guo, M.~Lipshteyn, Y.~Liron, J.~Padhye,
  S.~Raindel, M.~H. Yahia, and M.~Zhang, ``{Congestion Control for Large-Scale
  RDMA Deployments},'' \emph{Proceedings of the 2015 ACM Conference on Special
  Interest Group on Data Communication}, pp. 523--536, 2015. [Online].
  Available: \url{http://doi.acm.org/10.1145/2785956.2787484}
\BIBentrySTDinterwordspacing

\bibitem{hull}
M.~Alizadeh, A.~Kabbani, T.~Edsall, B.~Prabhakar, A.~Vahdat, and M.~Yasuda,
  ``{Less is more: trading a little bandwidth for ultra-low latency in the data
  center},'' \emph{Presented as part of the 9th USENIX Symposium on Networked
  Systems Design and Implementation (NSDI 12)}, pp. 253--266, 2012.

\bibitem{fattree}
\BIBentryALTinterwordspacing
M.~Al-Fares, A.~Loukissas, and A.~Vahdat, ``{A Scalable, Commodity Data Center
  Network Architecture},'' \emph{Proceedings of the ACM SIGCOMM 2008 Conference
  on Data Communication}, pp. 63--74, 2008. [Online]. Available:
  \url{http://doi.acm.org/10.1145/1402958.1402967}
\BIBentrySTDinterwordspacing

\bibitem{hyperx}
J.~H. Ahn, N.~Binkert, A.~Davis, M.~McLaren, and R.~S. Schreiber, ``{HyperX:
  topology, routing, and packaging of efficient large-scale networks},''
  \emph{Proceedings of the Conference on High Performance Computing Networking,
  Storage and Analysis}, p.~41, 2009.

\bibitem{dcell}
\BIBentryALTinterwordspacing
C.~Guo, H.~Wu, K.~Tan, L.~Shi, Y.~Zhang, and S.~Lu, ``{Dcell: A Scalable and
  Fault-tolerant Network Structure for Data Centers},'' \emph{SIGCOMM Comput.
  Commun. Rev.}, vol.~38, no.~4, pp. 75--86, August 2008. [Online]. Available:
  \url{http://doi.acm.org/10.1145/1402946.1402968}
\BIBentrySTDinterwordspacing

\bibitem{leaf-spine}
M.~Alizadeh and T.~Edsall, ``{On the data path performance of leaf-spine
  datacenter fabrics},'' \emph{2013 IEEE 21st Annual Symposium on
  High-Performance Interconnects}, pp. 71--74, 2013.

\bibitem{xpander}
A.~Valadarsky, G.~Shahaf, M.~Dinitz, and M.~Schapira, ``{Xpander: Towards
  Optimal-Performance Datacenters},'' \emph{Proceedings of the 12th
  International on Conference on emerging Networking EXperiments and
  Technologies}, pp. 205--219, 2016.

\bibitem{fbtopology}
\BIBentryALTinterwordspacing
\emph{{Introducing data center fabric, the next-generation Facebook data center
  network}}. [Online]. Available:
  \url{https://code.facebook.com/posts/360346274145943/introducing-data-center-fabric-the-next-generation-facebook-data-center-network/}
\BIBentrySTDinterwordspacing

\bibitem{jellyfish}
A.~Singla, C.-Y. Hong, L.~Popa, and P.~B. Godfrey, ``{Jellyfish: Networking
  data centers randomly},'' \emph{Presented as part of the 9th USENIX Symposium
  on Networked Systems Design and Implementation (NSDI 12)}, pp. 225--238,
  2012.

\bibitem{clos}
C.~Clos, ``{A Study of Non-Blocking Switching Networks},'' \emph{Bell Labs
  Technical Journal}, vol.~32, no.~2, pp. 406--424, 1953.

\bibitem{wcmp}
J.~Zhou, M.~Tewari, M.~Zhu, A.~Kabbani, L.~Poutievski, A.~Singh, and A.~Vahdat,
  ``{WCMP: Weighted cost multipathing for improved fairness in data centers},''
  \emph{Proceedings of the Ninth European Conference on Computer Systems},
  p.~5, 2014.

\bibitem{f10}
V.~Liu, D.~Halperin, A.~Krishnamurthy, and T.~Anderson, ``{F10: A
  Fault-Tolerant Engineered Network},'' \emph{10th {USENIX} Symposium on
  Networked Systems Design and Implementation ({NSDI} 13)}, pp. 399--412, 2013.

\bibitem{evolve}
R.~Govindan, I.~Minei, M.~Kallahalla, B.~Koley, and A.~Vahdat, ``{Evolve or
  Die: High-Availability Design Principles Drawn from Google's Network
  Infrastructure},'' \emph{Proceedings of the 2016 conference on ACM SIGCOMM
  2016 Conference}, pp. 58--72, 2016.

\bibitem{circuit_dc}
\BIBentryALTinterwordspacing
G.~Porter, R.~Strong, N.~Farrington, A.~Forencich, P.~Chen-Sun, T.~Rosing,
  Y.~Fainman, G.~Papen, and A.~Vahdat, ``{Integrating Microsecond Circuit
  Switching into the Data Center},'' \emph{Proceedings of the ACM SIGCOMM 2013
  Conference on SIGCOMM}, pp. 447--458, 2013. [Online]. Available:
  \url{http://doi.acm.org/10.1145/2486001.2486007}
\BIBentrySTDinterwordspacing

\bibitem{firefly}
N.~Hamedazimi, Z.~Qazi, H.~Gupta, V.~Sekar, S.~R. Das, J.~P. Longtin, H.~Shah,
  and A.~Tanwer, ``{FireFly: a reconfigurable wireless data center fabric using
  free-space optics},'' \emph{ACM SIGCOMM Computer Communication Review},
  vol.~44, no.~4, pp. 319--330, 2015.

\bibitem{tm_circuit}
S.~Bojja~Venkatakrishnan, M.~Alizadeh, and P.~Viswanath, ``{Costly Circuits,
  Submodular Schedules and Approximate Caratheodory Theorems},''
  \emph{Proceedings of the 2016 ACM SIGMETRICS International Conference on
  Measurement and Modeling of Computer Science}, pp. 75--88, 2016.

\bibitem{projector}
M.~Ghobadi, R.~Mahajan, A.~Phanishayee, N.~Devanur, J.~Kulkarni, G.~Ranade,
  P.-A. Blanche, H.~Rastegarfar, M.~Glick, and D.~Kilper, ``{Projector: Agile
  reconfigurable data center interconnect},'' \emph{Proceedings of the 2016
  conference on ACM SIGCOMM 2016 Conference}, pp. 216--229, 2016.

\bibitem{fatfree}
A.~Singla, ``{Fat-FREE Topologies},'' \emph{Proceedings of the 15th ACM
  Workshop on Hot Topics in Networks}, pp. 64--70, 2016.

\bibitem{cisco-growth}
\BIBentryALTinterwordspacing
\emph{{Cisco Global Cloud Index: Forecast and Methodology, 2015-2020 White
  Paper}}. [Online]. Available:
  \url{http://www.cisco.com/c/dam/en/us/solutions/collateral/service-provider/global-cloud-index-gci/white-paper-c11-738085.pdf}
\BIBentrySTDinterwordspacing

\bibitem{nature}
S.~Kandula, S.~Sengupta, A.~Greenberg, P.~Patel, and R.~Chaiken, ``{The nature
  of data center traffic: measurements \& analysis},'' \emph{Proceedings of the
  9th ACM SIGCOMM conference on Internet measurement conference}, pp. 202--208,
  2009.

\bibitem{d2tcp}
B.~Vamanan, J.~Hasan, and T.~Vijaykumar, ``{Deadline-aware Datacenter TCP
  (D2TCP)},'' \emph{ACM SIGCOMM Computer Communication Review}, vol.~42, no.~4,
  pp. 115--126, 2012.

\bibitem{dctcp}
\BIBentryALTinterwordspacing
M.~Alizadeh, A.~Greenberg, D.~A. Maltz, J.~Padhye, P.~Patel, B.~Prabhakar,
  S.~Sengupta, and M.~Sridharan, ``{Data Center TCP (DCTCP)},'' \emph{SIGCOMM
  Comput. Commun. Rev.}, vol.~41, no.~4, pp. 63--74, August 2010. [Online].
  Available: \url{http://dl.acm.org/citation.cfm?id=2043164.1851192}
\BIBentrySTDinterwordspacing

\bibitem{detail}
D.~Zats, T.~Das, P.~Mohan, D.~Borthakur, and R.~Katz, ``{DeTail: reducing the
  flow completion time tail in datacenter networks},'' \emph{ACM SIGCOMM
  Computer Communication Review}, vol.~42, no.~4, pp. 139--150, 2012.

\bibitem{mapred}
J.~Dean and S.~Ghemawat, ``{MapReduce: simplified data processing on large
  clusters},'' \emph{Communications of the ACM}, vol.~51, no.~1, pp. 107--113,
  2008.

\bibitem{dryad}
\BIBentryALTinterwordspacing
M.~Isard, M.~Budiu, Y.~Yu, A.~Birrell, and D.~Fetterly, ``{Dryad: Distributed
  Data-parallel Programs from Sequential Building Blocks},'' \emph{SIGOPS Oper.
  Syst. Rev.}, vol.~41, no.~3, pp. 59--72, March 2007. [Online]. Available:
  \url{http://doi.acm.org/10.1145/1272998.1273005}
\BIBentrySTDinterwordspacing

\bibitem{survey-lowlatency-tcp}
S.~Liu, H.~Xu, and Z.~Cai, ``{Low latency datacenter networking: A short
  survey},'' \emph{arXiv preprint arXiv:1312.3455}, 2013.

\bibitem{survey-incast}
\BIBentryALTinterwordspacing
Y.~Ren, Y.~Zhao, P.~Liu, K.~Dou, and J.~Li, ``{A survey on TCP Incast in data
  center networks},'' \emph{International Journal of Communication Systems},
  vol.~27, no.~8, pp. 1160--1172, 2014. [Online]. Available:
  \url{http://dx.doi.org/10.1002/dac.2402}
\BIBentrySTDinterwordspacing

\bibitem{survey-bandwidth-allocation}
\BIBentryALTinterwordspacing
L.~Chen, B.~Li, and B.~Li, ``Allocating bandwidth in datacenter networks: A
  survey,'' \emph{Journal of Computer Science and Technology}, vol.~29, no.~5,
  pp. 910--917, September 2014. [Online]. Available:
  \url{https://doi.org/10.1007/s11390-014-1478-x}
\BIBentrySTDinterwordspacing

\bibitem{survey-transport-ieee}
J.~Zhang, F.~Ren, and C.~Lin, ``{Survey on transport control in data center
  networks},'' \emph{IEEE Network}, vol.~27, no.~4, pp. 22--26, July 2013.

\bibitem{survey-tcp-india}
R.~K. Yadav, ``{TCP ISSUES IN DATA CENTER SYSTEM-SURVEY},'' \emph{International
  Journal of Advance Foundation and Research in Computer (IJAFRC)}, vol.~1,
  no.~1, pp. 21--25, 2014.

\bibitem{survey-dcn}
\BIBentryALTinterwordspacing
B.~Wang, Z.~Qi, R.~Ma, H.~Guan, and A.~V. Vasilakos, ``A survey on data center
  networking for cloud computing,'' \emph{Computer Networks}, vol.~91, pp. 528
  -- 547, 2015. [Online]. Available:
  \url{http://www.sciencedirect.com/science/article/pii/S138912861500300X}
\BIBentrySTDinterwordspacing

\bibitem{survey-congestion-india}
R.~P. Joglekar and P.~Game, ``{Managing Congestion in Data Center Network using
  Congestion Notification Algorithms},'' \emph{International Research Journal
  of Engineering and Technology (IRJET)}, vol.~3, no.~5, pp. 195--200, 2016.

\bibitem{survey-transport-springer}
\BIBentryALTinterwordspacing
P.~Sreekumari and J.-i. Jung, ``Transport protocols for data center networks: a
  survey of issues, solutions and challenges,'' \emph{Photonic Network
  Communications}, vol.~31, no.~1, pp. 112--128, February 2016. [Online].
  Available: \url{https://doi.org/10.1007/s11107-015-0550-y}
\BIBentrySTDinterwordspacing

\bibitem{bcube}
C.~Guo, G.~Lu, D.~Li, H.~Wu, X.~Zhang, Y.~Shi, C.~Tian, Y.~Zhang, and S.~Lu,
  ``{BCube: a high performance, server-centric network architecture for modular
  data centers},'' \emph{ACM SIGCOMM Computer Communication Review}, vol.~39,
  no.~4, pp. 63--74, 2009.

\bibitem{folded-clos}
\BIBentryALTinterwordspacing
K.~S. Solnushkin, \emph{{Fat-Tree Design}}. [Online]. Available:
  \url{http://clusterdesign.org/fat-trees/}
\BIBentrySTDinterwordspacing

\bibitem{portland}
\BIBentryALTinterwordspacing
R.~Niranjan~Mysore, A.~Pamboris, N.~Farrington, N.~Huang, P.~Miri,
  S.~Radhakrishnan, V.~Subramanya, and A.~Vahdat, ``{PortLand: A Scalable
  Fault-tolerant Layer 2 Data Center Network Fabric},'' \emph{Proceedings of
  the ACM SIGCOMM 2009 Conference on Data Communication}, pp. 39--50, 2009.
  [Online]. Available: \url{http://doi.acm.org/10.1145/1592568.1592575}
\BIBentrySTDinterwordspacing

\bibitem{leaf-spine-cisco}
\BIBentryALTinterwordspacing
\emph{{Cisco Data Center Spine-and-Leaf Architecture: Design Overview White
  Paper}}. [Online]. Available:
  \url{http://www.cisco.com/c/en/us/products/collateral/switches/nexus-7000-series-switches/white-paper-c11-737022.html}
\BIBentrySTDinterwordspacing

\bibitem{jellyfish-topo}
\BIBentryALTinterwordspacing
\emph{{Jellyfish: Networking Data Centers Randomly}}, {Originally presented in
  NSDI 2012, visited on Aug 3, 2017}. [Online]. Available:
  \url{http://web.eecs.umich.edu/~sugih/courses/eecs589/f13/28-Jellyfish.pdf}
\BIBentrySTDinterwordspacing

\bibitem{mptcp}
A.~Ford, C.~Raiciu, M.~Handley, and O.~Bonaventure, \emph{{TCP extensions for
  multipath operation with multiple addresses}}, 2013.

\bibitem{expander-graphs}
S.~Hoory, N.~Linial, and A.~Wigderson, ``{Expander graphs and their
  applications},'' \emph{Bulletin of the American Mathematical Society},
  vol.~43, no.~4, pp. 439--561, 2006.

\bibitem{twitter}
\BIBentryALTinterwordspacing
M.~Hashemi, \emph{{The Infrastructure Behind Twitter: Scale}}. [Online].
  Available:
  \url{https://blog.twitter.com/engineering/en\%5Fus/topics/infrastructure/2017/the-infrastructure-behind-twitter-scale.html}
\BIBentrySTDinterwordspacing

\bibitem{traffic_dc_char}
T.~Benson, A.~Anand, A.~Akella, and M.~Zhang, ``{Understanding data center
  traffic characteristics},'' \emph{ACM SIGCOMM Computer Communication Review},
  vol.~40, no.~1, pp. 92--99, 2010.

\bibitem{wild}
T.~Benson, A.~Akella, and D.~A. Maltz, ``{Network traffic characteristics of
  data centers in the wild},'' \emph{Proceedings of the 10th ACM SIGCOMM
  conference on Internet measurement}, pp. 267--280, 2010.

\bibitem{social_inside}
\BIBentryALTinterwordspacing
A.~Roy, H.~Zeng, J.~Bagga, G.~Porter, and A.~C. Snoeren, ``{Inside the Social
  Network's (Datacenter) Network},'' \emph{Proceedings of the 2015 ACM
  Conference on Special Interest Group on Data Communication}, pp. 123--137,
  2015. [Online]. Available: \url{http://doi.acm.org/10.1145/2785956.2787472}
\BIBentrySTDinterwordspacing

\bibitem{d3}
\BIBentryALTinterwordspacing
C.~Wilson, H.~Ballani, T.~Karagiannis, and A.~Rowtron, ``{Better Never Than
  Late: Meeting Deadlines in Datacenter Networks},'' \emph{SIGCOMM Comput.
  Commun. Rev.}, vol.~41, no.~4, pp. 50--61, August 2011. [Online]. Available:
  \url{http://doi.acm.org/10.1145/2043164.2018443}
\BIBentrySTDinterwordspacing

\bibitem{dctcp_better}
S.~Joy and A.~Nayak, ``{Improving flow completion time for short flows in
  datacenter networks},'' \emph{2015 IFIP/IEEE International Symposium on
  Integrated Network Management (IM)}, pp. 700--705, May 2015.

\bibitem{pfabric}
\BIBentryALTinterwordspacing
M.~Alizadeh, S.~Yang, M.~Sharif, S.~Katti, N.~McKeown, B.~Prabhakar, and
  S.~Shenker, ``{pFabric: Minimal Near-optimal Datacenter Transport},''
  \emph{SIGCOMM Comput. Commun. Rev.}, vol.~43, no.~4, pp. 435--446, August
  2013. [Online]. Available: \url{http://doi.acm.org/10.1145/2534169.2486031}
\BIBentrySTDinterwordspacing

\bibitem{tm_joint}
Z.~Cao, M.~Kodialam, and T.~Lakshman, ``{Joint static and dynamic traffic
  scheduling in data center networks},'' \emph{IEEE INFOCOM 2014-IEEE
  Conference on Computer Communications}, pp. 2445--2453, 2014.

\bibitem{tm_estimation}
Z.~Hu, Y.~Qiao, and J.~Luo, ``{Coarse-grained traffic matrix estimation for
  data center networks},'' \emph{Computer Communications}, vol.~56, pp. 25--34,
  2015.

\bibitem{karuna}
L.~Chen, K.~Chen, W.~Bai, and M.~Alizadeh, ``{Scheduling Mix-flows in Commodity
  Datacenters with Karuna},'' \emph{Proceedings of the 2016 conference on ACM
  SIGCOMM 2016 Conference}, pp. 174--187, 2016.

\bibitem{speed-google}
J.~Brutlag, ``Speed matters for {Google} web search,'' \emph{Google. June},
  2009.

\bibitem{amazon_100_ms}
\BIBentryALTinterwordspacing
G.~Linden, \emph{{Make Data Useful}}, 2006. [Online]. Available:
  \url{http://sites.google.com/site/glinden/Home/StanfordDataMining.2006-11-28.ppt}
\BIBentrySTDinterwordspacing

\bibitem{calendaring}
S.~Kandula, I.~Menache, R.~Schwartz, and S.~R. Babbula, ``{Calendaring for wide
  area networks},'' \emph{ACM SIGCOMM Computer Communication Review}, vol.~44,
  no.~4, pp. 515--526, 2015.

\bibitem{amoeba}
H.~Zhang, K.~Chen, W.~Bai, D.~Han, C.~Tian, H.~Wang, H.~Guan, and M.~Zhang,
  ``{Guaranteeing deadlines for inter-datacenter transfers},''
  \emph{Proceedings of the Tenth European Conference on Computer Systems},
  p.~20, 2015.

\bibitem{dynamo}
G.~DeCandia, D.~Hastorun, M.~Jampani, G.~Kakulapati, A.~Lakshman, A.~Pilchin,
  S.~Sivasubramanian, P.~Vosshall, and W.~Vogels, ``{Dynamo: Amazon's highly
  available key-value store},'' \emph{ACM SIGOPS Operating Systems Review},
  vol.~41, no.~6, pp. 205--220, 2007.

\bibitem{bing}
\BIBentryALTinterwordspacing
V.~Jalaparti, P.~Bodik, S.~Kandula, I.~Menache, M.~Rybalkin, and C.~Yan,
  ``{Speeding Up Distributed Request-response Workflows},'' \emph{Proceedings
  of the ACM SIGCOMM 2013 Conference on SIGCOMM}, pp. 219--230, 2013. [Online].
  Available: \url{http://doi.acm.org/10.1145/2486001.2486028}
\BIBentrySTDinterwordspacing

\bibitem{bullet}
R.~Kapoor, A.~C. Snoeren, G.~M. Voelker, and G.~Porter, ``{Bullet trains: a
  study of NIC burst behavior at microsecond timescales},'' \emph{Proceedings
  of the ninth ACM conference on Emerging networking experiments and
  technologies}, pp. 133--138, 2013.

\bibitem{bursty_is_bad}
L.~Kleinrock, ``{Queueing theory},'' \emph{Vol. i, Wiley, New York}, 1975.

\bibitem{ictcp}
H.~Wu, Z.~Feng, C.~Guo, and Y.~Zhang, ``{ICTCP: Incast congestion control for
  TCP in data-center networks},'' \emph{IEEE/ACM Transactions on Networking
  (TON)}, vol.~21, no.~2, pp. 345--358, 2013.

\bibitem{interrupt_moderation}
\BIBentryALTinterwordspacing
\emph{{Interrupt Moderation}}. [Online]. Available:
  \url{https://msdn.microsoft.com/en-us/windows/hardware/drivers/network/interrupt-moderation}
\BIBentrySTDinterwordspacing

\bibitem{juggler}
Y.~Geng, V.~Jeyakumar, A.~Kabbani, and M.~Alizadeh, ``{Juggler: a practical
  reordering resilient network stack for datacenters},'' \emph{Proceedings of
  the Eleventh European Conference on Computer Systems}, p.~20, 2016.

\bibitem{mptcp-hard}
C.~Raiciu, C.~Paasch, S.~Barre, A.~Ford, M.~Honda, F.~Duchene, O.~Bonaventure,
  and M.~Handley, ``{How hard can it be? designing and implementing a
  deployable multipath TCP},'' \emph{Proceedings of the 9th USENIX conference
  on Networked Systems Design and Implementation}, pp. 29--29, 2012.

\bibitem{rfc2001}
W.~Stevens, ``{RFC 2001: TCP Slow Start, Congestion Avoidance, Fast Retransmit,
  and Fast Recovery Algorithms},'' \emph{Request for Comments, IETF}, 1997.

\bibitem{LRO}
\BIBentryALTinterwordspacing
\emph{{Large receive offload}}. [Online]. Available:
  \url{https://lwn.net/Articles/243949/}
\BIBentrySTDinterwordspacing

\bibitem{RSS}
\BIBentryALTinterwordspacing
\emph{{Receive Side Scaling (RSS)}}. [Online]. Available:
  \url{https://technet.microsoft.com/en-us/library/hh997036.aspx}
\BIBentrySTDinterwordspacing

\bibitem{GRO}
\BIBentryALTinterwordspacing
\emph{{Generic receive offload}}. [Online]. Available:
  \url{https://lwn.net/Articles/358910/}
\BIBentrySTDinterwordspacing

\bibitem{eyeq}
V.~Jeyakumar, M.~Alizadeh, D.~Mazieres, B.~Prabhakar, and C.~Kim, ``{EyeQ:
  Practical network performance isolation for the multi-tenant cloud},''
  \emph{Presented as part of the}, 2012.

\bibitem{seawall}
\BIBentryALTinterwordspacing
A.~Shieh, S.~Kandula, A.~Greenberg, C.~Kim, and B.~Saha, ``{Sharing the Data
  Center Network},'' \emph{Proceedings of the 8th USENIX Conference on
  Networked Systems Design and Implementation}, pp. 309--322, 2011. [Online].
  Available: \url{http://dl.acm.org/citation.cfm?id=1972457.1972489}
\BIBentrySTDinterwordspacing

\bibitem{elasticswitch}
L.~Popa, P.~Yalagandula, S.~Banerjee, J.~C. Mogul, Y.~Turner, and J.~R. Santos,
  ``{ElasticSwitch: practical work-conserving bandwidth guarantees for cloud
  computing},'' \emph{ACM SIGCOMM Computer Communication Review}, vol.~43,
  no.~4, pp. 351--362, 2013.

\bibitem{acdc}
K.~He, E.~Rozner, K.~Agarwal, Y.~J. Gu, W.~Felter, J.~Carter, and A.~Akella,
  ``{AC/DC TCP: Virtual congestion control enforcement for datacenter
  networks},'' \emph{Proceedings of the 2016 conference on ACM SIGCOMM 2016
  Conference}, pp. 244--257, 2016.

\bibitem{vcc}
B.~Cronkite-Ratcliff, A.~Bergman, S.~Vargaftik, M.~Ravi, N.~McKeown,
  I.~Abraham, and I.~Keslassy, ``{Virtualized congestion control},''
  \emph{Proceedings of the 2016 conference on ACM SIGCOMM 2016 Conference}, pp.
  230--243, 2016.

\bibitem{iatcp}
J.~Hwang, J.~Yoo, and N.~Choi, ``{IA-TCP: a rate based incast-avoidance
  algorithm for TCP in data center networks},'' \emph{2012 IEEE International
  Conference on Communications (ICC)}, pp. 1292--1296, 2012.

\bibitem{sfs}
J.~Zhang, D.~Zhang, K.~Huang, and Z.~Qin, ``{Minimizing datacenter flow
  completion times with server-based flow scheduling},'' \emph{Computer
  Networks}, vol.~94, pp. 360--374, 2016.

\bibitem{phost}
P.~X. Gao, A.~Narayan, G.~Kumar, R.~Agarwal, S.~Ratnasamy, and S.~Shenker,
  ``{pHost: Distributed near-optimal datacenter transport over commodity
  network fabric},'' \emph{Proceedings of the CoNEXT}, 2015.

\bibitem{japan_incast_problem}
Y.~Yang, H.~Abe, K.~i.~Baba, and S.~Shimojo, ``{A Scalable Approach to Avoid
  Incast Problem from Application Layer},'' \emph{2013 IEEE 37th Annual
  Computer Software and Applications Conference Workshops}, pp. 713--718, July
  2013.

\bibitem{rto_min_random}
\BIBentryALTinterwordspacing
V.~Vasudevan, A.~Phanishayee, H.~Shah, E.~Krevat, D.~G. Andersen, G.~R. Ganger,
  G.~A. Gibson, and B.~Mueller, ``{Safe and Effective Fine-grained TCP
  Retransmissions for Datacenter Communication},'' \emph{Proceedings of the ACM
  SIGCOMM 2009 Conference on Data Communication}, pp. 303--314, 2009. [Online].
  Available: \url{http://doi.acm.org/10.1145/1592568.1592604}
\BIBentrySTDinterwordspacing

\bibitem{dibs}
K.~Zarifis, R.~Miao, M.~Calder, E.~Katz-Bassett, M.~Yu, and J.~Padhye, ``{DIBS:
  Just-in-time congestion mitigation for data centers},'' \emph{Proceedings of
  the Ninth European Conference on Computer Systems}, p.~6, 2014.

\bibitem{outcast}
P.~Prakash, A.~Dixit, Y.~C. Hu, and R.~Kompella, ``{The TCP outcast problem:
  exposing unfairness in data center networks},'' \emph{Proceedings of the 9th
  USENIX conference on Networked Systems Design and Implementation}, pp.
  30--30, 2012.

\bibitem{rcp}
N.~Dukkipati and N.~McKeown, ``{Why flow-completion time is the right metric
  for congestion control},'' \emph{ACM SIGCOMM Computer Communication Review},
  vol.~36, no.~1, pp. 59--62, 2006.

\bibitem{pdq}
C.-Y. Hong, M.~Caesar, and P.~Godfrey, ``{Finishing flows quickly with
  preemptive scheduling},'' \emph{ACM SIGCOMM Computer Communication Review},
  vol.~42, no.~4, pp. 127--138, 2012.

\bibitem{racs}
A.~Munir, I.~A. Qazi, and S.~B. Qaisar, ``{On achieving low latency in data
  centers},'' \emph{2013 IEEE International Conference on Communications
  (ICC)}, pp. 3721--3725, 2013.

\bibitem{fairness_diff_types}
T.~Bonald, L.~Massoulie, A.~Proutiere, and J.~Virtamo, ``{A queueing analysis
  of max-min fairness, proportional fairness and balanced fairness},''
  \emph{Queueing systems}, vol.~53, no. 1-2, pp. 65--84, 2006.

\bibitem{fairness-axiomatic}
T.~Lan, D.~Kao, M.~Chiang, and A.~Sabharwal, ``{An Axiomatic Theory of Fairness
  in Network Resource Allocation},'' \emph{2010 Proceedings IEEE INFOCOM}, pp.
  1--9, March 2010.

\bibitem{mmf}
D.~Nace and M.~Pioro, ``{Max-min fairness and its applications to routing and
  load-balancing in communication networks: a tutorial},'' \emph{IEEE
  Communications Surveys \& Tutorials}, vol.~10, no.~4, pp. 5--17, 2008.

\bibitem{drf}
\BIBentryALTinterwordspacing
A.~Ghodsi, M.~Zaharia, B.~Hindman, A.~Konwinski, S.~Shenker, and I.~Stoica,
  ``{Dominant Resource Fairness: Fair Allocation of Multiple Resource Types},''
  \emph{Proceedings of the 8th USENIX Conference on Networked Systems Design
  and Implementation}, pp. 323--336, 2011. [Online]. Available:
  \url{http://dl.acm.org/citation.cfm?id=1972457.1972490}
\BIBentrySTDinterwordspacing

\bibitem{mr-fe}
\BIBentryALTinterwordspacing
C.~Joe-Wong, S.~Sen, T.~Lan, and M.~Chiang, ``{Multiresource Allocation:
  Fairness-efficiency Tradeoffs in a Unifying Framework},'' \emph{IEEE/ACM
  Trans. Netw.}, vol.~21, no.~6, pp. 1785--1798, Dec. 2013. [Online].
  Available: \url{http://dx.doi.org/10.1109/TNET.2012.2233213}
\BIBentrySTDinterwordspacing

\bibitem{elastictree}
\BIBentryALTinterwordspacing
B.~Heller, S.~Seetharaman, P.~Mahadevan, Y.~Yiakoumis, P.~Sharma, S.~Banerjee,
  and N.~McKeown, ``{ElasticTree: Saving Energy in Data Center Networks},''
  \emph{Proceedings of the 7th USENIX Conference on Networked Systems Design
  and Implementation}, pp. 17--17, 2010. [Online]. Available:
  \url{http://dl.acm.org/citation.cfm?id=1855711.1855728}
\BIBentrySTDinterwordspacing

\bibitem{energy-aware-routing}
\BIBentryALTinterwordspacing
Y.~Shang, D.~Li, and M.~Xu, ``{Energy-aware Routing in Data Center Network},''
  \emph{Proceedings of the First ACM SIGCOMM Workshop on Green Networking}, pp.
  1--8, 2010. [Online]. Available:
  \url{http://doi.acm.org/10.1145/1851290.1851292}
\BIBentrySTDinterwordspacing

\bibitem{energy-aware-routing-link-disjoint}
\BIBentryALTinterwordspacing
R.~Wang, S.~Gao, W.~Yang, and Z.~Jiang, ``{Energy aware routing with link
  disjoint backup paths},'' \emph{Computer Networks}, vol. 115, no. Supplement
  C, pp. 42 -- 53, 2017. [Online]. Available:
  \url{http://www.sciencedirect.com/science/article/pii/S1389128617300270}
\BIBentrySTDinterwordspacing

\bibitem{energy-proportional-trees}
\BIBentryALTinterwordspacing
D.~Abts, M.~R. Marty, P.~M. Wells, P.~Klausler, and H.~Liu, ``{Energy
  Proportional Datacenter Networks},'' \emph{Proceedings of the 37th Annual
  International Symposium on Computer Architecture}, pp. 338--347, 2010.
  [Online]. Available: \url{http://doi.acm.org/10.1145/1815961.1816004}
\BIBentrySTDinterwordspacing

\bibitem{greendcn}
L.~Wang, F.~Zhang, J.~A. Aroca, A.~V. Vasilakos, K.~Zheng, C.~Hou, D.~Li, and
  Z.~Liu, ``{GreenDCN: A General Framework for Achieving Energy Efficiency in
  Data Center Networks},'' \emph{IEEE Journal on Selected Areas in
  Communications}, vol.~32, no.~1, pp. 4--15, January 2014.

\bibitem{presto}
K.~He, E.~Rozner, K.~Agarwal, W.~Felter, J.~Carter, and A.~Akella, ``{Presto:
  Edge-based load balancing for fast datacenter networks},'' \emph{ACM SIGCOMM
  Computer Communication Review}, vol.~45, no.~4, pp. 465--478, 2015.

\bibitem{timely}
\BIBentryALTinterwordspacing
R.~Mittal, V.~T. Lam, N.~Dukkipati, E.~Blem, H.~Wassel, M.~Ghobadi, A.~Vahdat,
  Y.~Wang, D.~Wetherall, and D.~Zats, ``{TIMELY: RTT-based Congestion Control
  for the Datacenter},'' \emph{Proceedings of the 2015 ACM Conference on
  Special Interest Group on Data Communication}, pp. 537--550, 2015. [Online].
  Available: \url{http://doi.acm.org/10.1145/2785956.2787510}
\BIBentrySTDinterwordspacing

\bibitem{codel}
K.~Nichols and V.~Jacobson, ``{Controlling queue delay},'' \emph{Communications
  of the ACM}, vol.~55, no.~7, pp. 42--50, 2012.

\bibitem{afd}
R.~Pan, L.~Breslau, B.~Prabhakar, and S.~Shenker, ``{Approximate fairness
  through differential dropping},'' \emph{ACM SIGCOMM Computer Communication
  Review}, vol.~33, no.~2, pp. 23--39, 2003.

\bibitem{reno}
J.~Padhye, V.~Firoiu, D.~F. Towsley, and J.~F. Kurose, ``{Modeling TCP Reno
  performance: a simple model and its empirical validation},'' \emph{IEEE/ACM
  Transactions on Networking (ToN)}, vol.~8, no.~2, pp. 133--145, 2000.

\bibitem{ndp}
\BIBentryALTinterwordspacing
M.~Handley, C.~Raiciu, A.~Agache, A.~Voinescu, A.~W. Moore, G.~Antichi, and
  M.~W\'{o}jcik, ``{Re-architecting Datacenter Networks and Stacks for Low
  Latency and High Performance},'' \emph{Proceedings of the Conference of the
  ACM Special Interest Group on Data Communication}, pp. 29--42, 2017.
  [Online]. Available: \url{http://doi.acm.org/10.1145/3098822.3098825}
\BIBentrySTDinterwordspacing

\bibitem{credit-based-cc}
\BIBentryALTinterwordspacing
I.~Cho, K.~Jang, and D.~Han, ``{Credit-Scheduled Delay-Bounded Congestion
  Control for Datacenters},'' \emph{Proceedings of the Conference of the ACM
  Special Interest Group on Data Communication}, pp. 239--252, 2017. [Online].
  Available: \url{http://doi.acm.org/10.1145/3098822.3098840}
\BIBentrySTDinterwordspacing

\bibitem{conga}
\BIBentryALTinterwordspacing
M.~Alizadeh, T.~Edsall, S.~Dharmapurikar, R.~Vaidyanathan, K.~Chu,
  A.~Fingerhut, V.~T. Lam, F.~Matus, R.~Pan, N.~Yadav, and G.~Varghese,
  ``{CONGA: Distributed Congestion-aware Load Balancing for Datacenters},''
  \emph{Proceedings of the 2014 ACM Conference on SIGCOMM}, pp. 503--514, 2014.
  [Online]. Available: \url{http://doi.acm.org/10.1145/2619239.2626316}
\BIBentrySTDinterwordspacing

\bibitem{expeditus}
P.~Wang, H.~Xu, Z.~Niu, D.~Han, and Y.~Xiong, ``{Expeditus: Congestion-aware
  Load Balancing in Clos Data Center Networks},'' \emph{Proceedings of the
  Seventh ACM Symposium on Cloud Computing}, pp. 442--455, 2016.

\bibitem{rackcc}
\BIBentryALTinterwordspacing
D.~Zhuo, Q.~Zhang, V.~Liu, A.~Krishnamurthy, and T.~Anderson, ``{Rack-level
  Congestion Control},'' \emph{Proceedings of the 15th ACM Workshop on Hot
  Topics in Networks}, pp. 148--154, 2016. [Online]. Available:
  \url{http://doi.acm.org/10.1145/3005745.3005772}
\BIBentrySTDinterwordspacing

\bibitem{ecn}
S.~Floyd, K.~Ramakrishnan, and D.~L. Black, ``{RFC 3168: The Addition of
  Explicit Congestion Notification (ECN) to IP},'' \emph{Request for Comments,
  IETF}, 2001.

\bibitem{tcn}
W.~Bai, K.~Chen, L.~Chen, C.~Kim, and H.~Wu, ``{Enabling ECN over Generic
  Packet Scheduling},'' \emph{Proceedings of the 12th International on
  Conference on emerging Networking EXperiments and Technologies}, pp.
  191--204, 2016.

\bibitem{cut-payload}
P.~Cheng, F.~Ren, R.~Shu, and C.~Lin, ``{Catch the whole lot in an action:
  Rapid precise packet loss notification in data center},'' \emph{11th USENIX
  Symposium on Networked Systems Design and Implementation (NSDI 14)}, pp.
  17--28, 2014.

\bibitem{tdma}
B.~C. Vattikonda, G.~Porter, A.~Vahdat, and A.~C. Snoeren, ``{Practical TDMA
  for datacenter Ethernet},'' \emph{Proceedings of the 7th ACM european
  conference on Computer Systems}, pp. 225--238, 2012.

\bibitem{fastpass}
\BIBentryALTinterwordspacing
J.~Perry, A.~Ousterhout, H.~Balakrishnan, D.~Shah, and H.~Fugal, ``{Fastpass: A
  Centralized "Zero-queue" Datacenter Network},'' \emph{Proceedings of the 2014
  ACM Conference on SIGCOMM}, pp. 307--318, 2014. [Online]. Available:
  \url{http://doi.acm.org/10.1145/2619239.2626309}
\BIBentrySTDinterwordspacing

\bibitem{flowtune}
\BIBentryALTinterwordspacing
J.~Perry, H.~Balakrishnan, and D.~Shah, ``Flowtune: Flowlet control for
  datacenter networks,'' \emph{14th {USENIX} Symposium on Networked Systems
  Design and Implementation ({NSDI} 17)}, pp. 421--435, 2017. [Online].
  Available:
  \url{https://www.usenix.org/conference/nsdi17/technical-sessions/presentation/perry}
\BIBentrySTDinterwordspacing

\bibitem{otcp}
S.~Jouet, C.~Perkins, and D.~Pezaros, ``{OTCP: SDN-managed congestion control
  for data center networks},'' \emph{NOMS 2016 - 2016 IEEE/IFIP Network
  Operations and Management Symposium}, pp. 171--179, April 2016.

\bibitem{fibbing}
\BIBentryALTinterwordspacing
S.~Vissicchio, O.~Tilmans, L.~Vanbever, and J.~Rexford, ``{Central Control Over
  Distributed Routing},'' \emph{Proceedings of the 2015 ACM Conference on
  Special Interest Group on Data Communication}, pp. 43--56, 2015. [Online].
  Available: \url{http://doi.acm.org/10.1145/2785956.2787497}
\BIBentrySTDinterwordspacing

\bibitem{hedera}
M.~Al-Fares, S.~Radhakrishnan, B.~Raghavan, N.~Huang, and A.~Vahdat, ``{Hedera:
  Dynamic Flow Scheduling for Data Center Networks.}'' \emph{NSDI}, vol.~10,
  pp. 19--19, 2010.

\bibitem{mahout}
A.~R. Curtis, W.~Kim, and P.~Yalagandula, ``{Mahout: Low-overhead datacenter
  traffic management using end-host-based elephant detection},'' \emph{2011
  Proceedings IEEE INFOCOM}, pp. 1629--1637, April 2011.

\bibitem{sdn}
N.~McKeown, T.~Anderson, H.~Balakrishnan, G.~Parulkar, L.~Peterson, J.~Rexford,
  S.~Shenker, and J.~Turner, ``{OpenFlow: enabling innovation in campus
  networks},'' \emph{ACM SIGCOMM Computer Communication Review}, vol.~38,
  no.~2, pp. 69--74, 2008.

\bibitem{l2dct}
A.~Munir, I.~A. Qazi, Z.~A. Uzmi, A.~Mushtaq, S.~N. Ismail, M.~S. Iqbal, and
  B.~Khan, ``{Minimizing flow completion times in data centers},''
  \emph{INFOCOM, 2013 Proceedings IEEE}, pp. 2157--2165, 2013.

\bibitem{mcp}
L.~Chen, S.~Hu, K.~Chen, H.~Wu, and D.~H. Tsang, ``{Towards minimal-delay
  deadline-driven data center TCP},'' \emph{Proceedings of the Twelfth ACM
  Workshop on Hot Topics in Networks}, p.~21, 2013.

\bibitem{daq}
C.~Ding and R.~Rojas-Cessa, ``{DAQ: deadline-aware queue scheme for scheduling
  service flows in data centers},'' \emph{2014 IEEE International Conference on
  Communications (ICC)}, pp. 2989--2994, 2014.

\bibitem{pase}
\BIBentryALTinterwordspacing
A.~Munir, G.~Baig, S.~M. Irteza, I.~A. Qazi, A.~X. Liu, and F.~R. Dogar,
  ``{Friends, Not Foes: Synthesizing Existing Transport Strategies for Data
  Center Networks},'' \emph{SIGCOMM Comput. Commun. Rev.}, vol.~44, no.~4, pp.
  491--502, August 2014. [Online]. Available:
  \url{http://doi.acm.org/10.1145/2740070.2626305}
\BIBentrySTDinterwordspacing

\bibitem{tfrc_paper}
S.~Floyd, M.~Handley, J.~Padhye, and J.~Widmer, ``{Equation-based congestion
  control for unicast applications},'' \emph{ACM SIGCOMM Computer Communication
  Review}, vol.~30, no.~4, pp. 43--56, 2000.

\bibitem{tcppacing_old}
A.~Aggarwal, S.~Savage, and T.~Anderson, ``{Understanding the performance of
  TCP pacing},'' \emph{INFOCOM 2000. Nineteenth Annual Joint Conference of the
  IEEE Computer and Communications Societies. Proceedings. IEEE}, vol.~3, pp.
  1157--1165, 2000.

\bibitem{tcppacing}
M.~Ghobadi and Y.~Ganjali, ``{TCP pacing in data center networks},'' \emph{2013
  IEEE 21st Annual Symposium on High-Performance Interconnects}, pp. 25--32,
  2013.

\bibitem{flowlet}
S.~Sinha, S.~Kandula, and D.~Katabi, ``{Harnessing TCP's burstiness using
  flowlet switching},'' \emph{Proceedings of ACM Hot-Nets}, 2004.

\bibitem{per_packet_load_balancing}
J.~Cao, R.~Xia, P.~Yang, C.~Guo, G.~Lu, L.~Yuan, Y.~Zheng, H.~Wu, Y.~Xiong, and
  D.~Maltz, ``{Per-packet load-balanced, low-latency routing for Clos-based
  data center networks},'' \emph{Proceedings of the ninth ACM conference on
  Emerging networking experiments and technologies}, pp. 49--60, 2013.

\bibitem{tinyflow}
H.~Xu and B.~Li, ``{TinyFlow: Breaking elephants down into mice in data center
  networks},'' \emph{2014 IEEE 20th International Workshop on Local \&
  Metropolitan Area Networks (LANMAN)}, pp. 1--6, 2014.

\bibitem{nicpic}
S.~Radhakrishnan, V.~Jeyakumar, A.~Kabbani, G.~Porter, and A.~Vahdat,
  ``{NicPic: Scalable and Accurate End-Host Rate Limiting},'' \emph{Presented
  as part of the 5th USENIX Workshop on Hot Topics in Cloud Computing}, 2013.

\bibitem{carousel}
\BIBentryALTinterwordspacing
A.~Saeed, N.~Dukkipati, V.~Valancius, V.~The~Lam, C.~Contavalli, and A.~Vahdat,
  ``{Carousel: Scalable Traffic Shaping at End Hosts},'' \emph{Proceedings of
  the Conference of the ACM Special Interest Group on Data Communication}, pp.
  404--417, 2017. [Online]. Available:
  \url{http://doi.acm.org/10.1145/3098822.3098852}
\BIBentrySTDinterwordspacing

\bibitem{sf-rate-limiting}
\BIBentryALTinterwordspacing
K.~He, W.~Qin, Q.~Zhang, W.~Wu, J.~Yang, T.~Pan, C.~Hu, J.~Zhang, B.~Stephens,
  A.~Akella, and Y.~Zhang, ``{Low Latency Software Rate Limiters for Cloud
  Networks},'' \emph{Proceedings of the First Asia-Pacific Workshop on
  Networking}, pp. 78--84, 2017. [Online]. Available:
  \url{http://doi.acm.org/10.1145/3106989.3107005}
\BIBentrySTDinterwordspacing

\bibitem{gatekeeper}
H.~Rodrigues, J.~R. Santos, Y.~Turner, P.~Soares, and D.~Guedes, ``{Gatekeeper:
  Supporting Bandwidth Guarantees for Multi-tenant Datacenter Networks},''
  \emph{WIOV}, 2011.

\bibitem{dpp}
\BIBentryALTinterwordspacing
\emph{{Cisco Application Centric Infrastructure Fundamentals}}. [Online].
  Available:
  \url{http://www.cisco.com/c/en/us/td/docs/switches/datacenter/aci/apic/sw/1-x/aci-fundamentals/b\%5FACI-Fundamentals/b\%5FACI\%5FFundamentals\%5FBigBook\%5Fchapter\%5F0100.html}
\BIBentrySTDinterwordspacing

\bibitem{pias}
W.~Bai, L.~Chen, K.~Chen, D.~Han, C.~Tian, and W.~Sun, ``{PIAS: Practical
  information-agnostic flow scheduling for data center networks},''
  \emph{Proceedings of the 13th ACM Workshop on Hot Topics in Networks}, p.~25,
  2014.

\bibitem{jump}
M.~P. Grosvenor, M.~Schwarzkopf, I.~Gog, R.~N. Watson, A.~W. Moore, S.~Hand,
  and J.~Crowcroft, ``{Queues don't matter when you can JUMP them!}''
  \emph{12th USENIX Symposium on Networked Systems Design and Implementation
  (NSDI 15)}, pp. 1--14, 2015.

\bibitem{secondnet}
C.~Guo, G.~Lu, H.~J. Wang, S.~Yang, C.~Kong, P.~Sun, W.~Wu, and Y.~Zhang,
  ``{Secondnet: a data center network virtualization architecture with
  bandwidth guarantees},'' \emph{Proceedings of the 6th International
  Conference}, p.~15, 2010.

\bibitem{trinity}
S.~Hu, W.~Bai, K.~Chen, C.~Tian, Y.~Zhang, and H.~Wu, ``{Providing bandwidth
  guarantees, work conservation and low latency simultaneously in the cloud},''
  \emph{IEEE INFOCOM 2016 - The 35th Annual IEEE International Conference on
  Computer Communications}, pp. 1--9, April 2016.

\bibitem{pacing_isi}
\BIBentryALTinterwordspacing
\emph{{Rate Based Pacing for TCP}}. [Online]. Available:
  \url{http://www.isi.edu/lsam/publications/rate\%5Fbased\%5Fpacing/}
\BIBentrySTDinterwordspacing

\bibitem{mptcp-improving}
C.~Raiciu, S.~Barre, C.~Pluntke, A.~Greenhalgh, D.~Wischik, and M.~Handley,
  ``{Improving datacenter performance and robustness with multipath TCP},''
  \emph{ACM SIGCOMM Computer Communication Review}, vol.~41, no.~4, pp.
  266--277, 2011.

\bibitem{qcn}
M.~Alizadeh, A.~Kabbani, B.~Atikoglu, and B.~Prabhakar, ``{Stability analysis
  of QCN: the averaging principle},'' \emph{Proceedings of the ACM SIGMETRICS
  joint international conference on Measurement and modeling of computer
  systems}, pp. 49--60, 2011.

\bibitem{scc}
C.~Tian, A.~Munir, A.~X. Liu, Y.~Liu, Y.~Li, J.~Sun, F.~Zhang, and G.~Zhang,
  ``{Multi-tenant multi-objective bandwidth allocation in datacenters using
  stacked congestion control},'' \emph{IEEE INFOCOM 2017 - IEEE Conference on
  Computer Communications}, pp. 1--9, May 2017.

\bibitem{pacing_small_buffer_practical}
Y.~Cai, Y.~S. Hanay, and T.~Wolf, ``{Practical packet pacing in small-buffer
  networks},'' \emph{2009 IEEE International Conference on Communications}, pp.
  1--6, 2009.

\bibitem{epn}
Y.~Lu, G.~Chen, L.~Luo, K.~Tan, Y.~Xiong, X.~Wang, and E.~Chen, ``{One more
  queue is enough: Minimizing flow completion time with explicit priority
  notification},'' \emph{IEEE INFOCOM 2017 - IEEE Conference on Computer
  Communications}, pp. 1--9, May 2017.

\bibitem{openvswitch}
B.~Pfaff, J.~Pettit, T.~Koponen, E.~Jackson, A.~Zhou, J.~Rajahalme, J.~Gross,
  A.~Wang, J.~Stringer, P.~Shelar \emph{et~al.}, ``{The design and
  implementation of Open vSwitch},'' \emph{12th USENIX symposium on networked
  systems design and implementation (NSDI 15)}, pp. 117--130, 2015.

\bibitem{8021p}
N.~Ek, \emph{{IEEE 802.1 P, Q-QOS on the MAC level}}, 1999.

\bibitem{diffserv}
S.~Blake, D.~Black, M.~Carlson, E.~Davies, Z.~Wang, and W.~Weiss, \emph{{An
  architecture for differentiated services}}, 1998.

\bibitem{rdma_scale}
C.~Guo, H.~Wu, Z.~Deng, G.~Soni, J.~Ye, J.~Padhye, and M.~Lipshteyn, ``{RDMA
  over commodity Ethernet at scale},'' \emph{Proceedings of the 2016 conference
  on ACM SIGCOMM 2016 Conference}, pp. 202--215, 2016.

\bibitem{flyways}
S.~Kandula, J.~Padhye, and P.~Bahl, ``{Flyways To De-Congest Data Center
  Networks},'' \emph{HotNets}, 2009.

\bibitem{letitflow}
\BIBentryALTinterwordspacing
E.~Vanini, R.~Pan, M.~Alizadeh, P.~Taheri, and T.~Edsall, ``{Let It Flow:
  Resilient Asymmetric Load Balancing with Flowlet Switching},'' \emph{14th
  {USENIX} Symposium on Networked Systems Design and Implementation ({NSDI}
  17)}, pp. 407--420, 2017. [Online]. Available:
  \url{https://www.usenix.org/conference/nsdi17/technical-sessions/presentation/vanini}
\BIBentrySTDinterwordspacing

\bibitem{ecmp}
C.~E. Hopps, \emph{{Analysis of an equal-cost multi-path algorithm}}, 2000.

\bibitem{load_balancing_survey}
S.~Mahapatra and X.~Yuan, ``{Load balancing mechanisms in data center
  networks},'' \emph{the 7th Int. Conf. \& Expo on Emerging Technologies for a
  Smarter World (CEWIT)}, 2010.

\bibitem{planck}
J.~Rasley, B.~Stephens, C.~Dixon, E.~Rozner, W.~Felter, K.~Agarwal, J.~Carter,
  and R.~Fonseca, ``{Planck: Millisecond-scale monitoring and control for
  commodity networks},'' \emph{ACM SIGCOMM Computer Communication Review},
  vol.~44, no.~4, pp. 407--418, 2015.

\bibitem{why_lossless_ethernet}
\BIBentryALTinterwordspacing
\emph{{Benefits Of SAN/LAN Convergence}}. [Online]. Available:
  \url{http://i.dell.com/sites/doccontent/business/solutions/whitepapers/en/Documents/benefits-san-lan-convergence.pdf}
\BIBentrySTDinterwordspacing

\bibitem{pfc}
\emph{{IEEE 802.1Qbb Priority-based Flow Control}}, 2011.

\bibitem{pfc_cisco}
\BIBentryALTinterwordspacing
\emph{{Priority Flow Control: Build Reliable Layer 2 Infrastructure}}.
  [Online]. Available:
  \url{https://www.cisco.com/c/en/us/products/collateral/switches/nexus-7000-series-switches/white\%5Fpaper\%5Fc11-542809.pdf}
\BIBentrySTDinterwordspacing

\bibitem{load-balancing-wild}
\BIBentryALTinterwordspacing
H.~Zhang, J.~Zhang, W.~Bai, K.~Chen, and M.~Chowdhury, ``{Resilient Datacenter
  Load Balancing in the Wild},'' \emph{Proceedings of the Conference of the ACM
  Special Interest Group on Data Communication}, pp. 253--266, 2017. [Online].
  Available: \url{http://doi.acm.org/10.1145/3098822.3098841}
\BIBentrySTDinterwordspacing

\bibitem{flier}
A.~Kabbani and M.~Sharif, ``{Flier: Flow-level congestion-aware routing for
  direct-connect data centers},'' \emph{IEEE INFOCOM 2017 - IEEE Conference on
  Computer Communications}, pp. 1--9, May 2017.

\bibitem{vlb}
R.~Zhang-Shen and N.~McKeown, ``{Designing a fault-tolerant network using
  valiant load-balancing},'' \emph{INFOCOM 2008. The 27th Conference on
  Computer Communications. IEEE}, 2008.

\bibitem{spray}
A.~Dixit, P.~Prakash, Y.~C. Hu, and R.~R. Kompella, ``{On the impact of packet
  spraying in data center networks},'' \emph{INFOCOM, 2013 Proceedings IEEE},
  pp. 2130--2138, 2013.

\bibitem{fastlane}
D.~Zats, A.~P. Iyer, G.~Ananthanarayanan, R.~H. Katz, I.~Stoica, and A.~Vahdat,
  ``{FastLane: Agile Drop Notification for Datacenter Networks},'' DTIC
  Document, Tech. Rep., 2013.

\bibitem{drill}
\BIBentryALTinterwordspacing
S.~Ghorbani, Z.~Yang, P.~B. Godfrey, Y.~Ganjali, and A.~Firoozshahian,
  ``{DRILL: Micro Load Balancing for Low-latency Data Center Networks},''
  \emph{Proceedings of the Conference of the ACM Special Interest Group on Data
  Communication}, pp. 225--238, 2017. [Online]. Available:
  \url{http://doi.acm.org/10.1145/3098822.3098839}
\BIBentrySTDinterwordspacing

\bibitem{ceph}
S.~A. Weil, S.~A. Brandt, E.~L. Miller, D.~D. Long, and C.~Maltzahn, ``{Ceph: A
  scalable, high-performance distributed file system},'' \emph{Proceedings of
  the 7th symposium on Operating systems design and implementation}, pp.
  307--320, 2006.

\bibitem{hdfs}
K.~Shvachko, H.~Kuang, S.~Radia, and R.~Chansler, ``{The hadoop distributed
  file system},'' \emph{2010 IEEE 26th symposium on mass storage systems and
  technologies (MSST)}, pp. 1--10, 2010.

\bibitem{corral}
\BIBentryALTinterwordspacing
V.~Jalaparti, P.~Bodik, I.~Menache, S.~Rao, K.~Makarychev, and M.~Caesar,
  ``{Network-Aware Scheduling for Data-Parallel Jobs: Plan When You Can},''
  \emph{SIGCOMM Comput. Commun. Rev.}, vol.~45, no.~4, pp. 407--420, August
  2015. [Online]. Available: \url{http://doi.acm.org/10.1145/2829988.2787488}
\BIBentrySTDinterwordspacing

\bibitem{tetris}
\BIBentryALTinterwordspacing
R.~Grandl, G.~Ananthanarayanan, S.~Kandula, S.~Rao, and A.~Akella,
  ``{Multi-resource Packing for Cluster Schedulers},'' \emph{Proceedings of the
  2014 ACM Conference on SIGCOMM}, pp. 455--466, 2014. [Online]. Available:
  \url{http://doi.acm.org/10.1145/2619239.2626334}
\BIBentrySTDinterwordspacing

\bibitem{silo}
K.~Jang, J.~Sherry, H.~Ballani, and T.~Moncaster, ``{Silo: Predictable message
  completion time in the cloud},'' Tech. rep., Microsoft Research, 2013.
  MSR-TR-2013-95, Tech. Rep., 2013.

\bibitem{ali-INFOCOM}
Q.~Xie, A.~Yekkehkhany, and Y.~Lu, ``{Scheduling with multi-level data
  locality: Throughput and heavy-traffic optimality},'' \emph{IEEE INFOCOM 2016
  - The 35th Annual IEEE International Conference on Computer Communications},
  pp. 1--9, April 2016.

\bibitem{ali-near-data-locality-scheduling}
A.~Yekkehkhany, ``{Near Data Scheduling for Data Centers with Multi Levels of
  Data Locality},'' \emph{arXiv preprint arXiv:1702.07802}, 2017.

\bibitem{gb-pandas}
A.~Yekkehkhany, A.~Hojjati, and M.~H. Hajiesmaili, ``{GB-PANDAS: Throughput and
  heavy-traffic optimality analysis for affinity scheduling},'' \emph{arXiv
  preprint arXiv:1709.08115}, 2017.

\bibitem{neat}
A.~Munir, T.~He, R.~Raghavendra, F.~Le, and A.~X. Liu, ``{Network Scheduling
  Aware Task Placement in Datacenters},'' \emph{Proceedings of the 12th
  International on Conference on emerging Networking EXperiments and
  Technologies}, pp. 221--235, 2016.

\bibitem{bgp_better_igp}
P.~Lahiri, G.~Chen, P.~Lapukhov, E.~Nkposong, D.~Maltz, R.~Toomey, and L.~Yuan,
  ``{Routing design for large scale data centers: BGP is the better IGP},''
  \emph{NANOG, Jun}, 2012.

\bibitem{bgp_pic}
C.~Filsfils, P.~Mohapatra, J.~Bettink, P.~Dharwadkar, P.~De~Vriendt, Y.~Tsier,
  V.~Van Den~Schrieck, O.~Bonaventure, P.~Francois \emph{et~al.}, ``{BGP Prefix
  Independent Convergence (PIC)},'' UCL, Tech. Rep., 2007.

\bibitem{shadow_macs}
K.~Agarwal, C.~Dixon, E.~Rozner, and J.~Carter, ``{Shadow MACs: Scalable
  label-switching for commodity Ethernet},'' \emph{Proceedings of the third
  workshop on Hot topics in software defined networking}, pp. 157--162, 2014.

\bibitem{past}
B.~Stephens, A.~Cox, W.~Felter, C.~Dixon, and J.~Carter, ``{PAST: Scalable
  Ethernet for data centers},'' \emph{Proceedings of the 8th international
  conference on Emerging networking experiments and technologies}, pp. 49--60,
  2012.

\bibitem{quagga_flow}
\BIBentryALTinterwordspacing
M.~R. Nascimento, C.~E. Rothenberg, M.~R. Salvador, and M.~F. Magalh\~{a}es,
  ``{QuagFlow: Partnering Quagga with OpenFlow},'' \emph{Proceedings of the ACM
  SIGCOMM 2010 Conference}, pp. 441--442, 2010. [Online]. Available:
  \url{http://doi.acm.org/10.1145/1851182.1851252}
\BIBentrySTDinterwordspacing

\bibitem{monsoon}
A.~Greenberg, P.~Lahiri, D.~A. Maltz, P.~Patel, and S.~Sengupta, ``{Towards a
  next generation data center architecture: scalability and commoditization},''
  \emph{Proceedings of the ACM workshop on Programmable routers for extensible
  services of tomorrow}, pp. 57--62, 2008.

\bibitem{b4}
S.~Jain, A.~Kumar, S.~Mandal, J.~Ong, L.~Poutievski, A.~Singh, S.~Venkata,
  J.~Wanderer, J.~Zhou, M.~Zhu \emph{et~al.}, ``{B4: Experience with a
  globally-deployed software defined WAN},'' \emph{ACM SIGCOMM Computer
  Communication Review}, vol.~43, no.~4, pp. 3--14, 2013.

\bibitem{source_routing}
S.~A. Jyothi, M.~Dong, and P.~Godfrey, ``{Towards a flexible data center fabric
  with source routing},'' \emph{Proceedings of the 1st ACM SIGCOMM Symposium on
  Software Defined Networking Research}, p.~10, 2015.

\bibitem{hula}
\BIBentryALTinterwordspacing
N.~Katta, M.~Hira, C.~Kim, A.~Sivaraman, and J.~Rexford, ``{HULA: Scalable Load
  Balancing Using Programmable Data Planes},'' \emph{Proceedings of the
  Symposium on SDN Research}, pp. 10:1--10:12, 2016. [Online]. Available:
  \url{http://doi.acm.org/10.1145/2890955.2890968}
\BIBentrySTDinterwordspacing

\bibitem{mptcp_cc}
D.~Wischik, C.~Raiciu, A.~Greenhalgh, and M.~Handley, ``{Design, Implementation
  and Evaluation of Congestion Control for Multipath TCP},'' \emph{NSDI},
  vol.~11, pp. 8--8, 2011.

\bibitem{delayed_ack_mptcp}
M.~Li, A.~Lukyanenko, S.~Tarkoma, and A.~Yla-Jaaski, ``{The Delayed ACK
  evolution in MPTCP},'' \emph{2013 IEEE Global Communications Conference
  (GLOBECOM)}, pp. 2282--2288, 2013.

\bibitem{xmp}
Y.~Cao, M.~Xu, X.~Fu, and E.~Dong, ``{Explicit multipath congestion control for
  data center networks},'' \emph{Proceedings of the ninth ACM conference on
  Emerging networking experiments and technologies}, pp. 73--84, 2013.

\bibitem{mmptcp}
M.~Kheirkhah, I.~Wakeman, and G.~Parisis, ``{MMPTCP: A multipath transport
  protocol for data centers},'' \emph{IEEE INFOCOM 2016 - The 35th Annual IEEE
  International Conference on Computer Communications}, pp. 1--9, April 2016.

\bibitem{red}
S.~Floyd and V.~Jacobson, ``{Random early detection gateways for congestion
  avoidance},'' \emph{IEEE/ACM Transactions on networking}, vol.~1, no.~4, pp.
  397--413, 1993.

\bibitem{red_one_parameter}
V.~Jacobson, K.~Nichols, and K.~Poduri, ``{RED in a Different Light},''
  \emph{Draft, Cisco Systems, September}, 1999.

\bibitem{multipathing-issues}
T.~D. Wallace and A.~Shami, ``{A Review of Multihoming Issues Using the Stream
  Control Transmission Protocol},'' \emph{IEEE Communications Surveys
  Tutorials}, vol.~14, no.~2, pp. 565--578, Second 2012.

\bibitem{mptcp_honda}
M.~Honda, Y.~Nishida, L.~Eggert, P.~Sarolahti, and H.~Tokuda, ``{Multipath
  congestion control for shared bottleneck},'' \emph{Proc. PFLDNeT workshop},
  pp. 19--24, 2009.

\bibitem{taps}
L.~Liu, D.~Li, and J.~Wu, ``{TAPS: Software Defined Task-Level Deadline-Aware
  Preemptive Flow Scheduling in Data Centers},'' \emph{Parallel Processing
  (ICPP), 2015 44th International Conference on}, pp. 659--668, 2015.

\bibitem{incast_cross_schedule}
H.-W. Tseng, W.-C. Chang, I.~Peng, P.-S. Chen \emph{et~al.}, ``{A Cross-Layer
  Flow Schedule with Dynamical Grouping for Avoiding TCP Incast Problem in Data
  Center Networks},'' \emph{Proceedings of the International Conference on
  Research in Adaptive and Convergent Systems}, pp. 91--96, 2016.

\bibitem{repflow_node_js}
S.~Liu, W.~Bai, H.~Xu, K.~Chen, and Z.~Cai, ``{Repflow on node.js: Cutting tail
  latency in data center networks at the applications layer},'' \emph{Computing
  Research Repository}, 2014.

\bibitem{rcd}
M.~Noormohammadpour, C.~S. Raghavendra, S.~Rao, and A.~M. Madni, ``{RCD: Rapid
  Close to Deadline Scheduling for datacenter networks},'' \emph{World
  Automation Congress (WAC), 2016}, pp. 1--6, 2016.

\bibitem{dcroute}
M.~Noormohammadpour, C.~S. Raghavendra, and S.~Rao, ``{DCRoute: Speeding up
  Inter-Datacenter Traffic Allocation while Guaranteeing Deadlines},''
  \emph{High Performance Computing, Data, and Analytics (HiPC)}, 2016.

\bibitem{varys}
\BIBentryALTinterwordspacing
M.~Chowdhury, Y.~Zhong, and I.~Stoica, ``{Efficient Coflow Scheduling with
  Varys},'' \emph{SIGCOMM Comput. Commun. Rev.}, vol.~44, no.~4, pp. 443--454,
  August 2014. [Online]. Available:
  \url{http://doi.acm.org/10.1145/2740070.2626315}
\BIBentrySTDinterwordspacing

\bibitem{mix-flow-mohammad}
\BIBentryALTinterwordspacing
M.~Noormohammadpour and C.~S. Raghavendra, ``{Comparison of Flow Scheduling
  Policies for Mix of Regular and Deadline Traffic in Datacenter
  Environments},'' \emph{{Department of Computer Science, University of
  Southern California, Report no. 17-973}}, 2017. [Online]. Available:
  \url{http://bit.ly/mix-traffic-scheduling}
\BIBentrySTDinterwordspacing

\bibitem{FCFS-tardiness}
H.~Leontyev and J.~H. Anderson, ``{Tardiness Bounds for FIFO Scheduling on
  Multiprocessors},'' \emph{19th Euromicro Conference on Real-Time Systems
  (ECRTS'07)}, pp. 71--71, July 2007.

\bibitem{fcfs}
\BIBentryALTinterwordspacing
A.~Wierman and B.~Zwart, ``{Is Tail-Optimal Scheduling Possible?}''
  \emph{Operations Research}, vol.~60, no.~5, pp. 1249--1257, 2012. [Online].
  Available: \url{https://doi.org/10.1287/opre.1120.1086}
\BIBentrySTDinterwordspacing

\bibitem{srpt}
L.~Schrage, ``{A proof of the optimality of the shortest remaining processing
  time discipline},'' \emph{Operations Research}, vol.~16, no.~3, pp. 687--690,
  1968.

\bibitem{las}
\BIBentryALTinterwordspacing
I.~A. Rai, G.~Urvoy-Keller, M.~K. Vernon, and E.~W. Biersack, ``{Performance
  Analysis of LAS-based Scheduling Disciplines in a Packet Switched Network},''
  \emph{SIGMETRICS Perform. Eval. Rev.}, vol.~32, no.~1, pp. 106--117, Jun
  2004. [Online]. Available: \url{http://doi.acm.org/10.1145/1012888.1005702}
\BIBentrySTDinterwordspacing

\bibitem{non-clair}
B.~Kalyanasundaram and K.~R. Pruhs, ``{Minimizing flow time
  nonclairvoyantly},'' \emph{Journal of the ACM (JACM)}, vol.~50, no.~4, pp.
  551--567, 2003.

\bibitem{proactive-cc}
\BIBentryALTinterwordspacing
L.~Jose, L.~Yan, M.~Alizadeh, G.~Varghese, N.~McKeown, and S.~Katti, ``{High
  Speed Networks Need Proactive Congestion Control},'' \emph{Proceedings of the
  14th ACM Workshop on Hot Topics in Networks}, pp. 1--7, 2015. [Online].
  Available: \url{http://doi.acm.org/10.1145/2834050.2834096}
\BIBentrySTDinterwordspacing

\bibitem{sdndef}
\BIBentryALTinterwordspacing
\emph{{Software-Defined Networking (SDN) Definition}}. [Online]. Available:
  \url{https://www.opennetworking.org/sdn-resources/sdn-definition}
\BIBentrySTDinterwordspacing

\bibitem{survey-sdn}
B.~A.~A. Nunes, M.~Mendonca, X.~N. Nguyen, K.~Obraczka, and T.~Turletti, ``{A
  Survey of Software-Defined Networking: Past, Present, and Future of
  Programmable Networks},'' \emph{IEEE Communications Surveys Tutorials},
  vol.~16, no.~3, pp. 1617--1634, Third 2014.

\bibitem{openflowspec}
\BIBentryALTinterwordspacing
\emph{{OpenFlow Switch Specification Version 1.0.0 (Wire Protocol 0x01)}}.
  [Online]. Available:
  \url{http://archive.openflow.org/documents/openflow-spec-v1.0.0.pdf}
\BIBentrySTDinterwordspacing

\bibitem{openflowspec11}
\BIBentryALTinterwordspacing
\emph{{OpenFlow Switch Specification Version 1.1.0 Implemented (Wire Protocol
  0x02)}}. [Online]. Available:
  \url{http://archive.openflow.org/documents/openflow-spec-v1.1.0.pdf}
\BIBentrySTDinterwordspacing

\bibitem{openflowspec15}
\BIBentryALTinterwordspacing
\emph{{OpenFlow Switch Specification Version 1.5.0 (Protocol version 0x06)}}.
  [Online]. Available:
  \url{https://www.opennetworking.org/images/stories/downloads/sdn-resources/onf-specifications/openflow/openflow-switch-v1.5.0.noipr.pdf}
\BIBentrySTDinterwordspacing

\bibitem{sdnbgp}
\BIBentryALTinterwordspacing
\emph{{Brain-Slug: a BGP-Only SDN for Large-Scale Data-Centers}}. [Online].
  Available:
  \url{https://www.nanog.org/sites/default/files/wed.general.brainslug.lapukhov.20.pdf}
\BIBentrySTDinterwordspacing

\bibitem{pdp-slides}
\BIBentryALTinterwordspacing
\emph{{Programming The Network Data Plane}}. [Online]. Available:
  \url{http://netseminar.stanford.edu/seminars/03\%5F31\%5F16.pdf}
\BIBentrySTDinterwordspacing

\bibitem{pdp-next}
\BIBentryALTinterwordspacing
\emph{{Data Plane Programmability, the next step in SDN}}. [Online]. Available:
  \url{http://sites.ieee.org/netsoft/files/2017/07/Netsoft2017\%5FKeynote\%5FBianchi.pdf}
\BIBentrySTDinterwordspacing

\bibitem{tofino}
\BIBentryALTinterwordspacing
\emph{{Barefoot Worlds Fastest and Most Programmable Networks}}. [Online].
  Available:
  \url{https://www.barefootnetworks.com/media/white\%5Fpapers/Barefoot-Worlds-Fastest-Most-Programmable-Networks.pdf}
\BIBentrySTDinterwordspacing

\bibitem{rmt}
\BIBentryALTinterwordspacing
P.~Bosshart, G.~Gibb, H.-S. Kim, G.~Varghese, N.~McKeown, M.~Izzard, F.~Mujica,
  and M.~Horowitz, ``{Forwarding Metamorphosis: Fast Programmable Match-action
  Processing in Hardware for SDN},'' \emph{Proceedings of the ACM SIGCOMM 2013
  Conference on SIGCOMM}, pp. 99--110, 2013. [Online]. Available:
  \url{http://doi.acm.org/10.1145/2486001.2486011}
\BIBentrySTDinterwordspacing

\bibitem{domino}
\BIBentryALTinterwordspacing
A.~Sivaraman, A.~Cheung, M.~Budiu, C.~Kim, M.~Alizadeh, H.~Balakrishnan,
  G.~Varghese, N.~McKeown, and S.~Licking, ``{Packet Transactions: High-Level
  Programming for Line-Rate Switches},'' \emph{Proceedings of the 2016 ACM
  SIGCOMM Conference}, pp. 15--28, 2016. [Online]. Available:
  \url{http://doi.acm.org/10.1145/2934872.2934900}
\BIBentrySTDinterwordspacing

\bibitem{p4}
P.~Bosshart, D.~Daly, G.~Gibb, M.~Izzard, N.~McKeown, J.~Rexford,
  C.~Schlesinger, D.~Talayco, A.~Vahdat, G.~Varghese \emph{et~al.}, ``{P4:
  Programming protocol-independent packet processors},'' \emph{ACM SIGCOMM
  Computer Communication Review}, vol.~44, no.~3, pp. 87--95, 2014.

\bibitem{p4-web}
\BIBentryALTinterwordspacing
\emph{{P4}}. [Online]. Available: \url{http://p4.org/}
\BIBentrySTDinterwordspacing

\bibitem{connectx}
\BIBentryALTinterwordspacing
\emph{{Ethernet Cards - Overview}}. [Online]. Available:
  \url{http://www.mellanox.com/page/ethernet\%5Fcards\%5Foverview}
\BIBentrySTDinterwordspacing

\bibitem{smart_nic}
D.~Firestone, ``{SmartNIC: FPGA Innovation in OCS Servers for Microsoft
  Azure},'' \emph{OCP US Summit}, 2016.

\bibitem{azure_smart_nic}
A.~Greenberg, ``{SDN for the Cloud},'' \emph{Keynote in the 2015 ACM Conference
  on Special Interest Group on Data Communication}, 2015.

\bibitem{ocp}
\BIBentryALTinterwordspacing
\emph{{Open Compute Project}}. [Online]. Available:
  \url{http://opencompute.org/}
\BIBentrySTDinterwordspacing

\bibitem{netmap}
L.~Rizzo and M.~Landi, ``{Netmap: Memory mapped access to network devices},''
  \emph{ACM SIGCOMM Computer Communication Review}, vol.~41, no.~4, pp.
  422--423, 2011.

\bibitem{netmap-github}
\BIBentryALTinterwordspacing
\emph{{Netmap - a framework for fast packet I/O}}. [Online]. Available:
  \url{https://github.com/luigirizzo/netmap}
\BIBentrySTDinterwordspacing

\bibitem{vpp}
\BIBentryALTinterwordspacing
\emph{{FD.io}}. [Online]. Available: \url{https://fd.io/technology}
\BIBentrySTDinterwordspacing

\bibitem{vpp-github}
\BIBentryALTinterwordspacing
\emph{{Vector Packet Processing}}. [Online]. Available:
  \url{https://github.com/chrisy/vpp}
\BIBentrySTDinterwordspacing

\bibitem{dpdk}
\BIBentryALTinterwordspacing
\emph{{DPDK}}. [Online]. Available: \url{http://dpdk.org/}
\BIBentrySTDinterwordspacing

\bibitem{sandstorm}
I.~Marinos, R.~N. Watson, and M.~Handley, ``{Network stack specialization for
  performance},'' \emph{ACM SIGCOMM Computer Communication Review}, vol.~44,
  no.~4, pp. 175--186, 2014.

\bibitem{mtcp}
E.~Jeong, S.~Wood, M.~Jamshed, H.~Jeong, S.~Ihm, D.~Han, and K.~Park, ``{mTCP:
  a highly scalable user-level TCP stack for multicore systems},'' \emph{11th
  USENIX Symposium on Networked Systems Design and Implementation (NSDI 14)},
  pp. 489--502, 2014.

\bibitem{softnic}
\BIBentryALTinterwordspacing
S.~Han, K.~Jang, A.~Panda, S.~Palkar, D.~Han, and S.~Ratnasamy, ``{SoftNIC: A
  Software NIC to Augment Hardware},'' EECS Department, University of
  California, Berkeley, Tech. Rep. UCB/EECS-2015-155, May 2015. [Online].
  Available:
  \url{http://www2.eecs.berkeley.edu/Pubs/TechRpts/2015/EECS-2015-155.html}
\BIBentrySTDinterwordspacing

\bibitem{ramcloud}
J.~Ousterhout, A.~Gopalan, A.~Gupta, A.~Kejriwal, C.~Lee, B.~Montazeri,
  D.~Ongaro, S.~J. Park, H.~Qin, M.~Rosenblum \emph{et~al.}, ``{The RAMCloud
  storage system},'' \emph{ACM Transactions on Computer Systems (TOCS)},
  vol.~33, no.~3, p.~7, 2015.

\bibitem{farm}
A.~Dragojevic, D.~Narayanan, M.~Castro, and O.~Hodson, ``{FaRM: fast remote
  memory},'' \emph{11th USENIX Symposium on Networked Systems Design and
  Implementation (NSDI 14)}, pp. 401--414, 2014.

\bibitem{infiniswap}
\BIBentryALTinterwordspacing
J.~Gu, Y.~Lee, Y.~Zhang, M.~Chowdhury, and K.~G. Shin, ``{Efficient Memory
  Disaggregation with Infiniswap},'' \emph{14th {USENIX} Symposium on Networked
  Systems Design and Implementation ({NSDI} 17)}, pp. 649--667, 2017. [Online].
  Available:
  \url{https://www.usenix.org/conference/nsdi17/technical-sessions/presentation/gu}
\BIBentrySTDinterwordspacing

\bibitem{roce}
\BIBentryALTinterwordspacing
\emph{{Supplement to Infiniband Architecture Specification Volume 1, Release
  1.2.1 Annex A16: RDMA over Converged Ethernet (RoCE)}}, {Infiniband Trade
  Association}, April 2010. [Online]. Available:
  \url{https://cw.infinibandta.org/document/dl/7148}
\BIBentrySTDinterwordspacing

\bibitem{roce2}
\BIBentryALTinterwordspacing
\emph{{Supplement to InfiniBand Architecture Specification Volume 1, Release
  1.2.2 Annex A17: RoCEv2 (IP routable RoCE)}}, {Infiniband Trade Association},
  September 2014. [Online]. Available:
  \url{https://cw.infinibandta.org/document/dl/7781}
\BIBentrySTDinterwordspacing

\bibitem{iwarp}
R.~Sharp, H.~Shah, F.~Marti, A.~Eiriksson, and W.~Noureddine, ``{RFC 7306:
  Remote Direct Memory Access (RDMA) Protocol Extensions},'' \emph{Request for
  Comments, IETF}, 2014.

\bibitem{iwarp_vs_roce}
\BIBentryALTinterwordspacing
\emph{{RoCE vs. iWARP Competitive Analysis}}. [Online]. Available:
  \url{http://www.mellanox.com/related-docs/whitepapers/WP\%5FRoCE\%5Fvs\%5FiWARP.pdf}
\BIBentrySTDinterwordspacing

\bibitem{qcn_original}
R.~Pan, B.~Prabhakar, and A.~Laxmikantha, \emph{{QCN: Quantized congestion
  notification an overview}}, 2007.

\bibitem{bolt}
B.~Stephens, A.~L. Cox, A.~Singla, J.~Carter, C.~Dixon, and W.~Felter,
  ``{Practical DCB for improved data center networks},'' \emph{IEEE INFOCOM
  2014-IEEE Conference on Computer Communications}, pp. 1824--1832, 2014.

\bibitem{facebook-dc-count-2016}
\BIBentryALTinterwordspacing
\emph{{Facebook in Los Lunas: Our newest data center}}. [Online]. Available:
  \url{https://code.facebook.com/posts/337730426564112/facebook-in-los-lunas-our-newest-data-center/}
\BIBentrySTDinterwordspacing

\bibitem{microsoft-azure-backbone}
\BIBentryALTinterwordspacing
M.~Russinovich, \emph{{Inside Microsoft Azure datacenter hardware and software
  architecture}}, 2017. [Online]. Available:
  \url{https://myignite.microsoft.com/videos/54962}
\BIBentrySTDinterwordspacing

\bibitem{swan}
C.-Y. Hong, S.~Kandula, R.~Mahajan, M.~Zhang, V.~Gill, M.~Nanduri, and
  R.~Wattenhofer, ``{Achieving high utilization with software-driven WAN},''
  \emph{ACM SIGCOMM Computer Communication Review}, vol.~43, no.~4, pp. 15--26,
  2013.

\bibitem{facebook-express-backbone}
\BIBentryALTinterwordspacing
\emph{{Building Express Backbone: Facebook's new long-haul network}}. [Online].
  Available:
  \url{https://code.facebook.com/posts/1782709872057497/building-express-backbone-facebook-s-new-long-haul-network/}
\BIBentrySTDinterwordspacing

\bibitem{95percentile}
\BIBentryALTinterwordspacing
\emph{{95th percentile bandwidth metering explained and analyzed}}. [Online].
  Available:
  \url{http://www.semaphore.com/blog/94-95th-percentile-bandwidth-metering-explained-and-analyzed}
\BIBentrySTDinterwordspacing

\bibitem{jetway}
Y.~Feng, B.~Li, and B.~Li, ``{Jetway: Minimizing costs on inter-datacenter
  video traffic},'' \emph{ACM international conference on Multimedia}, pp.
  259--268, 2012.

\bibitem{mbdt_initial}
Y.~Wang, S.~Su \emph{et~al.}, ``{Multiple bulk data transfers scheduling among
  datacenters},'' \emph{Computer Networks}, vol.~68, pp. 123--137, 2014.

\bibitem{mbdt}
S.~Su, Y.~Wang, S.~Jiang, K.~Shuang, and P.~Xu, ``{Efficient algorithms for
  scheduling multiple bulk data transfers in inter-datacenter networks},''
  \emph{International Journal of Communication Systems}, vol.~27, no.~12, 2014.

\bibitem{grease}
T.~Nandagopal and K.~P. Puttaswamy, ``{Lowering inter-datacenter bandwidth
  costs via bulk data scheduling},'' \emph{Cluster, Cloud and Grid Computing
  (CCGrid)}, pp. 244--251, 2012.

\bibitem{dtb}
N.~Laoutaris, G.~Smaragdakis, R.~Stanojevic, P.~Rodriguez, and R.~Sundaram,
  ``{Delay-tolerant Bulk Data Transfers on the Internet},'' \emph{IEEE/ACM
  TON}, vol.~21, no.~6, 2013.

\bibitem{postcard}
Y.~Feng, B.~Li, and B.~Li, ``{Postcard: Minimizing costs on inter-datacenter
  traffic with store-and-forward},'' \emph{International Conference on
  Distributed Computing Systems Workshops}, pp. 43--50, 2012.

\bibitem{ecoflow}
Y.~Lin, H.~Shen, and L.~Chen, ``{EcoFlow: An Economical and Deadline-Driven
  Inter-Datacenter Video Flow Scheduling System},'' \emph{Proceedings of the
  23rd ACM international conference on Multimedia}, pp. 1059--1062, 2015.

\bibitem{netstitcher}
N.~Laoutaris, M.~Sirivianos, X.~Yang, and P.~Rodriguez, ``{Inter-datacenter
  bulk transfers with netstitcher},'' \emph{ACM SIGCOMM Computer Communication
  Review}, vol.~41, no.~4, pp. 74--85, 2011.

\bibitem{dynamic_pricing}
\BIBentryALTinterwordspacing
V.~Jalaparti, I.~Bliznets, S.~Kandula, B.~Lucier, and I.~Menache, ``{Dynamic
  Pricing and Traffic Engineering for Timely Inter-Datacenter Transfers},''
  \emph{Proceedings of the 2016 Conference on ACM SIGCOMM 2016 Conference}, pp.
  73--86, 2016. [Online]. Available:
  \url{http://doi.acm.org/10.1145/2934872.2934893}
\BIBentrySTDinterwordspacing

\bibitem{swan-backbone}
\BIBentryALTinterwordspacing
\emph{{How Microsoft builds its fast and reliable global network}}. [Online].
  Available:
  \url{https://azure.microsoft.com/en-us/blog/how-microsoft-builds-its-fast-and-reliable-global-network/}
\BIBentrySTDinterwordspacing

\bibitem{bwe}
A.~Kumar, S.~Jain, U.~Naik, A.~Raghuraman, N.~Kasinadhuni, E.~C. Zermeno, C.~S.
  Gunn, J.~Ai, B.~Carlin, M.~Amarandei-Stavila \emph{et~al.}, ``{BwE: Flexible,
  hierarchical bandwidth allocation for WAN distributed computing},'' \emph{ACM
  SIGCOMM Computer Communication Review}, vol.~45, no.~4, pp. 1--14, 2015.

\bibitem{google-optical-network}
X.~Zhao, V.~Vusirikala, B.~Koley, V.~Kamalov, and T.~Hofmeister, ``{The
  prospect of inter-data-center optical networks},'' \emph{IEEE Communications
  Magazine}, vol.~51, no.~9, pp. 32--38, September 2013.

\bibitem{shapingvspolicing}
\BIBentryALTinterwordspacing
\emph{{Comparing Traffic Policing and Traffic Shaping for Bandwidth Limiting}}.
  [Online]. Available:
  \url{https://www.cisco.com/c/en/us/support/docs/quality-of-service-qos/qos-policing/19645-policevsshape.html}
\BIBentrySTDinterwordspacing

\bibitem{utube}
V.~K. Adhikari, S.~Jain, Y.~Chen, and Z.-L. Zhang, ``{Vivisecting youtube: An
  active measurement study},'' \emph{INFOCOM}, pp. 2521--2525, 2012.

\bibitem{netflix}
\BIBentryALTinterwordspacing
R.~Meshenberg, N.~Gopalani, and L.~Kosewski, \emph{{Active-Active for
  Multi-Regional Resiliency}}, 2013. [Online]. Available:
  \url{http://techblog.netflix.com/2013/12/active-active-for-multi-regional.html}
\BIBentrySTDinterwordspacing

\bibitem{dccast}
\BIBentryALTinterwordspacing
M.~Noormohammadpour, C.~S. Raghavendra, S.~Rao, and S.~Kandula, ``{DCCast:
  Efficient Point to Multipoint Transfers Across Datacenters},'' \emph{9th
  {USENIX} Workshop on Hot Topics in Cloud Computing (HotCloud 17)}, 2017.
  [Online]. Available:
  \url{https://www.usenix.org/conference/hotcloud17/program/presentation/noormohammadpour}
\BIBentrySTDinterwordspacing

\bibitem{ip_multicast}
M.~Cotton, L.~Vegoda, and D.~Meyer, ``{RFC 5771: IANA guidelines for IPv4
  multicast address assignments},'' \emph{Request for Comments, IETF}, 2010.

\bibitem{multicast-challenges}
B.~Quinn and K.~Almeroth, ``{RFC 3170: IP multicast applications: Challenges
  and solutions},'' \emph{Request for Comments, IETF}, 2001.

\bibitem{nice}
S.~Banerjee, B.~Bhattacharjee, and C.~Kommareddy, ``{Scalable Application Layer
  Multicast},'' \emph{SIGCOMM}, pp. 205--217, 2002.

\bibitem{narada}
Y.-h. Chu, S.~G. Rao, S.~Seshan, and H.~Zhang, ``{A case for end system
  multicast},'' \emph{IEEE Journal on selected areas in communications},
  vol.~20, no.~8, pp. 1456--1471, 2002.

\bibitem{avalanche}
A.~Iyer, P.~Kumar, and V.~Mann, ``{Avalanche: Data center multicast using
  software defined networking},'' \emph{COMSNETS}, pp. 1--8, 2014.

\bibitem{datacast}
J.~Cao, C.~Guo, G.~Lu, Y.~Xiong, Y.~Zheng, Y.~Zhang, Y.~Zhu, C.~Chen, and
  Y.~Tian, ``{Datacast: A Scalable and Efficient Reliable Group Data Delivery
  Service for Data Centers},'' \emph{IEEE Journal on Selected Areas in
  Communications}, vol.~31, no.~12, pp. 2632--2645, 2013.

\bibitem{failure-aware-routing}
\BIBentryALTinterwordspacing
M.~Ghobadi and R.~Mahajan, ``{Optical Layer Failures in a Large Backbone},''
  \emph{Proceedings of the 2016 Internet Measurement Conference}, pp. 461--467,
  2016. [Online]. Available: \url{http://doi.acm.org/10.1145/2987443.2987483}
\BIBentrySTDinterwordspacing

\bibitem{ffc}
H.~H. Liu, S.~Kandula, R.~Mahajan, M.~Zhang, and D.~Gelernter, ``{Traffic
  Engineering with Forward Fault Correction},'' \emph{SIGCOMM}, pp. 527--538,
  2014.

\end{thebibliography}

\begin{IEEEbiography}[{\includegraphics[width=1in,height=1.25in,clip,keepaspectratio]{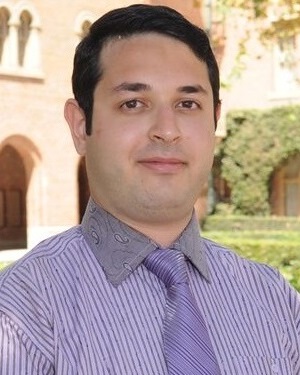}}]%
{Mohammad Noormohammadpour}
is currently a Ph.D candidate in Ming Hsieh Department of Electrical Engineering, University of Southern California (USC). His research is focused on improving the performance and reliability of data transfers across geographically dispersed datacenters. He was awarded with a Fellowship when he joined USC in 2014 soon after receiving his Bachelors degree in Electrical Engineering from Sharif University of Technology. He completed his Masters degree in Computer Science at USC in 2017. During summer of 2016, Mohammad interned at Cisco Innovation Center, located in Cambridge, Massachusetts, where he worked on distributed coded storage systems. Broadly speaking, his research interests are Networked Systems, Datacenters and Software Defined Networking.
\end{IEEEbiography}

\vskip 0pt plus -1fil

\begin{IEEEbiography}[{\includegraphics[width=1in,height=1.25in,clip,keepaspectratio]{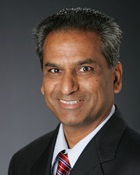}}]%
{Cauligi S. Raghavendra}
is a Professor of Electrical Engineering and Computer Science and is Vice Dean for Global Academic Initiatives for the Viterbi School of Engineering at the University of Southern California, Los Angeles. He was Chairman of EE-Systems Department from 2003-2005, Senior Associate Dean for Academic Affairs during 2005-2006, and Senior Associate Dean for Strategic Initiatives during 2006-2011. Previously, he was a faculty in the Department of Electrical Engineering-Systems at USC from 1982-1992, as Boeing Chair Professor of Computer Engineering in the School of Electrical Engineering and Computer Science at the Washington State University in Pullman, from 1992-1997, and with The Aerospace Corporation from August 1997-2001. He received the B.Sc physics degree from Bangalore University in 1973, the B.E and M.E degrees in Electronics and Communication from Indian Institute of Science, Bangalore in 1976 and 1978 respectively. He received the Ph.D degree in Computer Science from the University of California at Los Angeles in 1982. Dr. Raghavendra is a recipient of the Presidential Young Investigator Award for 1985 and is a Fellow of IEEE. His research interests include Wireless and Sensor Networks, Parallel and Distributed Systems, and Machine Learning.
\end{IEEEbiography}

\end{document}